\documentclass[aps, twocolumn, prx,showpacs, longbibliographye]{revtex4-2}
\usepackage{blkarray}
\usepackage{epsfig}
\usepackage{color}
\usepackage{mhchem}
\usepackage[makeroom]{cancel} 
\usepackage[normalem]{ulem} 
\usepackage{xcolor}
\pagecolor{white}
\usepackage{amssymb, graphics,graphicx, amsfonts, amsmath, empheq}
\usepackage[breakable,most,many]{tcolorbox}
\usepackage{varwidth}
\usepackage{capt-of}
\usepackage[export]{adjustbox}
\usepackage{hyperref}
\usepackage{bm}
\usepackage{comment}
\usepackage{float}

\def\d{\mathrm{d}}
\def\f{\mathfrak{f}}
\def\g{\mathfrak{g}}

\def\phic{\vec{\phi}^{\,(c)}}

\begin{document}
 
\title{Metastable Phase Separation and Information Retrieval in Multicomponent Liquids}

\author{Davide Marcato}
\thanks{These authors contributed equally to this work.}
\affiliation{Gulbenkian Institute of Molecular Medicine, 2780-156 Oeiras, Portugal}

\author{Rodrigo Braz Teixeira}
\thanks{These authors contributed equally to this work.}
\affiliation{Gulbenkian Institute of Molecular Medicine, 2780-156 Oeiras, Portugal }
\affiliation{Centro de Física Teórica e Computacional, Faculdade de Ciências, Universidade de Lisboa, 1749-016 Lisboa, Portugal}

\author{Izaak Neri}
\affiliation{ Department of Mathematics, King’s College London, 
 Strand, London, WC2R 2LS, United Kingdom}

\author{Pablo Sartori}
\affiliation{Gulbenkian Institute of Molecular Medicine, 2780-156 Oeiras, Portugal.}

\date{\today}

\begin{abstract}
Liquid mixtures can separate into phases with distinct composition. This phenomenon has recently come back to prominence due to its  role in complex biological liquids, such as the cytoplasm, which contain thousands of components. For simple two-component mixtures phase-separated states are global free energy minima. However, local free energy minima, i.e., metastable states, are known to play a dominant role in complex systems with many components. For example, Hopfield neural networks can  retrieve  stored information  from partial cues via relaxation to metastable states.
Under what conditions can phase separated states be metastable, and what are the implications for information retrieval in multicomponent liquids?
In this work we develop the general thermodynamic formalism of metastable phase separation. We then apply this formalism to an illustrative toy example inspired by recent experiments, binary mixtures with high-order interactions. Finally, as core application of the formalism, we study metastability in Hopfield liquids, a class of multicomponent mixtures capable of storing information on the composition of phases. We show that these phases can be nucleated from partial cues via metastable phase separation. Spatial simulations of liquids with a large number of components match our analytical solution.
Our work suggests that complex biological mixtures can retrieve information  through metastable phase separation.
\end{abstract}

\maketitle

\section{Introduction}

Phase separation is the phenomenon by which a mixture of multiple components spatially segregates into regions with different composition. Common intuition on  phase separation derives from results for binary mixtures~\cite{rubinstein2003polymer, samsafransbook, atkins2023atkins}. Yet, some of the liquids most studied today, such as the cellular cytoplasm, nucleoplasm, and plasma membrane, consist of thousands of different components~\cite{schaffer2025multimodal}. These liquids segregate into phases of distinct composition, called biomolecular condensates~\cite{banani2017biomolecular}, that coexist with each other. For example, in the nucleoplasm  a dozen condensates, including  Cajal bodies, paraspeckles, and nucleoli, have been identified~\cite{sabari2020biomolecular, you2020phasepdb}. These condensates can be composed of hundreds of distinct macromolecular components~\cite{wang2020organization,andersen2002directed}, and several of these components are shared among different condensates~\cite{machyna2013cajal}. Given that a mixture with $N\gg1$ components allows for a vast number of possible phases, it is remarkable that biological liquids reliably separate into phases with specific compositions~\cite{braz2024liquid, sear2003instabilities, jacobs2013predicting, thewes2023composition, zwicker2022evolved}. Due to the high complexity of this problem, a general theoretical framework capable of predicting the emergent phase separated states of a mixture with many components is lacking.

\begin{figure}[htbp!]\centering
\includegraphics[]{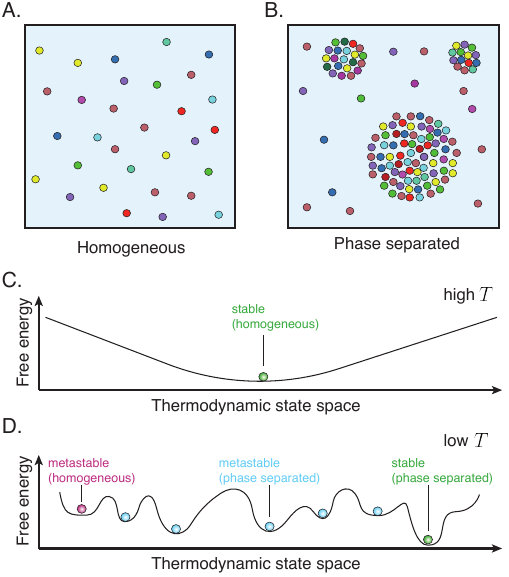}
\caption{{\it Metastable states in multicomponent mixtures.} {\bf A.} Schematic of a homogeneous state in a multicomponent mixture (solutes represented by colored circles, solvent by blue background). {\bf B.} Schematic of a  phase separated state in a multicomponent mixture. Several droplets of diverse sizes and compositions can coexist in such states. {\bf C.} Representation of the free energy landscape for a mixture with many components at fixed state parameters. At high temperature, there is only one stable state, which is the homogeneous state. {\bf D.} Same as C, but at lower temperatures. In this case, the homogeneous state is only one of many possible metastable states. All other metastable states are phase separated, including the global minimum corresponding with the thermodynamic equilibrium.}
\label{fig:scheme}
\end{figure}

The thermodynamics of a liquid mixture can be characterized by a free energy functional, which for $N\gg1$ is defined on a high-dimensional space~\cite{callen1998thermodynamics}. At high temperature this functional has a single global minimum, the homogeneous phase, which is a state of thermodynamic equilibrium (Fig.~\ref{fig:scheme}A and C). In contrast,  at low temperature the free energy functional can have multiple metastable states, as functions in high dimensions can have many minima. These states are stable towards small thermal fluctuations and can be long lived. Metastability in liquids is commonly associated with homogeneous states, because for simple binary mixtures the homogeneous states are metastable between the binodal and spinodal lines~\cite{rubinstein2003polymer, samsafransbook, atkins2023atkins}. However, generically, phase separated states can also be metastable, and therefore at low temperature mixtures with $N\gg1$ can exhibit many metastable phase separated states (Fig.~\ref{fig:scheme}B and D). 

Besides the above general argument, there are multiple reasons why metastability is expected to play a central role in the physics of multicomponent mixtures. We highlight two. First, metastable states are abundant in  complex systems with a large number of components and/or heterogeneous interactions, such as proteins~\cite{frauenfelder1991energy,onuchic1997theory}, spin glasses~\cite{mezard1987spin,charbonneau2023spin}, granular materials~\cite{jaeger1996granular}, or Hopfield neural networks~\cite{hertz2018introduction}, and recent work shows that metastable states are also abundant in multicomponent mixtures~\cite{braz2024liquid, qiang2024generic}. Second, it has been experimentally observed that some biological liquid mixtures phase separate at short times, yet transition to a different equilibrium phase at longer times~\cite{patel2015liquid, lin2015formation, das2025tunable, michaels2023amyloid}. Therefore, based on analogy with other complex systems and recent experiments, we expect metastability to be a crucial ingredient of phase separation in multicomponent mixtures.

In this paper, we develop the general thermodynamic formalism of metastable phase separation in multicomponent mixtures.  We derive the conditions for stationarity and metastability of phase separated states in multicomponent mixtures. We then discuss two applications. First, we develop a simple model for metastable phase separation in binary mixtures, and we show that this model exhibits the same  metastable phase separation phenomenology as observed in experiments with prion-like proteins~\cite{patel2015liquid, lin2015formation, das2025tunable, michaels2023amyloid}. As a second   application of metastable phase separation,  we develop a liquid version of an associative memory, previously named liquid Hopfield model \cite{braz2024liquid, du_et_al}, that can demix into metastable phases that are enriched and depleted in a  prescribed set of components. Note that this is different from the analysis of the liquid Hopfield model in Ref.~\cite{braz2024liquid}, which did not involve phase separation but retrieval of homogeneous states. As we show, these target phases can be retrieved through the nucleation of seeds that are enriched in some but not all the  prescribed components, even though the system has a large number of metastable states. 
This mechanism of controlled  phase separation is similar to  memory retrieval in the Hopfield model~\cite{amari1972learning,  pastur1977exactly, hopfield1982neural, hertz2018introduction}. It implies that the number of metastable non-target states is small enough so that the system does not exhibit a glassy dynamics. 

The paper is structured as follows. In sections \ref{sec:thermo} to \ref{sec:stab} we develop the thermodynamic formalism of metastable phase separation. Section \ref{sec:nlin}  analyzes a simple but illustrative example, binary mixtures with higher-order interactions. 
Section \ref{sec:Hopfield} investigates the retrieval of phases of prescribed compositions through nucleation in Hopfield liquids, which are a liquid analog of associative memory.  Finally, section~\ref{sec:disc} discusses our results in the context of existing literature.

\section{Thermodynamics of multicomponent mixtures}
\label{sec:thermo}
We start by presenting the thermodynamics of an incompressible multicomponent liquid mixture.  We  focus on the canonical ensemble in which the system of interest is contained in a rigid vessel, in contact with a thermal reservoir, and prevented from any material exchanges with the outside (see Supplemental Material (SM) Sec. S1 for an equivalent analysis in the grand canonical ensemble). We  discuss two sets of states for the mixture: spatially homogeneous states and spatially heterogeneous states.   We call  the latter   phase separated states.   In the present section,  these states are  discussed without reference to their stationarity or stability, as stationarity and stability will be addressed in subsequent sections. 

\subsection{Spatially homogeneous states}
Consider a spatially homogeneous multicomponent mixture  of $N$ distinct solute components, labeled by $i=1,\ldots,N$, that are dissolved in a solvent component, labeled by $i=0$.  The mixture is  in contact with a thermal reservoir at  temperature $T$, and is confined within  a rigid vessel of volume $V$. The vessel is closed, and we assume that no chemical reactions occur: therefore, the number of molecules of the solvent, $M_0$, and of the $N$ solute components, $\vec{M}=(M_1, \ldots,M_N)$, are fixed. We additionally assume that the solvent and solutes are all incompressible, which implies that the number of molecules of one of the species, which we take to be $M_0$, is not needed to specify the state of the system, see Appendix~\ref{app:euler}. Thus, the $N+2$ thermodynamic state parameters are $\vec{M}$, $T$, and $V$.

The thermodynamics of this system is described by the Helmholtz free energy~\cite{callen1998thermodynamics, fermi2012thermodynamics}.
\begin{align}\label{eq:freeenergy}
F= F(T,V,\vec{M}),
\end{align}
 The free energy density, $f=F/V$, depends only on $N+1$ variables
\begin{align}
f=f(T,\vec{\phi}), \label{eq:free}
\end{align}
where we have defined the volume fractions of the solute components  as  $\phi_i = \nu_iM_i/V$, with $\nu_i$ the molecular volumes.  As we focus on the incompressible case, the $\nu_i$ are constant (i.e. independent of $T$, $V$ and $\vec{M}$), see  Appendix~\ref{app:euler}. The volume fraction of the solvent is given by $\phi_0=1-\sum^N_{i=1}\phi_i$. Therefore, the parameters $\vec{\phi}$ are subject to the constraint $\sum_{i=1}^N\phi_i\le1$. In what follows, we will refer to $\vec{\phi}$ as the composition vector, unless otherwise specified.

The dependent intensive thermodynamic quantities can be computed in terms of partial derivatives of $f(T,\vec{\phi})$. For example, the exchange chemical potentials of the solutes are
\begin{align}\label{eq:chem}
\bar{\mu}_i= {\nu_i}\left(\frac{\partial f}{\partial \phi_i}\right)_{T,\vec{\phi}'},
\end{align}
where $'$ denotes that the vector excludes component $i$; and the osmotic pressure is
\begin{align}\label{eq:pressM}
\Pi=-f+\vec{\mu}\cdot\vec{\rho},
\end{align}
where we have introduced the number densities of the solutes, $\rho_i=\phi_i/\nu_i$.  In (\ref{eq:pressM}), $\vec{\mu} = \vec{\overline{\mu}}$   is the  vector of exchange potentials.   The usage of osmotic pressure and exchange chemical potential arises from incompressibility. In the case of chemical potentials, a change in the abundance of a solute must be compensated by a change in the abundance of a solvent, so the total volume does not change. Analogously, the change in volume required to compute the pressure is associated with adding or removing solvent molecules. 

In sections \ref{sec:stat_cond}--\ref{sec:stab}, which develop the general formalism of phase separation, we will not consider any specific form of the free energy of homogeneous states. In later sections, we will consider specific functional forms of $f(T,\vec{\phi})$ that correspond to the continuous limit of a lattice gas \cite{rubinstein2003polymer}. These are best described by separating the free energy density into energetic and entropic contributions, i.e., $f=u-Ts$. Without loss of generality,  we take for the energy density  a function of the form 
\begin{align}
u(\vec{\phi})=-\frac{v_2}{2\nu_0} \sum_{ij}J_{ij}\phi_i\phi_j+\frac{v_3}{6\nu_0}\sum_{ijk}K_{ijk}\phi_i\phi_j\phi_k+\ldots \label{eq:genInt}
\end{align}
where the affinity matrix $J_{ij}$ determines the pairwise interactions, the tensor $K_{ijk}$ determines the three-body interactions, etc. The strength of the interactions is determined by $v_2$, $v_3$, etc., which have units of energy. The entropy density  is taken to be
\begin{align}
s(\vec{\phi}) &=-\frac{k_{\rm B}}{\nu_0}\sum^N_{i=1}\phi_i\log(\phi_i)-\frac{k_{\rm B}}{\nu_0}\phi_0\log(\phi_0), \label{eq:entropy}
\end{align}
where $k_{\rm B}$ is Boltzmann’s constant and  the second term corresponds to the entropic contribution of the solvent. In this expression we have used the simplifying assumption that the molecular volumes of all solutes are equal to that of the solvent, i.e. $\nu_i=\nu_0$. In what follows we will measure energy in units of $k_{\rm B}T$. We do not set $\nu_0=1$, as later we shall consider a spatially resolved model.

\subsection{Phase separated states}
We now describe the thermodynamics of  multicomponent mixtures  in phase separated states. Such states can be described through multiple compartments, labeled by  $c=1,\ldots,C$, each representing a homogeneous state of the mixture.  Each compartment is characterized by a composition vector, $\phic$, and the volume of the compartment, $V^{(c)}$. Note that multiple compartments can share the same composition vector. As will become clear later, this redundancy is necessary because the number of different phases is not known a priori.

We associate to phase separated states the  free energy functional  $\mathfrak{F} = \sum_{c=1}^C V^{(c)}f(T,\phic)$.   Here, ``fraktur'' fonts are used for spatially heterogeneous free energies. Defining the intensive free energy functional as $\f=\mathfrak{F}/V$, we obtain 
\begin{align}\label{eq:f_comp}
\f(T, \{w^{(c)}\},\{\phic\}) = \sum_{c=1}^C w^{(c)}f(T,\phic),
\end{align}
where we have defined the volume fractions of the compartments as $w^{(c)}=V^{(c)}/V$. For simplicity, we refer in what follows to the $w^{(c)}$ as the  compartment volumes.  Above, and in what follows, we use curly brackets to denote vectors for which the entries are associated with the  compartments. Furthermore, with slight abuse of notation, we will use the super-index $\,^{(c)}$ to denote functions evaluated at $\phic$, e.g. $f^{(c)}=f(T,\phic)$.    Notice that the functional \eqref{eq:f_comp} does not include a contribution due to surface tension. In  Secs.~\ref{sec:nlin} and \ref{sec:Hopfield} we will include surface tension into the model, as we will consider the effect of space on phase separation.

As we are  working in the canonical ensemble with parameters $(T,V,\vec{M})$, the variables $w^{(c)}$ satisfy the 
volume constraint
\begin{align}\label{eq:volfrac}
\sum^C_{c=1}w^{(c)}=1,
\end{align}
which we will often use to express the volume of compartment $C$ as $w^{(C)}=1-\sum_{c<C}w^{(c)}$. The variables $\{\vec{\phi}^{(c)}\}$ satisfy the constraints 
\begin{align}\label{eq:const}
\sum^C_{c=1} w^{(c)} \phi^{(c)}_i = \phi_i.
\end{align}
As the variables, $\{w^{(c)},\phic\}$, span an $(N+1)C$ dimensional space, and Eqs.~\eqref{eq:volfrac} and \eqref{eq:const} constitute $N+1$ constraints, the allowed subspace for the variables has $(C-1)(N+1)$ dimensions.

The thermodynamic equilibrium state of the liquid mixture is obtained by minimizing the functional $\f$ with respect to its variables $\{w^{(c)}\}$, $\{\phic\}$ and $C$, subject to the constraints in Eqs.~\eqref{eq:volfrac} and \eqref{eq:const}. The  minima of $\f$ represent the equilibrium state, which is a function of the external parameters.   The number of phases $P$ in the equilibrium configuration equals the number of compartments with different compositions.   Notice that there exist redundant representations of the equilibrium configurations for which $C>P$, and thus $\f$ has multiple minima representing the equilibrium configuration. As already mentioned, since a priori the number of phases $P$ in the equilibrium state is  not known, it is necessary to allow for this redundancy.      

\section{Stationarity conditions}
\label{sec:stat_cond}
As anticipated in the Introduction,   metastable states are expected to play an important role in the physics of phase separation in multicomponent mixtures.    To initiate the study of metastable states, we derive in this Section conditions for stationarity in  multicomponent mixtures. Note that a  stationary state can be either a minimum, maximum, or saddle point of the free energy functional.  

Phase separated states are stationary if  conditions of chemical and mechanical balance are satisfied.   Chemical balance implies that the 
exchange chemical potentials of any given species $i$ are the same across all compartments, i.e.,
\begin{align}\label{eq:chem_bal}
\bar{\mu}_i^{(c)}=\bar{\mu}_i^{(c')} 
\end{align}
for all pairs of compartments $c$ and $c'$. 
Mechanical balance implies that the osmotic pressures of all compartments are identical, and so
\begin{align}\label{eq:press_bal}
\Pi^{(c)} = \Pi^{(c')}
\end{align}
for all pairs of compartments $c$ and $c'$. 

We remark that the above conditions are necessary conditions for minimizing the free energy in Eq.~\eqref{eq:f_comp}, but they are not sufficient. This crucial distinction is sometimes overlooked in the literature \cite{zwicker2025physics, zwicker2022intertwined}, where it is  stated that when Eqs.~\eqref{eq:chem_bal} and~\eqref{eq:press_bal} are satisfied the free energy is minimized. However, it is entirely possible that Eqs.~\eqref{eq:chem_bal} and~\eqref{eq:press_bal} are satisfied in a maximum or a saddle of the free energy. To guarantee that a stationary point is a minimum additional conditions must be tested, as we discuss later.

To derive the above conditions, we determine how  the free energy $\f$ changes in response to a small arbitrary perturbation that complies with the constraints of the system.    In particular, consider a state $(\{w^{(c)}\},\{\phic\})$, and a perturbed state near it with compartment volumes $\tilde{w}^{(c)}={w}^{(c)}+{\epsilon}^{(c)}$ and compositions $\tilde{\phi}_i^{(c)}={\phi}_i^{(c)} + \delta_i^{(c)}$, where  $|\epsilon^{(c)}|\ll 1$ and $|\delta_i^{(c)}|\ll 1$ are perturbation parameters. The perturbation must comply with the constraints in Eqs.~\eqref{eq:volfrac} and \eqref{eq:const}, which render $\delta_i^{(C)}$ and $\epsilon^{(C)}$  perturbation dependent functions, see Appendix~\ref{app:quadform} for explicit expressions. Then, we evaluate the change in free energy, $\Delta \f = \tilde{\f}-\f$, to linear order in the perturbation, where $\tilde{\f}$ is the free energy of the perturbed state.   The stationarity conditions are obtained by setting all linear terms in the expression of  $\Delta\f$ to zero, see Appendix~\ref{app:stat}. Setting to zero the terms in $\delta_i^{(c)}$ returns the chemical balance conditions \eqref{eq:chem_bal}, and  setting to zero the terms in $\epsilon^{(c)}$ returns the mechanical balance conditions \eqref{eq:press_bal}.   In conclusion, the stationary states of the liquid mixture are found by solving Eqs.~\eqref{eq:chem_bal} and \eqref{eq:press_bal} together with  the constraints ~\eqref{eq:volfrac} and \eqref{eq:const}. Notice that a homogeneous state is characterized by the same composition in each compartment: hence, the chemical and mechanical balance conditions are always satisfied, and all homogeneous states are  stationary (see SM Sec. S2 for a more detailed discussion).

Equations~\eqref{eq:chem_bal} and~\eqref{eq:press_bal} have a useful geometrical interpretation: solving them is equivalent to finding a set of points of the free energy surface that share a common tangent hyperplane (see Appendix~\ref{app:stat}). Different tangent hyperplanes correspond  to different stationary states, see Fig.~\ref{fig:Z4}.

Finally, notice how the fact that Eqs.~\eqref{eq:chem_bal} and~\eqref{eq:press_bal} depend only on the composition vectors $\{\vec{\phi}^{(c)}\}$ ultimately imposes an upper bound to the possible values of $P$. As we count $(N+1)(P-1)$ equations and $PN$ variables, and as the number of equations should be smaller or equal than the number of variables in order to avoid an overconstrained system, we conclude that
\begin{equation}
    P \leq N + 1,
\end{equation} 
which is the celebrated Gibbs phase rule~\cite{atkins2023atkins, fermi2012thermodynamics}.

\section{Stability conditions}
\label{sec:stab}
We next derive stability conditions for  stationary states. These  determine whether a particular stationary state is a maximum, saddle, or minimum.  We are particularly interested in minima, of which there are two kinds: local minima, which represent metastable states; and global minima, which represent thermodynamic equilibrium states. Although metastability is commonly associated with homogeneous states, phase-separated states may also be metastable, see Refs.~\cite{Cahn1969,Evans1997,evans1997diffusive,Olmsted1998} for examples.   Also, as briefly mentioned in the Introduction, metastable liquid droplets have been observed in {\it in vitro} experiments involving prion-like proteins or peptides at physiological concentrations~\cite{patel2015liquid, lin2015formation, das2025tunable, michaels2023amyloid}.   Here we derive the conditions for the metastability of phase separated states in multicomponent mixtures.
 
The key result of this section is a proof of the following statement: a  phase separated state  with $P$ phases is metastable if and only if  the hessians $\mathbf{h}^{(c)} = \mathbf{h}(\vec{\phi}^{(c)})$  for each of the compartments corresponding to different phases are positive definite. Hence, metastability of a phase separated state is equivalent (necessary and sufficient) to metastability of the phases it is composed of.  Note that for phase separated states there are two kinds   of  perturbations that respect the constraints of the canonical ensemble: those  that change  the volumes and compositions of the existing phases, and those that nucleate new phases. 

We first show that if the hessians $\mathbf{h}^{(c)}$ are all positive definite, then the free energy  of the mixture cannot decrease by implementing one  of these perturbations. This demonstrates that metastability of all phases is sufficient for metastability of the phase separated state. To this purpose, we analyze the convexity of the free energy, $\f$, in terms of its variables $\{w^{(c)}\}$ and $\{\vec{\phi}^{(c)}\}$. As in the case of stationarity, this convexity analysis is subject to the constraints in Eqs.~\eqref{eq:volfrac} and \eqref{eq:const}. We therefore compute the free energy difference $\Delta\f$  to second order in the perturbation (see Appendix \ref{app:quadform}).   Due to stationarity the linear terms vanish, and the remaining quadratic term can be written as  (see  Appendix~\ref{app:stab})
\begin{align}
    \Delta \f(\{\vec{\delta}^{\,(c)}\}',\{\epsilon^{(c)}\}') = \frac12\sum^C_{c=1} w^{(c)}\sum^N_{i,j=1}\delta_i^{(c)}\delta_j^{(c)} h_{ij}^{(c)}, \label{eq:freeEnergyChange}
\end{align} 
where the prime indicates that there is no explicit dependence on the $C$-th volume or composition perturbations. This latter fact is due to the   constraints in Eqs.~\eqref{eq:volfrac} and \eqref{eq:const}, which allows us to write
\begin{align}\label{eq:deltaiCM}
\delta_i^{(C)} 
= -\sum^{C-1}_{c=1} \delta^{(c)}_{i}\frac{w^{(c)}}{w^{(C)}} 
+ \sum^{C-1}_{c=1} \epsilon^{(c)} \frac{(\phi_i^{(C)}-\phi_i^{(c)})}{w^{(C)}}, 
\end{align}
and $\epsilon^{(C)}=-\sum_{c=1}^{C-1}\epsilon^{(c)}$.
From Eq.~\eqref{eq:freeEnergyChange} it follows that, in general, $\Delta \f$ is positive if all the hessians $\mathbf{h}^{(c)}$ are positive definite.   

 Note that  Eq.~\eqref{eq:freeEnergyChange} does not rule out the possibility of  
 \textit{soft modes}.   These are perturbations  that alter the volume fractions  without altering the composition of the phases, so that $\delta_{i}^{(c)} = 0$ for every $i$ and every $c$, while some of the $\epsilon^{(c)}\neq 0$. These soft modes   leave  the  free energy unaltered, $\Delta \f = 0$,  even if  the hessians are positive definite, and for such liquid mixtures the stable state is degenerate.    Such soft modes occur in the liquid Hopfield model of Sec.~\ref{sec:Hopfield}.

We next show that if one phase has a hessian with negative eigenvalues, then there will always be a perturbation that decreases the free energy. This demonstrates that metastability of all phases is not just sufficient, but also necessary for metastability of the phase separated state. A naive look at Eq.~(\ref{eq:freeEnergyChange}) may suggest that  a state can be metastable, even if one phase is unstable, viz., because the Hessians of the other phases have positive eigenvalues.   However,  this argument does not consider perturbations that nucleate a new phase.  If one phase has a negative eigenvalue, then the free energy of the liquid mixture can  be reduced by nucleating a new phase inside the unstable phase, a type of instability that requires using $C>P$. This instability is similar to that of  homogeneous states when it crosses the spinodal boundary, but here it applies for phase separated states (see Appendix~\ref{app:necessary} for details).     

\begin{figure*}[]\centering
\includegraphics[width=1\textwidth]{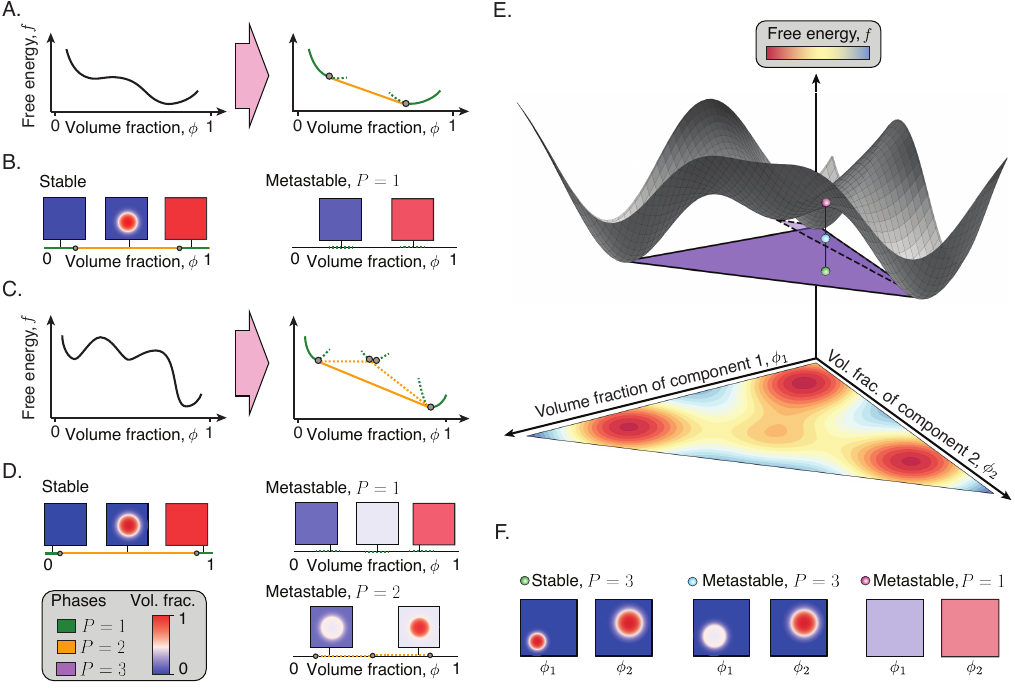}\caption{
\textit{Schematic illustration of the geometry underlying metastable phase separation.} \textbf{A.} Standard binary mixtures admit a single phase separated state, which can be found by the common tangent construction (orange line connecting two gray circles). The only metastable states are homogeneous states (dashed green line) {\bf B.} The geometry of the state space of a binary mixture is a straight line. The phase separated state is stable at intermediate solute volume fractions, with high and low volume fractions corresponding to stable homogeneous phases. In addition, there are two metastable regions corresponding to homogeneous phases. The square images show schematics of 2D spatial simulations of the system.  {\bf C.} In presence of non-linear interactions, binary mixtures may admit multiple phase separated states, corresponding to multiple common tangent constructions (three in this example, corresponding to the orange solid line and the two orange dashed lines). {\bf D} Besides globally stable phase separated states and homogeneous metastable states, the system in {\bf C} also admits two metastable phase separated states. Gray box contains color legend. {\bf E.} For a ternary mixture the state space geometry is a triangle. Due to the increase in dimensionality the system now admits a large variety of common tangent lines and planes (three examples are shown, corresponding to the violet planes and the red point). Of these, only those belonging to the convex hull correspond to stable equilibrium states. {\bf F.} Examples of a stable phase separated state, a metastable phase separated state, and a metastable homogeneous state. The three states correspond to the same thermodynamic parameters, as can be seen by the corresponding points (green, blue, red) that appear in panel E.   This figure is analogous to Fig.~\ref{fig:scheme}D.   However, now we have two species, $N=2$, and thus we need to specify the spatial distribution of both $\phi_1$ and $\phi_2$. 
}
\label{fig:Z4} 
\end{figure*}

Metastable states in multicomponent mixtures have a simple geometric interpretation, as illustrated in 
Fig.~\ref{fig:Z4}. Indeed, metastable states represent simplexes located on  affine, subspaces that are tangent to the free energy function $f(\vec{\phi})$.  The dimension of the tangent affine subspace equals the number of phases that compose the spatially heterogeneous state minus one.   For example, a heterogeneous state  of three phases is represented by a triangle on a  plane tangent to $f(\vec{\phi})$, as illustrated in Fig.~\ref{fig:Z4} (see also Appendix~\ref{app:stat}).

\section{Metastable phase separation in binary mixtures}
\label{sec:nlin}

We revisit metastable phase separation in binary mixtures consisting of a solvent and a single solute~\cite{evans1997diffusive,Evans1997}, as a simple introductory example, before turning to metastable phase separation in multicomponent mixtures in the following section. 
We begin by applying the formalism developed in Sections~\ref{sec:thermo}--\ref{sec:stab} to identify the metastable states. This analytical approach is then validated using continuum-space simulations that incorporate both diffusion and surface tension.

\subsection{Theoretical results}
We consider a mixture with one solute component, $N=1$, that can phase separate from the solvent. The free energy takes the form 
\begin{align}\label{eq:free_bin}
f(\phi) = u(\phi) + \frac{1}{\nu_0}\phi \log \phi + \frac{1}{\nu_0}(1-\phi) \log (1-\phi),
\end{align}
where $\phi=\phi_1$ is the composition vector, here a scalar, and we removed the subindex  $1$ \cite{samsafransbook}.  In this section we will refer to $\phi$ as the solute volume fraction. This model can be found in textbooks for the case in which $u(\phi)$ is quadratic \cite{samsafransbook, rubinstein2003polymer}, of which the Flory-Huggins model is a particular example \cite{flory1942thermodynamics} (see SM Sec. S4). Here we will show that phase separated metastable states can arise when $u(\phi)$ has a higher order non-linearity, as illustrated in Fig.~\ref{fig:Z4}C and D (see also SM Sec. S5  for mathematical details). As an example we consider 
\begin{align}\label{eq:ubin}
u(\phi) = - \frac{c}{12\nu_0}\left(\phi - \frac{1}{2}\right)^4,
\end{align}
where the constant $c$, which has units of energy, quantifies the strength of the interaction.

\begin{figure}[htbp!]\centering
\includegraphics[width=0.5\textwidth]{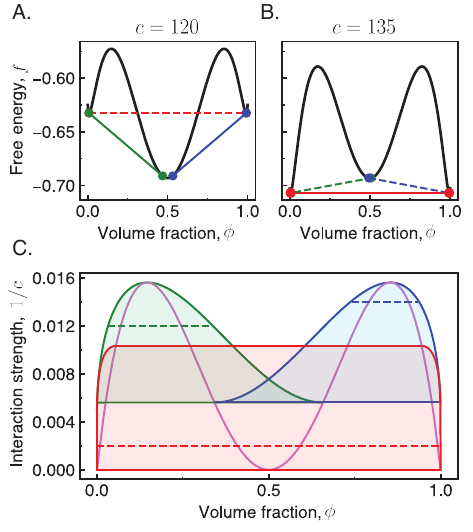}
\caption{{\it Free energy and stability diagram for a binary mixture with quartic potential}. {\bf A}. Free energy density $f(\phi)$, in units of $\nu_0^{-1},$ from Eq.~\eqref{eq:free_bin} with the energy function in Eq.~\eqref{eq:ubin} for $c=120$. Common tangent lines correspond to phase separated states, solid lines are stable and dashed metastable (see SM Sec. S5 for more details). {\bf B}. Same as in A, but for $c=135$. Stability is reversed: metastable states in A become stable, and the stable state in A is now metastable. {\bf C}. Stability diagram 
in the $(c^{-1},\phi)$ parameter space, with   $c^{-1}$
the inverse interaction energy and  $\phi$ the volume fraction of the solute. The spinodal line of the homogeneous state is plotted in pink. There are three additional lines delineating the regions where the three phase separated states are stable:  high and low volume fraction (red - family I), low and intermediate volume fraction (green - family II), and high and intermediate volume fraction (blue - family III).  The corresponding binodal lines can be found in SM Fig. S3. Tie lines are given by horizontal lines connecting coexisting volume fractions in the phase separated states, three examples are shown: one for $c = 71$ (blue dashed line), one for $c =83$ (green dashed line) and another for $c=500$ (red dashed line). A phase diagram can be constructed by selecting the phase of lowest free-energy for each parameter, see SM Fig. S3.}\label{fig:binary_phase_diag} 
\end{figure}

Fig.~\ref{fig:binary_phase_diag}A and B  displays representative plots of $f(\phi)$ for two different values of $c$. Metastable homogeneous states are obtained by identifying convex regions in the graph $f(\phi)$ (represented by dashed lines). Metastable phase separated states are obtained from constructing tangent lines to the graph $f(\phi)$ that connect two points in which the function is convex. Unlike in the case of binary mixtures with a quadratic energy function, this graph allows for three different tangent constructions, given by the  straight lines in Fig.~\ref{fig:binary_phase_diag}A and B, corresponding with three families of phase separated states. One line connects low and high volume fractions (family I, in red), another connects low and intermediate volume fractions (family II, in green), and another connects intermediate and high volume fractions (family III, in blue). Depending on the value of the interaction parameter $c$ and the thermodynamic control parameter $\phi$, the globally stable state will be either a homogeneous state or a phase separated state from one of the families. For example, at $c=120$ and $\phi=0.25$, the equilibrium state is a phase separated state from family II, while phase separated states from family I are metastable. On the other hand, for $c=135$ and $\phi=0.25$, the equilibrium state is a member of family I, and family II is metastable.

Panel C of Fig.~\ref{fig:binary_phase_diag}  shows lines delimiting the regions where the different phases are stable. The spinodal line of the homogeneous phase (solid pink line) separates the region of low $c$ for which the homogeneous state is metastable from the region at high $c$ where it is unstable.  The spinodal of the homogeneous phase is obtained from solving  the equation $f''(\phi_{\rm sp})=0$ towards $\phi_{\rm sp}$, which admits an analytical solution (see SM Sec. S5).
The regions that are enclosed by the other three solid lines  denote the regions for which the  phase separated states are stable (red, green, and blue solid lines, and the corresponding shaded regions, refer to the three families).   These lines are obtained by finding all solutions to the common tangent conditions for different values of $c$.  The three dashed lines correspond to three tie lines that connect the volume fractions of the two phases that compose phase separated states of each family (see SM Sec. S5 for explicit expressions). Observe that at intermediate values of $c$ multiple states are metastable.  In these regions the thermodynamic equilibrium state is found by comparing the free energies of the different metastable states (see Fig.~\ref{fig:binary_phase_diag}).

\begin{figure*}[ht!]\centering
\includegraphics{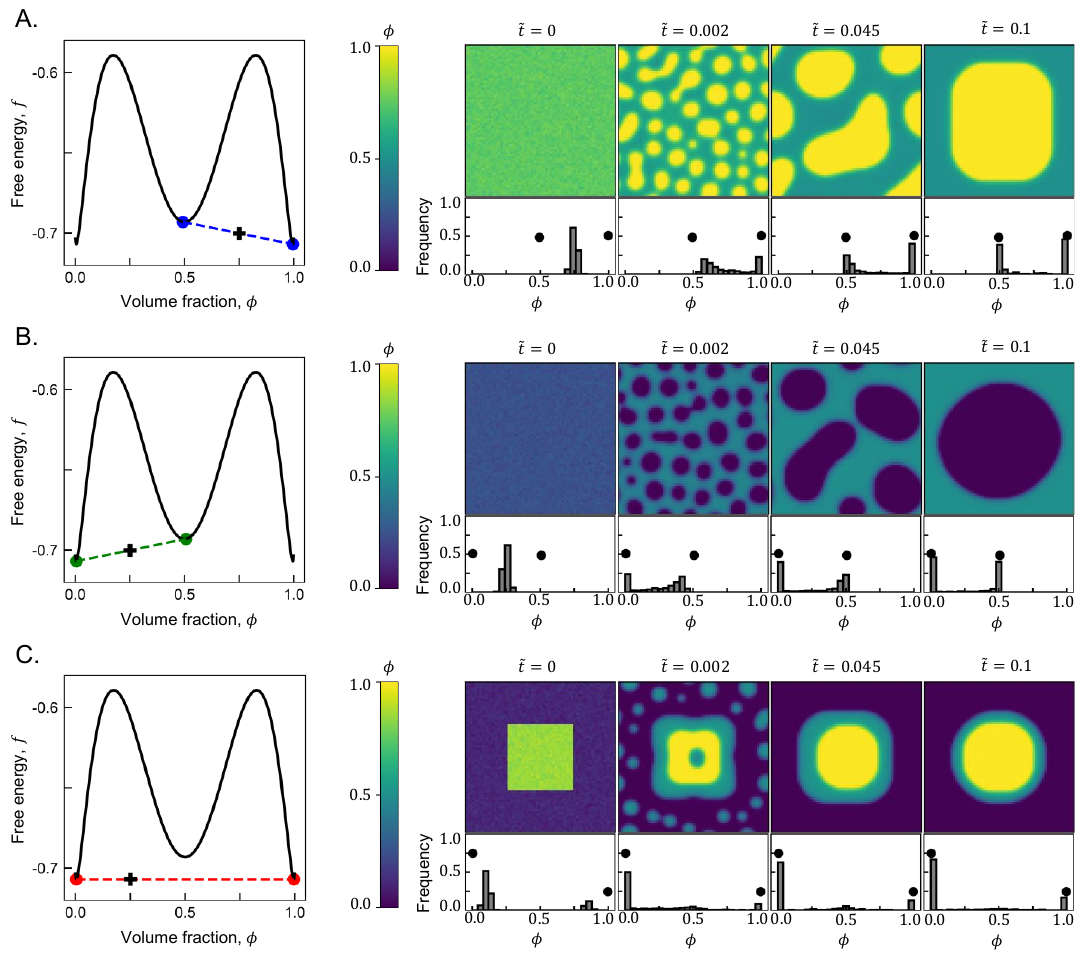}
\caption{{\it Phase separation dynamics for a binary mixture with quartic potential}.  Each Panel (A, B, or C) shows the evolution of the system towards a phase separated state belonging to one of the three possible families (families III, II and I as described in the main text). Left subpanels show the theoretical predictions, and the right subpanels show numerical solutions of the Cahn-Hilliard Eq.~\eqref{eq:dyn}.  \textbf{A.} Phase separation into a family III state from an initial homogeneous state with $\phi = 0.75$. Left panel shows the free energy density $f(\phi)$, in units of $\nu_0^{-1}$, with a cross noting the solute volume fraction. Right panels show snapshots of the time evolution of the Cahn-Hilliard equation at different dimensionless times $\tilde{t}$, with histograms of the solute volume fraction below.
\textbf{B.} Same as A, but the initial state has $\phi = 0.25$, which returns a family II solution. \textbf{C.} Same as A, but in this case solute is concentrated in a patch despite an overall value $\phi = 0.25$, which results in a family I solution. Throughout the figure $c = 135$. See also SM Sec. S6, S7b and S7d for further details.}
\label{fig:sim_quartic}
\end{figure*}

\subsection{Phase separation in continuous space}

The approach we have followed so far treats space as a discrete set of compartments. We now numerically investigate the dynamics of phase separation for the model at hand in continuous space by solving a set of equations analogous to the Cahn-Hilliard equations~\cite{cahn1958free} (see Appendix~\ref{app:dynamics}). 
In brief, we generalize the composition vectors of the solutes across  compartments into vector fields in space, ${\bm r}$, and time, $t$, i.e., $\vec{\phi}^{(c)}\to\vec{\phi}({\bm r},t)$. For the case at hand we simply have one composition field, and its time evolution is given by
\begin{align}\label{eq:dyn}
    \frac{\partial \phi({\bm r},t)}{\partial t} = \ell \nabla^2 \bar{\mu}({\bm r},t),
\end{align}
where $\nabla$ is the gradient towards ${\bf r}$. Here,  $\ell$ is a response coefficient, sometimes called mobility, and $\bar{\mu}({\bm r},t)$ is  a generalization of the exchange chemical potential into continuous space, which takes the form

\begin{align}\label{eq:mu_tilde}
    \bar{\mu}({\bf r},t) = -\frac{c}{3}\left(\phi-\frac{1}{2}\right)^3 + \log\left(\frac{{\phi}}{1-{\phi}}\right)-k\nabla^2\phi. 
\end{align}
The coefficient $k$  determines the width of the interfaces  and  is related to the surface tension (see SM Sec. S6 for a non-dimensional formulation).

Figure~\ref{fig:sim_quartic} shows the dynamics of phase separation for a fixed value of the interaction parameter $c$.   In Panels A and B, the liquid mixture  is  initially prepared into a homogeneous state with some spatial white noise.   For such general initial conditions the system relaxes towards a   phase separated state that is metastable and not globally stable.   In Panel A, with high solute volume fraction, the phase separated state belongs to family III; and in Panel B, with low solute volume fraction, the system relaxes towards a phase separated state of family II.  Note that in both cases the thermodynamic equilibrium is a phase separated state of family I.

Panel C shows an example of an initial state that relaxes towards the equilibrium phase separated state, which is a state of family I.  Note that the initial condition is a specific one for which the solute is concentrated in one region. The transient dynamics is interesting. Initially, different spatial regions of the liquid mixture appear to  evolve towards phase separated states in the families II and III.   However, ultimately the regions of intermediate volume fraction shrink, and the system reaches a phase separated state that is a global free energy minimum.

To conclude, we have shown the emergence of metastable phase separation in a simple model of a binary mixture with  a quartic interaction energy. The numerical results in Fig.~\ref{fig:sim_quartic} indicate that metastable phase separated states are easier to access than the global equilibrium state.   This example thus shows the relevance of metastable phase separated states, in accordance with previous studies~\cite{Cahn1969, Olmsted1998, Evans1997, Cheng2008}. This phenomenology is similar to the one observed in in-vitro experiments with prion-like proteins~\cite{patel2015liquid, lin2015formation, das2025tunable, michaels2023amyloid}, and we elaborate on this in the   discussion further.

\section{Metastable Phase separation in Hopfield liquids}\label{sec:Hopfield}
We next apply the framework for metastable phase separation  to the liquid Hopfield model. This model represents a liquid mixture with a large number of components that can encode in their interactions a large number of phases with prescribed compositions.  The study of such mixtures is interesting from an engineering point of view with the aim of building synthetic liquids capable of information retrieval~\cite{chalk2024learning, chaderjian2025diverse}, and as a toy model for biological mixtures, such as the cytoplasm, that have many components and demix into condensates with specific composition~\cite{banani2017biomolecular}. The liquid Hopfield model has  been studied in  the grand-canonical ensemble~\cite{braz2024liquid}, and the conditions for the stability of a single phase with a prescribed composition have been determined. However, the problem of multi-phase coexistence and separation dynamics in such a model has not yet been discussed. Here,  using the formalism derived in this paper, we study metastable phase separation in the liquid Hopfield model.  

\subsection{Definition of the Liquid Hopfield model}
We begin the model definition by specifying the  compositions in the target phases.   We  consider $p$  target vectors  $\vec{\xi}^{(\alpha)}$ with $\alpha=1,\ldots,p$ a label identifying the different targets. The target vectors prescribe     whether component $i$ should be enriched ($\xi_i^{(\alpha)} = 1$) or depleted ($\xi_i^{(\alpha)} = 0$) in phase $\alpha$.   For simplicity, we assume that the total number of components enriched in each target phase is the same, i.e., $\sum^N_{i=1}\xi_i^{(\alpha)} = Q$ with $Q$ an $\alpha$-independent constant. In this case, we can define normalized target vectors  $\vec{\gamma}^{(\alpha)} = (\vec{\xi}^{(\alpha)} - q)/n$, where  $q = Q/N$  is the sparsity parameter and $n = \sqrt{q(1 - q)}$ is a normalization factor. By construction, the normalized target vectors  satisfy $\vec{\gamma}^{(\alpha)} \cdot \vec{1} = 0$ and   $\vec{\gamma}^{(\alpha)}\cdot  \vec{\gamma}^{(\alpha)}=N$, where $\vec{1}$ is the vector of all unit entries.

In the liquid Hopfield model the   affinity matrix $J_{ij}$, which specifies the energy $u$ in Eq.~\eqref{eq:genInt},  takes a particular form,   so that the liquid demixes into phases of prescribed compositions. Following~\cite{hertz2018introduction, braz2024liquid, giorgio2}, $J_{ij}$ is defined as the  orthogonal projector on the linear space spanned by the $p$ normalized target vectors $\vec{\gamma}^{(\alpha)}$: 
\begin{equation}\label{eq:hebb}
    J_{ij} = \sum_{\alpha=1}^p \gamma_i^{(\alpha)} M_{\alpha j}, 
\end{equation} 
where  $M_{\alpha j}$ is the pseudo-inverse (also known as the Moore-Penrose inverse) of the matrix $\gamma^{(\alpha)}_i$~\cite{Moore,penrose}.   In the specific case when  the $p$ target vectors  $\vec{\gamma}^{(\alpha)}$  are linearly independent, we have that:
\begin{equation}\label{eq:pseudo}
M_{\alpha j} =  \sum^p_{\beta=1} c_{\alpha\beta}^{-1} \gamma_j^{(\beta)} ,
\end{equation}
where $c_{\alpha\beta} = \sum_{i=1}^N \gamma_i^{(\alpha)} \gamma_i^{(\beta)} / N$ is the pattern covariance matrix~\cite{kohonen1972correlation, dotsenko1991statistical}. Inserting Eq.~(\ref{eq:pseudo}) into Eq.~(\ref{eq:hebb}) we recover the form in~\cite{braz2024liquid} for $J_{ij}$. Notice that the inverse of $c_{\alpha\beta}$ does not exist when  the targets are not linearly independent.

Besides the pairwise affinity term, in Ref.~\cite{braz2024liquid} it was shown that the presence of higher order non-linearities enables stability of target states. We therefore use the simple diagonal form
\begin{align}\label{eq:Kijk}
K_{ijk}=N^2\delta_{ij}\delta_{jk}\quad
\end{align}
for the cubic term in Eq.~\eqref{eq:genInt}, and set all remaining terms to zero. Note that the $N-$scaling of $J_{ij}$ and $K_{ijk}$ is such that the quadratic and cubic contributions to the free energy are of the same order of magnitude.   

For simplicity we consider the case for which the total volume fraction occupied by each of the $i$ solute components are identical, i.e.,  $\phi_i=\phi/N$.

\begin{figure}\centering
\includegraphics[width=0.5\textwidth]{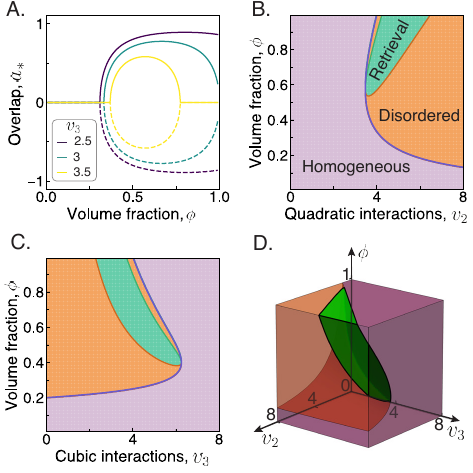}
\caption{{\it Stationarity and stability diagrams of the liquid Hopfield model.} \textbf{A.} Overlap \( a_\ast \) as a function of the volume fraction \( \phi \), for different values of the cubic interaction \( v_3 \) and at fixed  \( \ v_2 = 4 \). The onset of non-zero overlap indicates the emergence of stationary target states.  The quantity $a_\ast$ is obtained by solving Eq.~\eqref{eq:easy1}. \textbf{B.} Spinodal lines in the \( (v_2, \phi) \)-plane, delimiting regions where either the homogeneous state (violet) or all retrieval phases (green) are stable and for fixed \( v_3=3\). In the orange region neither the homogeneous nor  the retrieval states are stable. The curves drawn correspond to solving the equations obtained by replacing the inequalities in  Eqs.~\eqref{eq:stab_targets} and \eqref{eq:phiv2v3} by equalities. \textbf{C.} Same as B. but in the \( (v_3, \phi) \) plane for fixed \( v_2 = 4 \). \textbf{D.} Schematic representation of the stability diagram in the \( v_2, v_3, \phi \) space. For clarity, the small region between the homogeneous and retrieval region was omitted.  } \label{fig:hopf_stab} 
\end{figure}

\subsection{Stationarity of phase separated retrieval states}
\label{sec:stat_PhaseSeparatedHopf}

To investigate metastable phase separation in the liquid Hopfield model, we first consider  phase separated states  with two compartments ($C=2$). Later,  at the end of the section, we discuss the case with $C>2$.   

Let us consider a mixture for which one compartment has an enrichment profile that is similar to the target state $\alpha$. Thus, in this compartment components with $\xi_i^{(\alpha)}=1$ are enriched and components with $\xi_i^{(\alpha)}=0$ are depleted, where $\alpha$ is one of the $p$ prescribed target compositions. We label the phase corresponding with this compartment by $\alpha$, and its compartment volume is denoted by $w^{(\alpha)}$. Due to conservation of component abundances, the  second compartment, labeled by $-\alpha$,  is conversely enriched in the components with $\xi_i^{(\alpha)}=0$ and depleted in those with $\xi_i^{(\alpha)}=1$.  In what follows, we call $\alpha$ the ``{\it $\alpha$-th retrieval phase}''   and we refer to  $-\alpha$  as its ``{\it anti-retrieval phase}''.  We call the  phase separated state consisting of a pair of phases, one retrieval and the other anti-retrieval, {\it the retrieval state}.  

We focus for now on the simple case for which $q=1/2$, while the general case of arbitrary $q$ is discussed in  Appendix~\ref{app:hopfield_stat}.  For $q=1/2$,  if a certain amount of a  component is enriched in a retrieval phase, the exact same amount is depleted in its anti-retrieval phase.  Thus, the compositions of an  $\alpha$-th retrieval phase are given by
\begin{align}\label{eq:drop1}
\phi_i^{(\alpha)} &=\frac{\phi}{N}\left(1+a\gamma_i^{(\alpha)}\right),
\end{align}
and those of its anti-retrieval phase are 
\begin{align}
\phi_i^{(-\alpha)} &=\frac{\phi}{N}\left(1-a \gamma_i^{ (\alpha)}\right)\quad.\label{eq:drop2}
\end{align}

In the expressions \eqref{eq:drop1} and \eqref{eq:drop2}, $a$ is  the {\it overlap parameter} of the retrieval phase that quantifies the quality of retrieval.  The quality of retrieval of an arbitrary  composition vector $\vec{\phi}$ is determined by  the $p$ {\it overlap observables}  
\begin{equation}\label{eq:obs_ov}
a^{(\nu)}(\vec{\phi}) = \frac{\vec{\gamma}^{(\nu)}\cdotp\vec{\phi}}{\vec{1}\cdotp\vec{\phi}}
\end{equation}
with $\nu=1,2,\ldots, p$.  If the targets $\gamma^{(\alpha)}$ are orthogonal, then the retrieval phase $\vec{\phi}^{(\alpha)}$ satisfies  
\begin{equation}
a^{(\nu)}(\vec{\phi}^{(\alpha)}) = a \: \delta_{\nu,\alpha} .\label{eq:longtime}
\end{equation}
Therefore, $a=0$ corresponds to a homogeneous state, i.e., no retrieval; $a\in(0,1)$ to retrieval, i.e., components present in the target have higher densities than those absent in the target; and $a=1$ corresponds to full retrieval, i.e., the components with  $\xi^{(\alpha)}_i= 0$ have been completely removed from the phase.   If the $a^{(\nu)}$ are nonzero for multiple values of $\nu$, then we speak of a {\it spurious state}.

Using Eqs.~\eqref{eq:drop1} and \eqref{eq:drop2} in the constraint for the compositions, Eq.~\eqref{eq:const}, we find that both phases have equal volume,  
\begin{equation}
w^{(\alpha)} = w^{(-\alpha)} = 1/2, 
\end{equation}
which is a consequence of the symmetry of the model.  Instead, for  $q<1/2$ the two compartments have different volumes (see  Appendix~\ref{app:hopfield_stat}).

Next, we demonstrate that a phase separated  state consisting of a pair of phases, $\alpha$ and $-\alpha$, that have equal volumes is stationary if the overlap parameter $a$ takes on a specific value $a^\ast$.   To this aim, we apply the stationarity conditions, Eqs.~\eqref{eq:chem_bal} and~\eqref{eq:press_bal}, to the phase separated state. The exchange chemical potentials of the components in each of the compartments are given by
\begin{align} \label{eq:chem_pot_drop}
 \bar{\mu}_{i}^{(\pm\alpha)} &=
 \mp v_2 \phi a   \gamma_i^{(\alpha)}+\frac{v_3\phi^2}{2}(1 \pm a \gamma_i^{(\alpha)})^{2} \nonumber\\
 &+\log({\phi (1\pm  a \gamma_i^{(\alpha)})})-\log(1-\phi)-\log(N),
\end{align}
and the corresponding osmotic pressures are 
\begin{align}\label{eq:press_a}
\Pi^{(\pm\alpha)} = \frac{  v_2 \phi^2 a^2}{2\nu_0}-\frac{v_3\phi^3}{3\nu_0}(1+3 a^2) + \frac{1}{\nu_0}\log(1-\phi).
\end{align} 
One can then show that if  $a=a_\ast$, with $a_\ast$ the solution to 
\begin{align}
a_\ast = \tanh( a_\ast \phi (v_2 -v_3\phi) ), \label{eq:easy1}
\end{align}
then the stationarity conditions $\bar{\mu}_{i}^{(\alpha)} = \bar{\mu}_{i}^{(-\alpha)}$ and $\Pi^{(\alpha)} =  \Pi^{(-\alpha)}$ are satisfied for the above expressions of $\bar{\mu}_{i}^{(\pm\alpha)}$ and $\Pi^{(\pm\alpha)}$.

Notice that the stationary compositions of the two phases $\alpha$ and $-\alpha$  are the same as those found previously  in  the grand canonical setup~\cite{braz2024liquid}.  However, here we consider a spatially heterogeneous state consisting of two phases, while Ref.~\cite{braz2024liquid} studied homogeneous states consisting of one phase. Furthermore, here the volume fraction $\phi$ is an external control parameter, whereas in the grand canonical ensemble it is a variable that is  determined by the   chemical potentials of the reservoirs. The coexistence of phases $\alpha$ and $-\alpha$ is here a consequence of particle conservation. Nevertheless, the fact that these two phases are local minima of the grand potential is a consequence of the symmetry of   $J_{ij}$, and also occurs in the Hopfield model for associative memory \cite{hertz2018introduction}.

Figure~\ref{fig:hopf_stab}A shows the solutions to Eq.~\eqref{eq:easy1} as a function of $\phi$.  The plot shows that the liquid mixture has stationary phases that successfully retrieve the prescribed phases if 
\begin{align}
\phi(v_2-v_3\phi)
> 1. \label{eq:phiv2v3}
\end{align}
There is an intermediate value of the volume fraction $\phi$ for which the overlap $a$, and thus the quality of retrieval, is maximal.    Notice that Fig.~\ref{fig:hopf_stab}A does not show the typical first order phase transition of liquid mixtures, as it is a plot of the stationary state.   To study   phase transitions we should analyze  the stability of the resultant phase, and as will become evident, near the transition line  $\phi(v_2-v_3\phi)\approx 1$ the stationary phase is not stable.

We finally remark that the liquid Hopfield model also admits stationary states with $C>2$ that consist of multiple pairs of retrieval and anti-retrieval phases.  In particular, a phase-separated state composed of multiple pairs of stationary phases, with compositions $\vec{\phi}^{(\alpha)}$ and $\vec{\phi}^{(-\alpha)}$ as given by Eqs.~\eqref{eq:drop1} and~\eqref{eq:drop2}, and with compartment volumes satisfying $w^{(\alpha)} = w^{(-\alpha)}$ (but otherwise arbitrary), is stationary provided that $a = a^\ast$ holds for all pairs of retrieval–antiretrieval phases.    Next, we study the stability of these states.  

\begin{figure*}\centering
\includegraphics{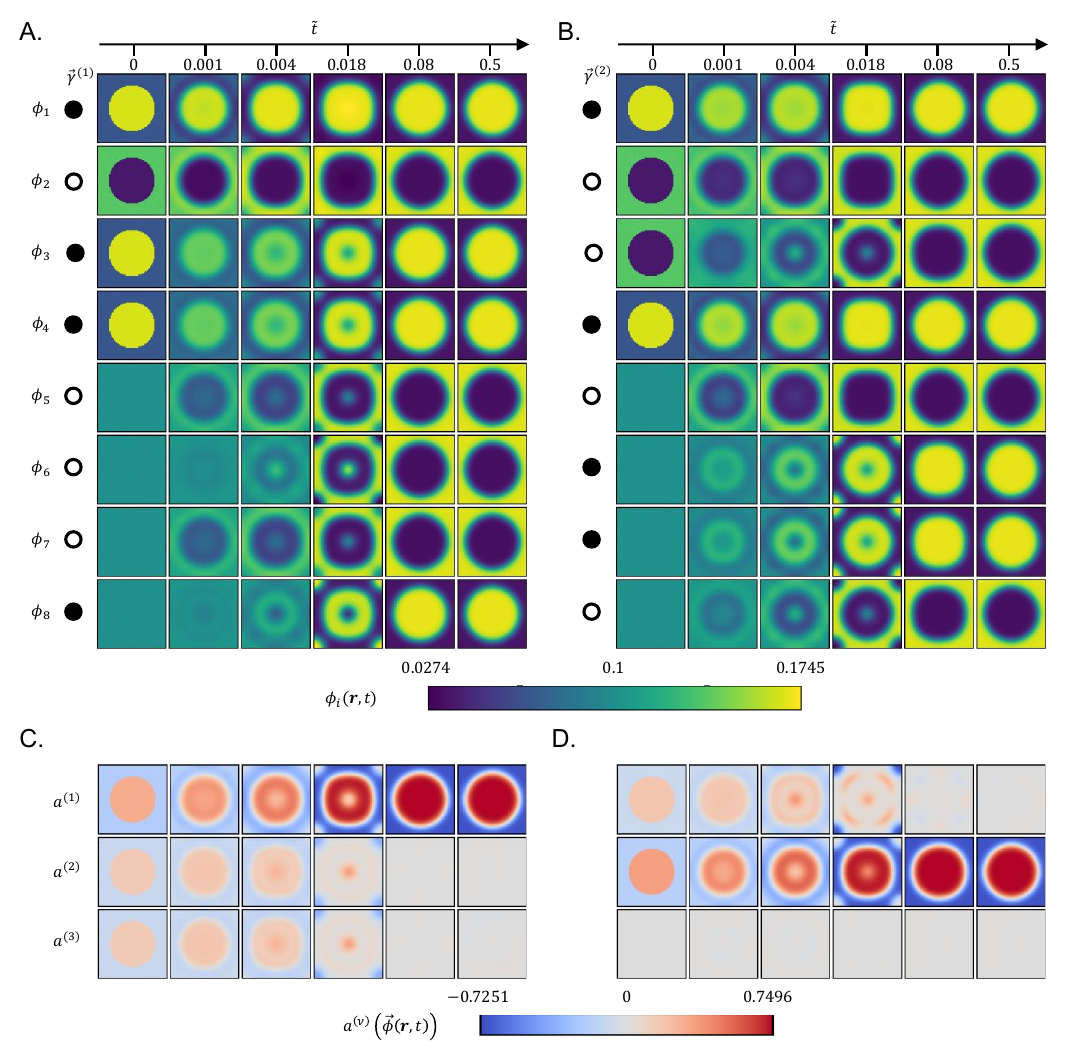}
\caption{{\it Retrieval of information from a partial cue in the liquid Hopfield model}. {\bf A.} Snapshots of the retrieval dynamics of the compositions, $\phi_i({\bf r},t)$,  obtained from  solving the Cahn-Hilliard Eqs.~\eqref{eq:CH1}-\eqref{eq:CH2} for the liquid Hopfield model when it stores three targets. The initial state has a small seed that  overlaps with target $\alpha=1$ only in components $i=1,\ldots,4$. As time progresses the system relaxes to the target phase separated state. {\bf B.} Same as A, including same targets encoded in the interaction matrix $J_{ij}$, but the initial condition is set so that the first four components match target  $\alpha = 2$. The system thus relaxes to the retrieval phase separated state $\alpha=2$. {\bf C.} The three local overlaps $a^{(\nu)}(\vec{\phi}({\bf r},t))$ with $a^{(\nu)}$ defined in Eq.~\eqref{eq:obs_ov} and for $\vec{\phi}$ from Panel A. {\bf D.} Same as C, but with the compositions of Panel B. The model parameters used in the figure are $v_2 = 4$, $v_3 = 3$, $\phi = 0.8$. Further details can be found in SM Sec. S7d. Three targets are encoded in $J_{ij}$, from which two are shown, and the third is $\vec{\gamma}^{(3)}=(+-+-++--)$ (see SM Fig. S5).} \label{fig:hopf_MPS}
\end{figure*}

\subsection{Metastability of homogeneous and phase separated  retrieval states}
\label{sec:meta_hopf}
The  homogeneous state is metastable if  the Hessian of its free energy is positive.  As derived in~\cite{braz2024liquid},  the spinodal of the homogeneous state is attained when   Eq.~\eqref{eq:phiv2v3}  becomes an equality.  The parameter region for which the homogeneous state is metastable is indicated with a  purple shade in   Figs.~\ref{fig:hopf_stab}B-D.  For fixed parameters $\phi$, the homogeneous state gets unstable at large enough values of $v_2$, which represents the onset of phase separation (see Figs.~\ref{fig:hopf_stab}B).   A metastable retrieval region exists for large enough values of $\phi$ and for a certain parameter range of the interactions $v_2$ and $v_3$, which have to be positive but not too large.  

As  we demonstrated in Sec.~\ref{sec:stab},  stationary phase separated states are metastable provided all constituent phases are stable.  For the liquid Hopfield model sufficient conditions for the stability of any retrieval phase were derived in Ref.~\cite{braz2024liquid} and are given by
\begin{align}\label{eq:stab_targets}
\phi(1+a_\ast)(v_2-v_3\phi(1+a_\ast))&\le1\\
\phi(1-a_\ast)(v_2-v_3\phi(1-a_\ast))&\le1,\label{eq:stab_targets2}
\end{align}
where $a_\ast$ solves Eq.~\eqref{eq:easy1} (see Appendix~\ref{app:hopfield_stab} for a generalization to arbitrary $q$). Therefore,  a retrieval state consisting of two phases, $P=2$, is metastable provided these conditions are met for the  pair of retrieval and anti-retrieval phase that it is composed of.  We emphasize that the conditions \eqref{eq:stab_targets} and \eqref{eq:stab_targets2} are sufficient, but not necessary, for metastability. In general, the stability of retrieval phases is different for each choice of the targets $\{\vec{\xi}^{(\mu)}\}_{\mu=1,\ldots,p}$ encoded by $J_{ij}$. Furthermore, the stability of a particular retrieval phase depends on the retrieved target $\alpha$. Nevertheless, as it was shown in Ref.~\cite{braz2024liquid}, the sufficient conditions above become also necessary  when $p$  is large enough.

Figures~\ref{fig:hopf_stab}B-D  show the region in parameter space for which  phase separated retrieval states are stable (indicated with the green shade). In this region, all $p$ encoded retrieval states are stable, provided that these retrieval states are linearly independent, and thus $p\leq N-1$. Note that stability of retrieval  states requires a repulsive cubic interaction,  in agreement with the findings in Ref.~\cite{braz2024liquid}.  Figures~\ref{fig:hopf_stab}B-D  also show a region in which neither the homogeneous nor the  retrieval states are stable.  In the previous work~\cite{braz2024liquid} this region was called the disordered region, as the stable configurations in this region have little resemblance with the prescribed compositions.

For $C>2$, the liquid Hopfield model can simultaneously retrieve multiple pairs of stationary retrieval, anti-retrieval phases, $(\vec{\phi}^{(\alpha)},\vec{\phi}^{(-\alpha)})$, corresponding with different target compositions $\alpha$.     In this case, the stationary configurations are not metastable, as they can have soft modes (see Sec.~\ref{sec:stab}). Indeed, if we  vary the volumes of the different phases with the constraint $\sum^P_{\alpha=1} w^{(\alpha)} = \sum^P_{\alpha=1} w^{(-\alpha)}$ maintained, where $P$ indicates the number of retrieved target compositions, then we obtain phase separated states with the same free energy as the original configuration.  Thus, the local minimum of the free energy is in fact a metastable plateau.  A possible way to  stabilize such a configuration is by introducing a nonzero  surface tension, as  we do in the following section.   Due to surface tension, the number of interfaces should be minimized, and therefore only the  state within the metastable plateau that has the  minimal surface tension will remain metastable when including surface tension into the model.  

\subsection{Information retrieval via nucleation of retrieval phases}
\label{sec:ass_MPS}
So far we have shown that the liquid Hopfield model admits a parameter regime in which all retrieval phases are metastable (green shaded area in Fig.~\ref{fig:hopf_stab}B-D).  However, this is not sufficient to demonstrate that Hopfield liquids can reliably nucleate the stored phases from a partial cue of information, i.e., from a region in space with compositions that have some similarity with a target state. Indeed, the free energy  of the liquid Hopfield model may have  metastable spurious states that disrupt the relaxation towards the desired target phases. Therefore, in this section we investigate the nucleation ability of Hopfield liquids by studying their dynamics in continuous space.

Our starting point is the Cahn-Hilliard equations for a multicomponent mixture,
\begin{equation}
\frac{\partial \phi_i(\mathbf{r},t)}{\partial t} = \ell \nabla^2 \bar{\mu}_i(\mathbf{r},t), \label{eq:CH1}
\end{equation}
with  the spatially dependent exchange chemical potentials given by
\begin{equation}
\bar{\mu}_i =\nu_0\left(\frac{\partial f}{\partial \phi_i } - k \nabla^2 \phi_i \right); \label{eq:CH2}
\end{equation} 
see  Appendix~\ref{app:dynamics} for a more detailed exposition. These equations describe the dynamics of a multicomponent mixture in continuous space. We then consider an initial condition  that is slightly biased towards one of the target compositions, and thus acts as a partial information cue.  Starting from this initial condition, we investigate whether the system nucleates the retrieval phases described by the Eqs.~\eqref{eq:drop1} and \eqref{eq:drop2}.

The partial cue, which acts as nucleation seed, consists of a small patch enriched (or depleted) in some (but not all) the components of the target composition. Besides this small patch the remaining space is a homogeneous mixture.  Figs.~\ref{fig:hopf_MPS}A and B show that the Hopfield liquid indeed evolves towards  a phase separated state with two phases ($P=2$).  The final stable phases are precisely the retrieval and anti-retrieval phases  that match with the initial cue provided.     We find successful nucleation for all the $p$ targets that are stored in the affinity matrix of  the Hopfield liquid (in this example $p=3$, see also SM Fig. S5).  Thus the different retrieval phases are simultaneously metastable and can be nucleated reliably by choosing a proper initial cue. Importantly, the initial cue does not have to be complete, which demonstrates that Hopfield liquids function properly as liquid analogues of associative memories.

The snapshots in Panels C and D of Fig.~\ref{fig:hopf_MPS} show the same dynamics as panels A and B, but the coloring scheme uses the local overlap observables,  $a^{(\nu)}(\vec{\phi}(\mathbf{r})\,)$ (see Eq.~(\ref{eq:obs_ov})). This allows us to visualize nucleation of  retrieval phases as the appearance  of a uniformly colored phase (see details of coloring scheme in the caption). At long times, the numerical solution approaches the analytically computed values for the overlaps, $a^{(\nu)}(\vec{\phi}^{(\alpha)}) = \pm a_\ast \delta_{\alpha,\nu}$, corroborating the theory.  Therefore, Hopfield liquids are capable of information retrieval from partial cues, in agreement with theoretical predictions.

\subsection{Metastable phase separation in  multicomponent mixtures}
\label{sec:ass_MPS_many}

We have shown that  Hopfield liquids are multicomponent mixtures that can reliably retrieve phases with prescribed compositions  via nucleation seeds that have partial information on the target phases.    In this section, we study how the liquid Hopfield model  spontaneously phase-separates into multiple phases with distinct, prescribed compositions, and thus serves as a  toy model  for how the cytoplasm demixes through liquid-liquid phase separation into multiple biomolecular condensates.  

\begin{figure}[]\centering
\includegraphics{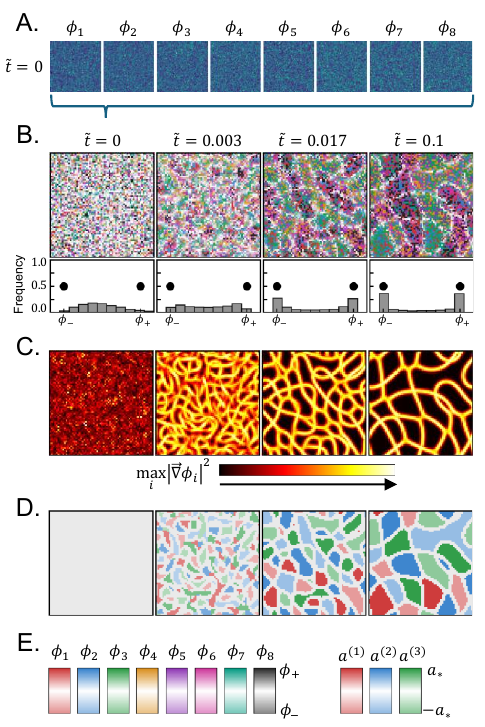}
\caption{{\it Nucleation and coexistence of multiple target phases from a random initial condition.} 
\textbf{A.} Initial compositions are spatially uniform with $\phi_i = 0.1$, plus a small noise perturbation. Color bar as in Figs.~\ref{fig:hopf_MPS}A, B.
\textbf{B.} Time evolution of the density fields. Each pixel is colored according to the largest density field at that point (color bars in Panel E). Visually, the emergence of different domains is apparent. The histograms show how the density fields approach the values $\phi_{+}$ and $\phi_{-}$, where we have defined $\phi_{\pm} = \phi/N(1\pm a^*)$.
\textbf{C.} Time evolution of interfaces. The panels show the heatmaps of $\max_{i}|\nabla\phi_{i}|^2$. With increasing time, the percentage of interface points decreases. The arrow below the color bar indicates increasing values of the gradient. \textbf{D.} Nucleation of target phases identified by the algorithm in SM Sec. S7c. Phases are colored according to the maximum overlap with the targets. Interface points are left gray.
\textbf{E.} Color bars used in Panels~C and D. Parameters as in Fig.~\ref{fig:hopf_MPS}, see also SM Sec. S7d.}
\label{fig:hopf_coex}
\end{figure}
Figure~\ref{fig:hopf_coex} shows the dynamics of a  Hopfield  liquid in the region of parameter space for which stationary retrieval states are stable.  As in the previous section, we consider here the case $N = 8$ and $p = 3$. The system is initialized in a state that  is spatially homogeneous, with a small spatially fluctuating perturbation (see Panel A).

As the system evolves over time, the components start to spatially segregate, so that the densities of particular components are enhanced in different spatial regions. This phase separation process can be seen in Panel B (see Panel E for the coloring scheme). Since $q=1/2$, the phases are not  pure,  but rather  complex mixtures enriched and depleted in specific components.

To systematically identify the different phases present in the system at a given time, we follow the approach developed in Refs.~\cite{mao2019phase, brenner}. In brief, the method uses the gradient of the compositions, shown in panel C, to determine domain interfaces, and then identifies the composition of phases among the domains (see SM Sec. S7c). In Panel D, each of the identified phases is  colored based on the value of the largest (or lowest) overlap with one of the targets. As one can see, as time progresses the amount of interface is reduced (see detailed analysis in SM Fig. S6), and the domains become larger. Importantly, the overlaps with the targets increase over time (the colors become more vivid).

\subsection{Retrieval capacity of Hopfield liquids}\label{sec:capacity}

\begin{figure*}\centering
\includegraphics{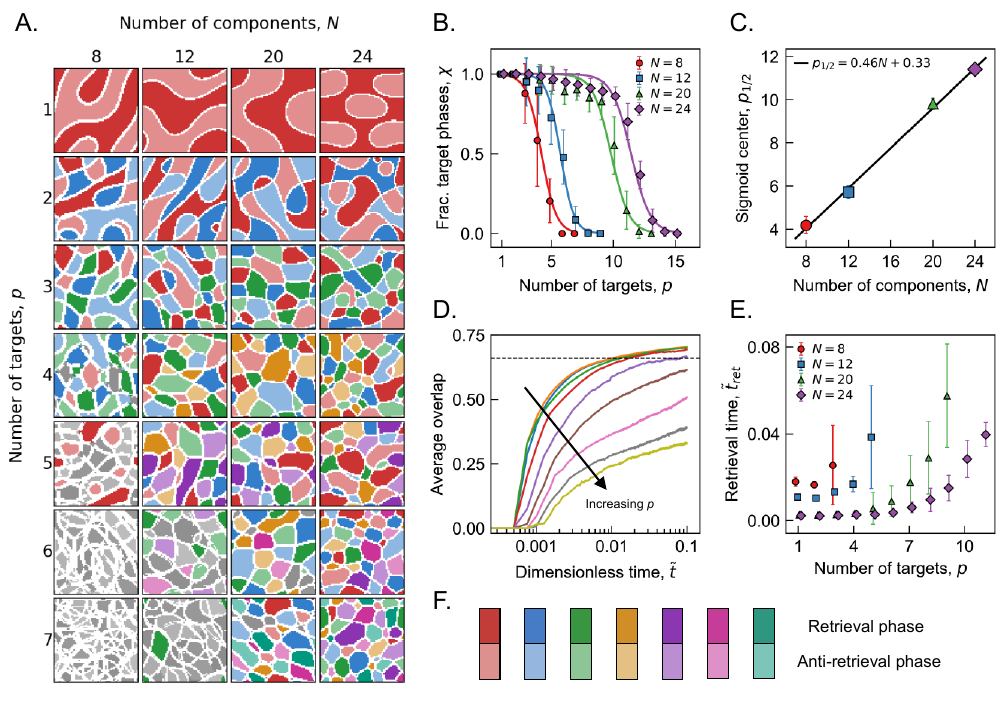}
\caption{{\it Retrieval capacity of a Hopfield liquid.}  \textbf{A. } Visualization of the  phases in  the liquid Hopfield mixture in the region where all retrieval phases are stable (green region of Fig.~\ref{fig:hopf_stab}). The figure data is obtained from solving Cahn-Hilliard equations for a given number of components $N$ and targets $p$ as indicated in the figure.  
The phases are identified as regions where the gradients of the  $\phi_i(\mathbf{r})$ are small  using  the procedure described in SM Sec. S7c.  Phases are visualized with chromatic colors if they are identified as target phases [if the absolute value of the overlap of a phase with a given target is larger than 0.66, \textit{i.e} $\simeq 0.9\,a_\ast$, the phase is considered a target phase and it is assigned a chromatic color (see Panel F)].  If a phase is not identified as a target phase, then it is given a random shade of gray.  Interface points are left white. \textbf{B. } The fraction of target phases, $\chi$, \textit{i.e.} the ratio between the number of target phases and the total number of phases in a given configuration, is plotted as a function of  $p$  for  various values of $N$ as indicated in the legend. Markers are sample averages  over the final configurations of 30 Cahn-Hilliard simulations with randomized initial conditions and targets.  The solid lines are  sigmoid functions $\chi(p) = 1/(1 + \exp[b (p - p_{1/2})])$, where $b$ and $p_{1/2}$ are  fitting parameters. \textbf{C. } The mid-points of the sigmoids in Panel B (indicated with $p_{1/2}$) are plotted against the number of components $N$ and fitted with a straight line. \textbf{D.} Average overlap as a function of $\tilde{t}$, for $N=12$
and increasing values of $p$ (from $1$ to $9$). For each phase, the overlap is computed with the best-matching stored target, and the average is taken over all phases weighted by their volume fractions (see SM Sec. S7d for details). The curves represent averages over $30$ independent simulations.
The dashed black line indicates the threshold used to identify (anti)retrieval phases in Panels A-C, \textit{i.e.} $0.9 a_\ast$. \textbf{E.} Retrieval time, $\tilde{t}_{ret}$, defined as the first time the average overlap exceeds $0.9 a_\ast$, as a function of $p$ for different values of $N$.  Markers are sample averages over $30$ Cahn-Hilliard simulations, with error bars representing the sample standard deviation.  The retrieval time $\tilde{t}_{ret}$ increases with $p$, implying that retrieval dynamics is hindered by the presence of a large number of metastable states.  \textbf{F.} Color scheme for Panel A. The values of the parameters are $\phi=0.8$, $v_2 = 4$ and $v_3 = 3$. For more figure details check SM Sec. S7d.}\label{fig:hopf_SN}
\end{figure*}

We now investigate the number of target phases that a liquid mixture with $N$ components can successfully retrieve. For associative neural networks it is known that the number of patterns that can be reliably retrieved increases proportionally with the number of neurons \cite{hopfield1982neural, amit1985}. In section~\ref{sec:meta_hopf} we showed that the liquid Hopfield model can store up to $N-1$ targets, with all of them being simultaneously stable. However,  Hopfield liquids also  admit spurious  phases, and therefore it remains to be addressed whether the target phases are  retrievable under physical dynamics.

In order to measure the retrieval capacity of the liquid Hopfield model, we numerically solve the Cahn-Hilliard equations for several values of $p$ and  $N$ starting from a random initial condition.  
We consider here orthogonal targets, so that the phase identification algorithm separates them easily (see SM Sec. S7d for details on how to generate orthogonal targets). Qualitatively similar results can be obtained for non-orthogonal targets (see SM Fig. S7).
 
In Fig.~\ref{fig:hopf_SN} (Panel A) we present the final snapshots of such Cahn-Hilliard  simulations.  Different phases are  identified through the method described in SM Sec. S7c, which works here well as we have chosen orthogonal targets. Target phases are highlighted with bright colors, whereas non-target phases are colored with different shades of gray. For $N = 8$, we observe that  for $p<4$ all phases in the  final configuration are either retrieval or anti-retrieval phases, while for  $p\geq 4$  the mixture evolves towards a phase separated state that contains non-target phases. For larger values of $N$, reliable retrieval is possible for a larger number of target phases, as the non-target phases appear at higher values of $p$.  Hence, the retrieval capacity of the liquid Hopfield model increases as a function of the number of components.   

To quantify the retrieval capacity of the liquid Hopfield model, we plot  in 
Panel B the fraction,  $\chi$, of demixed phases in the final configuration  that correspond to retrieval or anti-retrieval phases   as a function of $p$. For each $N$, $\chi$  decays like a sigmoid  as a function of $p$, and thus the fraction of phases that corresponds to the encoded targets decreases as a function of $p$.   Crucially, the mid-point of the sigmoid, $p_{1/2}$, increases linearly  with $N$ (see Panel C).  Fig.~\ref{fig:hopf_SN}D and E characterize the kinetics of retrieval, which show that as the number of encoded targets increases the dynamics of retrieval becomes slower (see SM Fig.~S6 for interface dynamics). Taken together, these results demonstrate that liquid mixtures can retrieve a number of prescribed phases that is extensive on the number of components in the mixture.

\section{Discussion}
\label{sec:disc}
\subsection{Summary}

In this paper, we have derived  the conditions for metastability of phase separated states in multicomponent liquid mixtures. We have found that stability of all constituent phases is necessary and sufficient for stability of the phase separated state, with only singular exceptions. To illustrate this, we developed a minimal model for metastable phase separation in a binary mixture with highly non-linear interactions. Furthermore, studying the  liquid Hopfield model in the canonical ensemble, we have shown that multicomponent liquid mixtures can phase separate into phases of prescribed compositions and thus perform reliable information retrieval through nucleation.  Therefore,  this latter model can be seen as a toy model for cytoplasmic organization through phase separation or phase separation in synthetic multicomponent mixtures.

\subsection{Relation to prior theoretical studies}

Theoretical research on multicomponent liquids can be broadly divided into two lines of research: the direct approach and the inverse approach.  In the direct approach, the affinity matrix $J_{ij}$  in Eq.~\eqref{eq:genInt} is  defined as a random matrix. Investigation then focuses on either characterizing the spinodal manifold, see Refs.~\cite{sear2003instabilities, carugno2022instabilities, thewes2023composition, parkavousi2025enhanced, parkavousi2025compositional}, which can be done analytically using random matrix theory; or on characterizing the low temperature phase separated states~\cite{jacobs2013predicting, jacobs2017phase, girard2022kinetics, brenner}, which can be done numerically via simulations.  In the inverse approach, the compositions of the different phases are prescribed, and  the choice of the  affinity matrix $J_{ij}$ is such that the prescribed phases are stable. This approach has been investigated through numerical optimization of $J_{ij}$ in Refs.~\cite{jacobs2017phase,  zwicker2022evolved, chen2023programmable}.

In contrast to these works, our paper solves the inverse problem analytically by studying the liquid Hopfield model in the canonical ensemble. We exploited the structure of a Hebbian $J_{ij}$ to analytically characterize  metastable phase separated states that correspond to the prescribed compositions. Crucially, such states can be enriched in a large number of components, $Q\sim N$, so that in the limit $N\gg1$  an arbitrary pair of targets share a finite fraction of components. In contrast, Refs.~\cite{jacobs2017phase, chen2023programmable} studied the case in which for $N\gg1$ a pair of targets do not share components, see also discussion in  Ref.~\cite{braz2024liquid}. Moreover, we were able to simulate the spatial dynamics of a mixture with a large number of components (up to $N=24$) and obtain results that precisely agree with our theoretical predictions.

In order to achieve such an accurate picture of demixing in multicomponent liquids, we first developed the general formalism of metastable phase separation. Metastability has been observed in experiments \cite{patel2015liquid, lin2015formation, das2025tunable, michaels2023amyloid}, discussed analytically in the grand canonical ensemble (in the case of a single phase)\cite{braz2024liquid}, explored in numerical studies of phase separation \cite{zwicker2022evolved, brenner}, and its role in the phase separation dynamics in a variety of contexts has been recognized since long ago \cite{Cahn1969, Evans1997, Olmsted1998, Cheng2008}. However, the mathematical conditions for metastable phase separation had so far not been derived.  These conditions are useful for verifying whether the  phase separated states obtained from numerical algorithms, such as the one in Ref.~\cite{zwicker2022evolved},  are metastable, and not saddles in a high-dimensional landscape. 
The result that metastability of a phase separated state is, excluding the (peculiar) case of emergent soft-modes, equivalent to metastability of all constituent phases is simple yet powerful.

\subsection{Relation to experimental work}

We discuss three lines of experimental research related to our work: metastability of biomolecular condensates, principles of cytoplasmic organization, and synthetic liquids.

Multiple papers have shown that solutions of prion-like proteins or peptides can rapidly phase separate into liquid droplets~\cite{patel2015liquid, lin2015formation, das2025tunable, michaels2023amyloid}. However, after several hours, these droplets transition to a distinct aggregate phase, indicating that the droplets are metastable. The binary mixture model in Sec.~\ref{sec:nlin} constitutes a toy-model for such experiments. In particular, Ref.~\cite{das2025tunable} shows that in the metastable state medium-density and high-density liquid phases coexist, whereas in the stable state a low-density liquid phase coexists with a high-density aggregate phase. Such results are consistent with those in Fig.~\ref{fig:sim_quartic}, although the dense phase in our model is liquid. Overall, our thermodynamic framework is a first step towards quantifying metastable phase separation in these experimental systems.

The cellular cytoplasm is a mixture with a large number of components that encodes  condensates of precise composition, see e.g. \cite{you2020phasepdb}. The same network of protein interactions that allows for  condensates to phase separate under particular conditions, also allows for multiple condensates to coexist with each other. The results in Sec.~\ref{sec:Hopfield} show that Hopfield liquids are multicomponent mixtures that exhibit both these features,  see Figs.~\ref{fig:hopf_MPS} and \ref{fig:hopf_coex} respectively, and therefore constitute a toy-model of the cytoplasm. It remains open how to relate this model to empirically determined protein interaction networks \cite{gavin2002functional, schaffer2025multimodal, saar2024protein}.

Our work  also relates to synthetic multicomponent mixtures, which can be designed using DNA technology \cite{sato2020sequence, biffi2013phase, chaderjian2025diverse}. In particular, recent work has demonstrated how DNA nanostar sequences can be designed so that the interaction matrix of nine different components enables phase separation into nine liquid phases, each enriched in a single component~\cite{chaderjian2025diverse}. A conceptually straightforward, but technically challenging, next step is to find sequences that encode the interaction matrix in Eq.~\ref{eq:hebb}, instead of the diagonal interaction matrix studied in \cite{chaderjian2025diverse}. This would allow the creation of a Hopfield liquid, in which multiple phases with prescribed composition share components with each other.

\subsection{Extensions of Hopfield liquids}

We have studied metastable phase separation in a minimal version of the liquid Hopfield model. In Ref.~\cite{braz2024liquid} we showed that several aspects of the model can be generalized, without altering its essential behavior. As an example of a model extension, in Appendix~\ref{app:hopfield_stat} and SM Sec. S8 we study the effect on phase separation of having the target compositions enriched in $Q<N/2$ components. We find that, unlike for the case $Q=N/2$,  for $Q<N/2$ the retrieval and anti-retrieval  phases have different volumes  ($w^{(\alpha)}<w^{(-\alpha)}$, see SM Fig. S8). We have run Cahn-Hilliard simulations to test these predictions for the particular case of $N = 20$, $p = 1$ and $q = 2/5$.   The simulations give results in good agreement with theoretical predictions (see SM Fig. S9).
Furthermore, we have numerically investigated the case of a non-equimolar mixture, \textit{i.e} a mixture in which the abundances of the solutes are not necessarily all equal. In particular, we have performed simulations for different values of the standard deviation of the $\phi_i$'s (up to roughly 67\% of the average value $\phi/N$). The results show how, with increasing heterogeneity in the abundances of the solutes, the probability of successful retrieval decreases, whereas the asymmetry between retrieval and anti-retrieval phases increases (see SM Fig. S10).
These phenomenologies are more similar to those of biomolecular condensates.  Model extensions can also be aimed at finding alternative ways of stabilizing Hopfield phases. In the main text we used a cubic repulsive interaction to stabilize phase-separated retrieval phases, an idea originally proposed in Ref.~\cite{braz2024liquid} and later also explored in \cite{luo2024beyond}. Next, we propose three alternative approaches to stabilize demixing of Hopfield phases.

First, it is possible that chemical reactions can be used to stabilize retrieval phases. Indeed, chemical dissipation is known to unfold a large class of different behaviors in liquid phase separation~\cite{zwicker2015suppression, kirschbaum2021controlling, parkavousi2025enhanced, klosin2020phase, Huberman1976}, including phase stabilization. Moreover, retrieval of target structures far from equilibrium has already been realized for solid assemblies~\cite{Benoist2025, Benoist2026}.

Second, the surface energy tensor, taken as diagonal in Appendix~\ref{app:dynamics}, may also be exploited to stabilize retrieval phases in Hopfield liquids, in the spirit of the Ref.~\cite{mao2020designing}. Indeed, as  we show in SM Sec. S10, at $v_3=0$ there exists a region of parameter space for which the only unstable mode of the phase separated target state is one that creates a new interface. Therefore, a strong and programmable surface energy tensor may stabilize retrieval phases in absence of cubic non-linearities.

Third, an interaction matrix $J_{ij}$ different from the Hebbian matrix proposed here may result in stable Hopfield phases. Alternatives to the Hebbian rule have been numerically studied in~\cite{jacobs2021self, chen2023programmable}, which did not include cubic interactions. However, the $N\to\infty$ scaling there studied applies  to the limit $q\to0$, in which phases do not share components with each other, see the discussion in Ref.~\cite{braz2024liquid}. Indeed, strong nonlinearities are a generic requirement of neural networks to adequately perform computations~\cite{hopfield1984neurons, bolle2003spherical}. It remains nonetheless an open mathematical question whether the entropic nonlinearities suffice to stabilize Hopfield phases for an alternative $J_{ij}$  in the limit of large $N$ and nonzero $q$.

\subsection{Perspective}

Biological systems are characterized by a high dimensional interaction space. The naif expectation for this scenario is that disorder dominates biological matter~\cite{sear2003instabilities, jacobs2013predicting}. Nevertheless, ordered structures emerge across biological scales: proteins fold into well defined conformations~\cite{finkelstein2016protein}, folded proteins bind specifically to form multimeric assemblies~\cite{marsh2015structure}, and assemblies are often concentrated into liquid condensates~\cite{brangwynne2009germline}. From the biological viewpoint this should not come as a surprise, since biological interactions are evolved so that ordered structures perform specific functions. However, from the physics point of view, finding the rules behind this  multifarious state of matter is an exciting and challenging endeavor~\cite{fink2001many, sartori2020lessons, braz2024liquid}.

From a physics point of view, we have shown how multicomponent liquids are capable of two fundamental information processing tasks: pattern completion and information storage. To do so, we have built upon the (imperfect) mathematical equivalence between multicomponent mixtures and neural networks. Given the outstanding flexibility that Hopfield networks have recently been shown to have~\cite{ramsauer2020hopfield}, it is easy to imagine how more general interaction matrices than the ones considered here can allow for other information processing tasks~\cite{Zentner2026}.

From the biology point of view, our work hints towards an analogy between the cellular cytoplasm and neural networks. Structured protein complexes  are the cytoplasmic analogues of neuronal spatial maps \cite{sartori2020lessons, hopfield2010neurodynamics, zhong2017associative}, and biomolecular condensates are the cytoplasmic analogues of associative memories \cite{braz2024liquid, hopfield1982neural}. A promising next step guided by this analogy is investigating whether the evolution of the cytoplasm can be understood as a neural learning rule. Indeed, Refs.~\cite{patel2015liquid, das2025tunable} demonstrate that point mutations are sufficient to change the phenomenology of protein phase separation, yielding aberrant condensates. This suggests that the interactions between proteins have been learned through evolutionary adaptation to optimize the formation of liquid phases. All in all, whether this ``cytoplasm--neural network analogy'' can be useful beyond a mathematical equivalence remains a stimulating direction for future inquiry.

\appendix

\begin{center}
    {\bf APPENDICES}
\end{center}

\section{Thermodynamics of incompressible multicomponent mixtures}
\label{app:euler}

We  describe the thermodynamics of an incompressible  mixture  consisting of $N$ solutes and a solvent.   We consider the setup for which the temperature, volume, and molar numbers of the components are  fixed.   This setup is often considered in the literature, even though  experiments typically are performed at constant pressure and real mixtures are not necessarily incompressible~\cite{atkins2023atkins}.   The advantage of this setup is that it is compatible with simple continuum models that are easily amenable to spatial simulations~\cite{rubinstein2003polymer}, such as the Flory-Huggins model and its variants~\cite{flory1942thermodynamics, huggins1941solutions, mao2019phase}.

The thermodynamic state parameters of the mixture in this setup are the temperature, $T$;  the volume, $V$; the number of molecules of the solvent, $M_0$; and the number of molecules of the $N$ solutes, $M_i$, with $i=1,\ldots, N$. The corresponding thermodynamic potential for these $N+3$ variables is the Helmholtz free energy
\begin{align}
    F=F(T,V,M_0,\vec{M}),
\end{align}
where $\vec{M}=(M_1,\ldots,M_N)$.  Its differential expression is given by 
\begin{align}
    \d F=-S\d T-p\d V+\sum_{i=0}^N\mu_i\d M_i,
\end{align}
where $S$ is the entropy, $p$ the pressure, and $\mu_i$ are the chemical potentials. Differential forms for the latter quantities can also be expressed in terms of the state variables. For  the pressure this gives
\begin{align}\label{eq:dp}
    \d p &= \left(\frac{\partial p}{\partial T}\right)_{V,\tilde{M}}\d T+\left(\frac{\partial p}{\partial V}\right)_{T,\tilde{M}}\d V\nonumber\\
    &+\sum_{i=0}^N\left(\frac{\partial p}{\partial M_i}\right)_{T, V,\tilde{M}'(i)}\d M_i,
\end{align}
where $\tilde{M}$ denotes the vector $\tilde{M}=(M_0,\vec{M})$, and  $\tilde{M}'(i)$ denotes all variables $M_j$ for which $j\neq i$.

Incompressibility implies that the pressure can only be changed by varying the temperature, while changes in volume and composition result in divergent pressure. Mathematically, this is the limit for which 
\begin{align}
    \left(\frac{\partial p}{\partial V}\right)_{T,\tilde{M}}&\to\infty,\\
    \left(\frac{\partial p}{\partial T}\right)_{V,\tilde{M}}\left(\frac{\partial p}{\partial V}\right)_{T,\tilde{M}}^{-1}&\to{0},\\
    \left(\frac{\partial p}{\partial M_i}\right)_{T,V,\tilde{M}'(i)}\left(\frac{\partial p}{\partial V}\right)_{T,\tilde{M}}^{-1} &\to {\rm constant}.
\end{align}
These conditions can be rendered into a more intuitive form by using the chain rule. Indeed, the first condition states that the isothermal compressibility, $\kappa_T=-V^{-1}(\partial V/\partial p)_{T,\tilde{M}}$, is null; the second condition states that the thermal expansion coefficient, $\alpha=V^{-1}(\partial V/\partial T)_{p,\tilde{M}}$, is null; and the third condition states that the molecular volumes of all components,  $\nu_i=(\partial V/\partial M_i)_{T,p,\tilde{M}'(i)}$, are constant (independent of state variables). From these conditions we can isolate the volume differential in Eq.~\eqref{eq:dp} as 
\begin{align}
    \d V &=\sum_{i=0}^N\nu_i\d M_i,
\end{align}
which can be integrated to give
\begin{align}\label{eq:constraintMo}
    V=\sum_{i=0}^N\nu_iM_i.
\end{align}
This equation is a constraint that allows us to remove one of the extensive thermodynamic parameters in the problem. We choose to remove $M_0$, without any loss of generality, and so the free energy of the incompressible system is 
\begin{equation}
F  = F_{\rm inc}(T,V,\vec{M}), 
\end{equation}
where $F_{\rm inc}(T,V,\vec{M})=F(T,V,M_0(V,\vec{M}), \vec{M})$ is the free energy in Eq.~\eqref{eq:freeenergy}, and with $M_0(V,\vec{M})$ obtained from  Eq.~\eqref{eq:constraintMo}. Hereafter, and in the main text, we drop the subindex ``inc''.

Next, we  develop the intensive formulation of this incompressible setting. First, we note that the free energy density $f$  is  defined  by 
\begin{eqnarray}
f(T, \rho_1,\ldots,\rho_N) &=&  \frac{F(T,V,M_1,\ldots,M_N)}{V}  \nonumber\\ 
&=& F(T,1,\rho_1,\ldots,\rho_N) .\label{eq:f}
\end{eqnarray} 
Here, $\rho_i = M_i/V$ are the number densities of the solutes.   For the second equality of \eqref{eq:f} we have used that $F$ is  a homogeneous first-order function of its extensive parameters, which follows from the postulates of thermodynamics~\cite{callen1998thermodynamics}. It is convenient to   express  densities in terms of volume fractions $\phi_i$ through 
\begin{equation}
\phi_i = \rho_i \nu_i .\label{eq:volumeFrac}
\end{equation}
The incompressibility constraint (\ref{eq:constraintMo}) reduces to  $\phi_0=1-\sum_{i=1}^N\phi_i$, which implies $\sum_{i=1}^N\phi_i\leq 1$. Using Eq.~\eqref{eq:volumeFrac} and the fact that the $\nu_i$ are constants,  we can define a free energy potential that is a function of volume fractions, viz., 
\begin{equation}
f_\phi(T,\phi_1,\ldots,\phi_N) = f(T,\phi_1/\nu_1,\ldots,\phi_N/\nu_N).
\end{equation}
The function $f_\phi$ is the free energy as it appears in \eqref{eq:free}, although in the main text we drop the $\phi$  sub-index.    

To finalize, we note that the exchange chemical potentials of the solutes can be written as
\begin{equation}
\bar{\mu}_i  = \left(\frac{\partial F}{\partial M_i}\right)_{T,V,\vec{M}'(i)} =  {\nu_i} \left(\frac{\partial f_\phi}{\partial \phi_i}\right)_{T,\vec{\phi}'(i)},
\end{equation}
where $M_0$ and $\phi_0$ are determined by the incompressibility constraints.   The osmotic pressure can be written as
\begin{equation}
\Pi = - \left(\frac{\partial F}{\partial V}\right)_{T,\vec{M}}  =  -f + \sum^N_{i=1}\phi_i\left(\frac{\partial f_\phi}{\partial \phi_i}\right)_{T,\vec{\phi}'(i)},
\end{equation}
where we have used that $F = Vf$ and $\phi_i = M_i \nu_i/V$, and as before $M_0$ is not fixed. The exchange chemical potentials and the osmotic pressure appear in the stationarity conditions for phase separation and thus play an important role in this paper.  In fact, we note that due to incompressibility, the chemical potentials, $\mu_i=(\partial F/\partial M_i)_{T,V,\tilde{M}'(i)}$, are not well defined, as any change in the number of components of the system leads to a change in volume, thus violating incompressibility. The pressure, $p=-(\partial F/\partial V)_{T,\tilde{M}}$, is also not well defined, since any change in volume must be associated with change in the number of components, which is inconsistent with the thermodynamic  definition of pressure.   
 
\section{Constrained perturbation of a phase separated state}\label{app:quadform}
To derive the stationarity and stability conditions of  spatially heterogeneous states, we study how the free energy $\f$ changes in response to a perturbation in the volume fractions and in the volume of the compartments. Therefore, considering the reference state $(\{w^{(c)}\},\{\vec{\phi}^{(c)}\})$, we study the perturbed state characterized by $\tilde{w}^{(c)}=w^{(c)}+\epsilon^{(c)}$ and ${\tilde{\phi}}_i^{(c)}={\phi}_i^{(c)}+\delta_i^{(c)}$, where $\epsilon^{(c)}$ and $\delta_i^{(c)}$ are the compartment volume and the compartment composition perturbation parameters. As the perturbed state must also be compatible with the constraints in Eqs.~\eqref{eq:volfrac} and \eqref{eq:const}, we have that
\begin{align}\label{eq:eps_wC}
\sum_{c=1}^C (w^{(c)}+\epsilon^{(c)})=1,
\end{align}
and also
\begin{align}\label{eq:dens_cons_C}
\phi_i&=\sum_{c=1}^C (w^{(c)}+\epsilon^{(c)})({\phi}_i^{(c)}+ \delta_i^{(c)}).
\end{align}
Using the constraint on the volumes of the reference state, Eq.~\eqref{eq:eps_wC} returns
\begin{align}\label{eq:eC}
\epsilon^{(C)}=-\sum_{c<C} \epsilon^{(c)}. 
\end{align}
Additionally,  using the volume and composition constraints of the reference state, we can write Eq.~\eqref{eq:dens_cons_C} as 
\begin{align}\label{eq:perturb_C}
\sum_{c<C} (w^{(c)}+\epsilon^{(c)})\delta_i^{(c)}+(1-\sum_{c<C}w^{(c)}+\epsilon^{(C)})\delta_i^{(C)}=-\Gamma_i,
\end{align} 
with $\Gamma_i=\sum_{c<C}\epsilon^{(c)}(\phi_i^{(c)}-\phi_i^{(C)})$. These two latter equations constitute constraints on the perturbation parameters. In particular, expanding Eq.~\eqref{eq:perturb_C} to quadratic order we obtain
\begin{align}\label{eq:diC}
\delta_i^{(C)} &= -\sum_{c<C} \delta^{(c)}_{i}\frac{w^{(c)}}{1-\sum_{c<C}w^{(c)}} + \sum_{c<C} \epsilon^{(c)} \frac{(\phi_i^{(C)}-\phi_i^{(c)})}{1-\sum_{c<C}w^{(c)}}\nonumber\\
&+ O(\epsilon \delta,\delta^2,\epsilon^2). 
\end{align}

We now move on to evaluate the free energy of the perturbed state, $\tilde{\f}=\f(\{\tilde{w}^{(c)}\},\{\tilde{\vec{\phi}}^{(c)}\})$.
Expanding to second order and using Eq.~\eqref{eq:perturb_C} to express the term $(1-\sum w^{(c)}-\sum\epsilon^{(c)})\sum^N_{i=1}\delta_i^{(C)}\partial_if^{(C)}$ in terms of $\Gamma_i$ and $\delta_i^{(c)}$, the free energy of the perturbed state can be written as
\begin{widetext}
\begin{align}
\tilde{\f}&=\f+\sum_{c<C}\epsilon^{(c)}\left[f(\phic)-f(\vec{\phi}^{(C)})-\sum_i(\phi_i^{(c)}-\phi_i^{(C)})\partial_if^{(c)}\right] + \sum_{c<C}\sum_i\delta_i^{(c)}\left[w^{(c)}(\partial_if^{(c)}-\partial_if^{(C)})\right] \nonumber \\ 
    &+\sum_{c<C} w^{(c)}\frac12\sum_{ij}\delta_i^{(c)}\delta_j^{(c)} h_{ij}^{(c)} + (1-\sum_{c<C}w^{(c)})\frac12\sum_{ij}\delta_i^{(C)}\delta_j^{(C)} h_{ij}^{(C)}-\sum_i\sum_{c<C}\epsilon^{(c)}\delta_i^{(c)}(\partial_if^{(C)}-\partial_if^{(c)}). \label{eq:generalPerturb}
 \end{align}
\end{widetext}
In the following sections, we study the first and second order terms of this perturbation to establish the stationarity and stability conditions of a phase separated state.

\section{Stationarity conditions of phase separated states}
\label{app:stat}
To obtain the stationarity conditions we set the first order terms in the perturbative expansion (\ref{eq:generalPerturb})  to zero. Since the perturbation variables themselves, $\epsilon^{(c)}$ and $\delta_i^{(c)}$, are arbitrary, the factors in the brackets must be null. Setting to zero the factor in the second  term  yields
\begin{align}\label{eq:difc_difC}
\partial_if^{(c)} = \partial_if^{(C)}
\end{align}
for any $i$ and any $c$.   Using the definition of exchange chemical potential,  Eq.~\eqref{eq:chem}, we obtain the chemical balance condition in Eq.~\eqref{eq:chem_bal}. Similarly, the factor in the brackets of the first term is null if and only if  
\begin{align}\label{eq:max}
f^{(c)} - \sum_i \partial_if^{(c)} \phi_i^{(c)} = f^{(C)} - \sum_i \partial_if^{(C)} \phi_i^{(C)},
\end{align}
where we have used the condition of chemical balance. Using the expression of osmotic pressure we thus derive the conditions of mechanical balance in Eq.~\eqref{eq:press_bal}.

The compositions of the phases define a plane that passes through all $P$ points $(\vec{\phi}^{(c)}, f^{(c)})$ and is tangent to the free energy surface $f(\vec{\phi})$. To see this, first note that the equation for a tangent plane that passes through a point $(\vec{\phi}^{(c)},f^{(c)})$ in $\mathbb{R}^{N+1}$ is given by
\begin{align}
z &= f^{(c)} + \sum_i\partial_if^{(c)}(\varphi_i-\phi_i^{(c)})
\end{align}
where $(z,\vec{\varphi})\in \mathbb{R}^{N+1}$. This can also be written as
\begin{align}
z =f^{(c)} - \sum_i\partial_if^{(c)}\phi_i^{(c)}+ \sum_i\partial_if^{(c)}\varphi_i.
\end{align}
Due to Eq.~\eqref{eq:difc_difC}, the last term in the right hand side of the expression above is the same for all $c$; and due to Eq.~\eqref{eq:max} the first two terms are also the same for all $c$. Therefore, the set of $P$ composition vectors that satisfies the stationarity conditions define a common tangent plane.  In addition, since $\vec{\phi}=\sum^P_{c=1}w^{(c)}\vec{\phi}^{(c)}$, we find that the compositions are restricted to a simplex within that plane.

\section{Stability conditions}\label{app:stab}
\label{app:necessary}
The stability conditions of phase separated states are obtained by imposing that any small change in free energy at a given stationary point is positive. We therefore evaluate the change in free energy $\Delta \f = \tilde{\f}-\f$. Stationarity implies that the linear terms and the cross-quadratic terms vanish, and so we have 
\begin{align}\label{eq:pp_dF}
    \Delta \f &= \sum_{c=1}^C w^{(c)}\frac12\sum_{ij}\delta_i^{(c)}\delta_j^{(c)} h_{ij}^{(c)},
\end{align}
where the volume in the last compartment is given by Eq.~\eqref{eq:volfrac} and the corresponding perturbation in composition by Eq.~\eqref{eq:diC}. From the expression (\ref{eq:pp_dF}), we find that if the Hessians of all compartments are positive definite, then the change in free energy must be positive, and the state is stable. Therefore, stability of all phases composing a phase separated states is sufficient for stability of the phase separated state.

To prove the converse, i.e., that stability of all phases is necessary for the stability of the phase separated state,  we  show that if one phase is  unstable, then there always exists a perturbation that lowers the free energy of the phase separated state.  Thus, consider  a system with $P$ phases, with one of them being unstable.  We represent the system by $C=P+1$ compartments, with the  compartments $c=1$ and $c=2$ representing the same phase, which is the unstable one.   We also set the compartment volumes in these two compartments equal to each other.   In this setup, we study the following perturbation
\begin{align}
\vec{\delta}^{(c)}=(\delta_{c,1} - \delta_{c,2})\vec{r}\quad,
\end{align}
with $\vec{r}$ the component perturbation vector. The perturbation $\vec{\delta}^{(c)}$ only affects the first two compartments, and leaves the remaining ones intact. Furthermore, the fact that $\delta_i^{(1)}=-\delta_i^{(2)}=r_i$ ensures that the constraint in Eq.~\eqref{eq:diC} is satisfied. For this perturbation, the change in free energy from Eq.~\eqref{eq:freeEnergyChange} takes the form
\begin{align}
\Delta \f
&= w^{(1)}\sum_{ij}h^{(1)}_{ij}r_ir_j.,
\end{align}
and so it follows that if the hessian of the phase has negative eigenvalues, a reduction in free energy of the phase separated states is  possible. Thus, we conclude that stability of all phases is necessary (and also sufficient) for stability of the phase separated state.

We finalize by showing how to write the change in free energy of a phase separated phase as a quadratic form. To do so, we use Eq.~\eqref{eq:diC} in Eq.~\eqref{eq:pp_dF}, which gives
\begin{align}
    \Delta \mathfrak{f} &= \frac12\sum_{ij}\sum_{c<C} \delta_i^{(c)}\delta_j^{(c)} w^{(c)}h_{ij}^{(c)}\nonumber\\
    & +\frac{1}{2w^{(C)}}\sum_{ij} \sum_{c<C}\delta^{(c)}_{i}w^{(c)}h_{ij}^{(C)} \sum_{d<C} \delta^{(d)}_{j}w^{(d)}  \nonumber\\ 
    & + \frac{1}{2w^{(C)}}\sum_{ij} \sum_{c<C} \epsilon^{(c)} \Delta_i^{(c)} h_{ij}^{(C)} \sum_{d<C} \epsilon^{(d)}\Delta_j^{(d)} \nonumber\\ 
    & - \frac{1}{2w^{(C)}}\sum_{ij} \sum_{c<C}  \delta^{(c)}_{i}w^{(c)}h_{ij}^{(C)}\sum_{d<C}\epsilon^{(d)}\Delta_j^{(d)} \nonumber\\ 
    & - \frac{1}{2w^{(C)}}\sum_{ij} \sum_{c<C}  \delta^{(c)}_{j}w^{(c)}h_{ij}^{(C)}\sum_{d<C}\epsilon^{(d)}\Delta_i^{(d)},
\end{align}
and where we have defined $\Delta_i^{(c)}=(\phi_i^{(C)}-\phi_i^{(c)})$. We can thus write $\Delta \mathfrak{f}$ in terms of a matrix ${\bf A}$ which has the block form
\begin{equation}\label{eq:quad_matrix_A}
{\bf A} = \left(\begin{array}{cc}{\bf P}& {\bf Q} \\ {\bf Q}^{\top} & {\bf R}\end{array}\right)\quad.
\end{equation}  
Here, the  matrix ${\bf R}$ is a square matrix  of size $C-1$ and with elements
\begin{align}
R_{cd} = \frac{1}{w^{(C)}}(\vec{\Delta}^{\,(c)})^{\top}\cdot {\bf h}^{(C)}\cdot \vec{\Delta}^{\,(d)}.
\end{align}
The matrix ${\bf Q}$ is  a rectangular matrix with dimensions $(C-1)N\times(C-1)$.   It can be defined through $(C-1)^{2}$ blocks, each corresponding to a vector of size $N$:
\begin{align}
\vec{Q}_{cd} = - \frac{w^{(c)}}{w^{(C)}}{\bf h}^{(C)}\cdot \vec{\Delta}^{\,(d)} .
\end{align}
Finally, the matrix ${\bf P}$ is square, and has size $N(C-1)$. It can be defined through  $(C-1)^{2}$ blocks each corresponding to a square matrix of size $N$:
\begin{align}
{P}_{cd} = \delta_{cd}w^{(c)}{\bf h}^{(c)} + \frac{w^{(c)}w^{(d)}}{w^{(C)}}{\bf h}^{(C)}.
\end{align}
This completes the definition of ${\bf A}$.

We note that stationarity and stability conditions of phase separated states of multicomponent mixtures can also be derived with an alternative, algebraic approach that  uses  Lagrange multipliers  and the bordered Hessian. The interested reader can find this algebraic method in SM Sec. S3.

\section{Phase separation dynamics}\label{app:dynamics}

In this section we provide a minimal framework to describe the spatio-temporal dynamics of phase separation of incompressible mixtures based on linearly irreversible thermodynamics~\cite{de2013non}. This approach, which is a straightforward generalization of Cahn-Hilliard dynamics~\cite{cahn1958free,mao2019phase}, generalizes the thermodynamic variables of multicomponent mixtures to continuous spatial fields (locally at equilibrium) that can vary over time, taking into account the role of  interfacial energy.

Consider the position vector $\bm{r}$  spanning the whole volume $V$ of the vessel that contains the mixture (hereafter we use bold-font to denote quantities that are spatial vectors). Symbolically, this can be expressed as $\int_V{\rm d}{\bm r}=V$,
which is analogous to Eq.~\eqref{eq:volfrac}. We aim at characterizing the time evolution of the solute composition fields $\phi_i(\bm{r}, t)$, with $i=1,\ldots,N$ (as in the main text, we will use arrows to denote vectors in composition space, e.g. $\vec{\phi}({\bm r}, t)$). The total abundances of components are conserved, and so 
\begin{align}
\frac1V\int_V\phi_i(\bm{r}, t){\rm d}{\bm r}=\phi_i,
\end{align}
which generalizes Eq.~\eqref{eq:const}, and shows that $\phi_i$ is dimensionless. To enforce these constraints, the composition fields follow the conservation equation
\begin{align}
\frac{\partial \phi_i(\bm{r}, t)}{\partial t} = -\nabla \cdot \bm{J}_i(\bm{r}, t).
\end{align}
where the vector field ${\bm J}_i$ is the volume flux of species $i$, and has units of volume per unit area and time, i.e. length per time.

Following the standard formalism of linearly irreversible thermodynamics~\cite{de2013non}, we take that these fluxes are linear in the gradients of the corresponding driving forces. The driving forces corresponding to the density fields are their thermodynamic conjugate variables, the exchange chemical potential fields $\bar{\mu}({\bm r},t)$, which have units of energy density. We thus  have 
\begin{align}\label{eq:flux}
\bm{J}_i(\bm{r}, t) = - \sum_{j=1}^{N} L_{ij} \nabla \bar{\mu}_j(\bm{r}, t),
\end{align}
where the transport matrix $L_{ij}$ is symmetric, which is a formulation of  Onsager's relations ($L_{ij}$ has units of area per enery and time). To determine the exchange chemical potential fields, we  define the free energy functional $\mathfrak{F}=\mathfrak{F}[\vec{\phi}({\bm r}, t)]$ as
\begin{align}
    \mathfrak{F}[\vec{\phi}(\bm{r}, t)]= \int_V {f}(\vec{\phi}(\bm{r}, t), \nabla \vec{\phi}(\bm{r}, t)) \, {\rm d} {\bm r},
\end{align}
where the free energy density $f(\vec{\phi}, \nabla \vec{\phi})$ includes a gradient dependence to account for interfacial tension, and thus generalizes the homogeneous free energy density in the main text. The exchange chemical potential fields are then given by the functional derivatives of $\mathfrak{F}[\vec{\phi}({\bm r}, t)]$ with respect to the composition fields, i.e.
\begin{align}\label{eq:local_chem_pot_definition}
\bar{\mu}_i({\bm r}, t) = \nu_i\frac{\delta \mathfrak{F}[\vec{\phi}({\bm r}, t)]}{\delta {\phi}_i({\bm r}, t)}.
\end{align}

We take as free energy density functional a simple generalization of Cahn-Hilliard expression to multicomponent mixtures \cite{cahn1958free, mao2019phase}. In this case, we have 
\begin{align}
f(\vec{\phi}, \nabla \vec{\phi}) = f(\vec{\phi}) +  \sum_{i = 1}^{N}\sum_{j = 1}^N \frac{K_{ij}}{2}(\nabla{\phi_i} )\cdot(\nabla{\phi_j}). 
\end{align}
Here $f(\vec{\phi})$ is the homogeneous free energy density of the main text, and the second term characterizes the free energy of the interface between domains with different composition, with the symmetric matrix $K_{ij}$ determining the energetic cost for gradients of different solutes  \cite{samsafransbook}, and measured in units of energy per unit length.

The equations above provide a closed dynamical system, which captures the dynamics of phase separation. To fully complete the system there are multiple model parameters that need to be provided: the total volume fractions $\vec{\phi}$, the interaction tensors, and the matrices of transport and interfacial energy. In this work, we exclusively focus on the particular case in which $L_{ij}=\ell\delta_{ij}$ and $K_{ij} = k\delta_{ij}$, where the parameter $\ell$ sets the time-scale of transport and $k$ the spatial-scale of interfaces, see SM Sec. S6 for a discussion of units.
With spatially uniform \(\ell,k\), the exchange chemical potentials and fluxes read
\begin{align}
\bar{\mu}_i(\bm r,t) &= \nu_i\!\left[\frac{\partial f}{\partial \phi_i}(\vec\phi)-k\,\nabla^2\phi_i\right],\\
\bm J_i(\bm r,t) &= -\,\ell\,\nabla \bar{\mu}_i(\bm r,t).
\end{align}
Inserting these into the continuity equation yields the multicomponent Cahn--Hilliard system of equations
\begin{align}
\frac{\partial \phi_i}{\partial t}
&= \ell\,\nu_i\,\nabla^2\!\left[\frac{\partial f}{\partial \phi_i}(\vec\phi)-k\,\nabla^2\phi_i\right],
\qquad i=1,\ldots,N.
\label{eq:CH_explicit_multi}
\end{align}
A non-dimensional version of this equation can be found in SM Sec. S6.

\section{Stationarity conditions for phase separated states in the liquid Hopfield model}\label{app:hopfield_stat}

We  derive the conditions for stationarity of a phase separated state that corresponds to a target and its complement, i.e. $C=2$ (see also SM Sec. S8 and S9 for details on calculations). Here, we generalize the results in the main text for the case of arbitrary $q$, and so the ansatz for the compositions is
\begin{align}\label{eq:ansatz_q_app}
\phi_i^{(\alpha)} &= \frac{\phi_+}{N} \left( 1 + a_+ \gamma_i^{(\alpha)} \right), \\
\phi_i^{(-\alpha)} &= \frac{\phi_-}{N} \left( 1 - a_- \gamma_i^{(\alpha)} \right),
\end{align}
where $\phi_+ = \sum_i\phi_i^{(\alpha)}$ and $\phi_- = \sum_i \phi_i^{(-\alpha)}$ are the total densities of the retrieval phase and the anti-retrieval phase, and $a_+ = \sum_i \phi_i^{(\alpha)} \gamma_i^{(\alpha)} /\phi_+$ and $a_- = -\sum_i \phi_i^{(-\alpha)} \gamma_i^{(\alpha)}/\phi_-$ are the overlap parameters that quantify the enrichment and depletion of components in these compartments in accordance to the given target. For the volume of the compartments we have that $w^{(\alpha)} = 1 - w^{(-\alpha)}$, and simplify notation by writing $w^{(\alpha)}=w$. Therefore, in order to obtain a stationary solution we need to determine five parameters: $w$, $a_+$, $a_-$, $\phi_+$ and $\phi_-$.

We start by considering the density constraints from Eq.~\eqref{eq:const}. Using the expressions in Eq.~\eqref{eq:ansatz_q_app} we obtain that the $N$ constraints reduce to two different constraints, one for components in which $\xi_i^{(\alpha)}=1$ and another for components in which $\xi_i^{(\alpha)}=0$. After some manipulations, the two constraints can be written as 
\begin{align}
\phi_+ &= \frac{\phi}{w} \, \frac{a_-}{a_+ + a_-},\label{eq:rhoP}\\
\phi_- &= \frac{\phi}{1-w} \, \frac{a_+}{a_+ + a_-}.\label{eq:rhoM}
\end{align}
These two equations allow us to compute $\phi_+$ and $\phi_-$ from the remaining three parameters.

Next, we consider the conditions of chemical balance from Eq.~\eqref{eq:chem_bal}. In each compartment the components have two possible values for the exchange chemical potential, depending on whether they are present or absent in the retrieval phase. Therefore, chemical balance provides two different equations: $\mu^{(\alpha)}_i=\mu^{(-\alpha)}_i$ for $\xi^{\alpha}_i=1$, and $\mu^{(\alpha)}_i=\mu^{(-\alpha)}_i$ for $\xi^{\alpha}_i=0$. Writing these two equations and further using Eq.~\eqref{eq:ansatz_q_app} results in two equations with three unknowns, $a_+$, $a_-$ and $w$. After some algebra, these two equations can be written as 
\begin{align}
a_+ &= n\frac{(1-w)(e^{f}-e^{g})}{(1-w)[(1-q)e^{g}+e^{f}q] +w e^{f+g}},\label{eq:aP}\\ 
a_- &=   n\frac{w(e^{f}-e^{g})}{1-w + w[(1-q)e^{f}+qe^{g}]}, \label{eq:aM}
\end{align}
where the functions $f$ and $g$ are given by
\begin{widetext}
\begin{align}
f(a_+,a_-,w) &=  v_2 \phi \frac{a_-a_+}{a_-+a_+}\left(\frac{1}{w(1-w)}\right)  \left(\frac{1-q}{n}  - \phi \frac{a_-a_+}{2(a_-+a_+)}\frac{1-2w}{(1-w)w}\right)\\   
 &- \frac{ v_3\phi^2}{2(a_-+a_+)^2} \left(\frac{a^2_-}{w^2}(1+a_+[1-q]/n)^2   - \frac{a^2_+}{(1-w)^2}(1-a_-[1-q]/n)^2 \right)\nonumber\\
 & +\frac{ v_3 \phi^3}{3(a_-+a_+)^3} \left( \frac{a^3_-}{w^3}\left(1 + 3a_+^2 + a_+^3 \frac{1-2q}{n} \right) -  \frac{a^3_+}{(1-w)^3} \left(1 + 3a_-^2 - a_-^3 \frac{1-2q}{n} \right) \right),\nonumber\\
g(a_+,a_-,w)  &=   v_2 \phi \frac{a_-a_+}{a_-+a_+}\left(\frac{1}{w(1-w)}\right)  \left(-\frac{q}{n} -  \phi \frac{a_-a_+}{2(a_-+a_+)}\frac{1-2w}{(1-w)w}\right)\\ 
&  - \frac{ v_3\phi^2}{2(a_-+a_+)^2}  \left(\frac{a^2_-}{w^2}(1-a_+q/n)^2   - \frac{a^2_+}{(1-w)^2}(1+a_-q/n)^2 \right)\nonumber\\ 
& +\frac{ v_3 \phi^3}{3(a_-+a_+)^3} \left( \frac{a^3_-}{w^3}\left(1 + 3a_+^2 + a_+^3 \frac{1-2q}{n} \right) -  \frac{a^3_+}{(1-w)^3} \left(1 + 3a_-^2 - a_-^3 \frac{1-2q}{n} \right) \right).\nonumber
\end{align}
\end{widetext}

Lastly, we consider the balance of osmotic pressures across the two compartments. This constitutes a single equation that, after some manipulations, can be written as
\begin{widetext}
\begin{align}
&-\frac{ v_2 \phi^2(a_-a_+)^2}{2(a_++a_-)^2} \left(\frac{1-2w}{(1-w)^2w^2}\right)+ \frac{ v_3 \phi^3}{3(a_-+a_+)^3} \left( \frac{a^3_-}{w^3}\left(1 + 3a_+^2 + a_+^3 \frac{1-2q}{n} \right) -  \frac{a^3_+}{(1-w)^3} \left(1 + 3a_-^2 - a_-^3 \frac{1-2q}{n} \right) \right) 
\nonumber\\ 
&= \log \left( \frac{1-w}{w}\frac{w(a_++a_-)-\phi a_-}{(1-w)(a_++a_-)-\phi a_+}\right). \label{eq:mechBalv2}
\end{align} 
\end{widetext}

Overall, Equations~\eqref{eq:rhoP}, \eqref{eq:rhoM}, \eqref{eq:aP}, \eqref{eq:aM}, and \eqref{eq:mechBalv2} constitute a system of five equations with five unknowns: $\phi_+$, $\phi_-$, $a_+$, $a_-$, and $w$. Solving this system towards the unknowns ensures that the phase separated state is stationary.  We remark that this system of equations, which is valid for arbitrary values of $q$, reduces to the one in the main text for $q=1/2$, as shown in SM Sec. S8.

\section{Stability conditions for phase separated retrieval phases in the liquid Hopfield model}\label{app:hopfield_stab}

Reference \cite{braz2024liquid} derives sufficient conditions for the stability of  stationary  phases of the form (\ref{eq:ansatz_q_app}). These conditions are given   in terms of the  coefficients 
\begin{align} 
c^{(\pm\alpha)}_1 &=  v_3 \phi_\pm (1\mp a_\pm q/n) + \frac{1}{\phi_\pm}  \frac{1}{1\mp a_\pm q/n},\\
c^{(\pm\alpha)}_2 &= \mp\frac{ v_3\phi_\pm a_\pm }{n} \pm \frac{a_\pm}{n\phi_\pm}\frac{1}{(1\pm a_\pm(1-q)/n)(1\mp a_\pm q/n)},
\end{align}
for the retrieval phase ($+\alpha$) and the anti-retrieval phase $-\alpha$.   
The conditions for stability take the form: 
\begin{align}
& - v_2 + c^{(\pm\alpha)}_1 - c^{(\pm\alpha)}_2 >0 \ \ {\rm if}\quad c^{(\pm\alpha)}_2 \geq 0 \label{eq:suf1}\\
& - v_2 + c^{(\pm\alpha)}_1 >0 \quad \quad {\rm if}\quad c^{(\pm\alpha)}_2 < 0 \quad\quad . \label{eq:suf2}
\end{align} 
Note that a phase separated state consists of pairs of retrieval, anti-retrieval phases, and both phases must satisfy the Eqs.~(\ref{eq:suf1}-\ref{eq:suf2}).

\begin{acknowledgments}
{\it Acknowledgements.} RBT acknowledges financial support from the Portuguese Foundation for Science and Technology (FCT) under the contract  2022.12272.BD. This work was financially supported by core funding from GIMM to P.S. and D.M. The authors would like to thank G.~Carugno for valuable contributions at the early stages of this work,  T.E.~Veenstra for feedback on Fig.~6, and T.~Rotheray for spotting some typos while proofreading the manuscript. We furthermore thank C.F.~Lee for valuable discussions. 
\end{acknowledgments}

\bibliography{liqhop}
\clearpage

\onecolumngrid

\section*{Supplemental Material}
\setcounter{subsection}{0}
\setcounter{equation}{0}
\setcounter{figure}{0}
\makeatletter
\renewcommand{\thesubsection}{S\arabic{subsection}}
\renewcommand{\p@subsection}{}
\renewcommand{\thesubsubsection}{\thesubsection\alph{subsubsection}}
\renewcommand{\p@subsubsection}{}
\renewcommand{\theequation}{S\arabic{equation}}
\renewcommand{\thetable}{S\arabic{table}}
\renewcommand{\thefigure}{S\arabic{figure}}
\renewcommand{\theHfigure}{S\arabic{figure}}
\renewcommand{\theHequation}{S\arabic{equation}}
\makeatother

\subsection{Stationarity and stability conditions in the grand canonical ensemble\label{app:grandcanonical_formalism}}
The mathematical framework developed in Appendices B, C and D of the main text to study the stationarity and the stability of phase separated states can be easily adapted to tackle the same problem in the grand canonical ensemble, \textit{i.e.} when the composition of the system is allowed to fluctuate. In this scenario, the central thermodynamic quantity is not the free energy $f$, but rather the grand potential

\begin{equation}
    g(\vec{\phi}) = f(\vec{\phi}) - \sum_{i = 1}^{N}\mu_i\phi_i.
\end{equation}

Our goal will then be to study the response of the grand potential of the phase separated system
\begin{equation}\label{eq:grandpotential_phaseseparated}
    \g = \sum_{c = 1}^{C}w^{(c)}g^{(c)}
\end{equation}
upon perturbation of the compartments' volume fractions ($\tilde{w}^{(c)} = w^{(c)} + \epsilon^{(c)}$) and compositions ($\tilde{\phi}_i^{(c)} = \phi_i^{(c)} + \delta^{(c)}_i$), and with $g^{(c)} = g(\vec{\phi}^{(c)})$.
The main technical difference with respect to the treatment for the canonical ensemble is that, since the system composition is allowed to fluctuate, we do not need to enforce any constraint on the $\delta^{(c)}$'s. The grand potential of the perturbed state reads (up to second order in perturbation)

\begin{multline}\label{eq:grandpotential_expansion}
\tilde{\g} \simeq \g + \sum_{c=1}^{C}w^{(c)}\sum_{i=1}^{N}\bigg(\frac{\partial f}{\partial \phi_{i}}(\vec{\phi}^{(c)}) - \mu_i\bigg)\delta^{(c)}_i + \sum_{c < C}\epsilon^{(c)}(g^{(c)} - g^{(C)})  \\
+\sum_{c=1}^{C}\epsilon^{(c)}\sum_{i=1}^{N}\bigg(\frac{\partial f}{\partial \phi_{i}}(\vec{\phi}^{(c)}) - \mu_i\bigg)\delta^{(c)}_i + \frac{1}{2}\sum_{c=1}^{C}w^{(c)}\sum^N_{i,j=1}\frac{\partial^2 f}{\partial \phi_i \partial \phi_j}(\vec{\phi}^{(c)})\delta^{(c)}_i\delta^{(c)}_j.
\end{multline}

Now, stationarity is achieved when the first order terms in the expansion equal to $0$. This leads to the conditions
\begin{equation}
\frac{\partial f}{\partial \phi_{i}}(\vec{\phi}^{(c)}) - \mu_i = 0, 
\end{equation}
for all $i=1,2,\ldots,N$, and 
\begin{equation}
    g^{(c)} - g^{(C)} = 0,
\end{equation}
for all $c=1,2,\ldots, C-1$.   Hence, the conditions to have a stationary state are that i) the $\vec{\phi}^{(c)}$ are stationary points of the grand potential $g(\vec{\phi})$ and ii) the grand potential is the same in all compartments.

For the stability of the stationary states, we need to check that every perturbation ultimately leads to an increase in the grand potential, \textit{i.e.} $\tilde{g} - g > 0$. 

For a stationary state,  the term of order $\epsilon\delta$ vanishes in Eq.~\eqref{eq:grandpotential_expansion}, and thus the stability condition reads
\begin{equation}  \sum_{i,j}\frac{\partial^2 f}{\partial \phi_i \partial \phi_j}(\vec{\phi}^{(c)})\delta^{(c)}_i\delta^{(c)}_j > 0
\end{equation}
Thus, a stationary state is metastable if the hessian of the grand potential in each compartment is positive definite (notice that the grand potential and the free energy have the same hessian).  Note  that \textit{every} perturbation with  $\delta_{i}^{(c)}=0$ leaves the grand potential of the stationary state unchanged. Such perturbations are the grand canonical equivalent of the soft modes (see Sec. IV of the main text). 

To summarize, a sufficient condition to have metastable phase separation in the grand canonical ensemble is that the grand potential $g(\vec{\phi})$ admits at least two minima and that the value of the grand potential in those minima is the same.  This typically happens for specific critical values of the chemical potentials that have to be finely tuned.

\subsection{Stationarity and stability conditions for spatially homogeneous states}
\label{app:homo_stab}
We show that (i) spatially homogeneous states are stationary, and (ii) that  a homogeneous state is metastable if and only if the Hessian of its free energy is positive definite.  

We verify both points explicitly by   analyzing how the free energy change of an arbitrary homogeneous state changes under all possible perturbations of this state that respect the constraints given by Eqs. (8) and (9) of the main text of the canonical ensemble. Note that due to these latter constraints, in a homogeneous state the only perturbations that are allowed for are those corresponding with the nucleation of one or more distinct phases. 

The simplest way to represent a homogeneous state is as a state consisting of a single compartment, $C=1$. However, for $C=1$ there exist no perturbations that are compatible with the constraints given by Eqs. (8) and (9) of the main text.  Therefore, to allow for perturbations we  represent the target, homogeneous state in terms of multiple compartments $C>1$.    Notably, if the perturbation corresponds with the nucleation of $P$ distinct phases, then we  represent the homogeneous state using  at least $P$ compartments.  

Let us therefore represent the homogeneous state as a configuration of the form $\{w^{(c)}\}$ and $\{\vec{\phi}^{(c)}\}$ with $\vec{\phi}^{(c)}=\vec{\phi}$ for all compartments $c$; the latter conditions ensures that the state is homogeneous.   The volumes $w^{(c)}$ may appear arbitrary, but in fact as we will see, they   determine the size of the nucleated phases in the perturbation.   

Next we consider a perturbed state for which the compartment volumes take the  form $\tilde{w}^{(c)} = w^{(c)} + \epsilon^{(c)}$ and the compositions are given by $\tilde{\phi}^{(c)}_i = \phi^{(c)}_i + \delta^{(c)}_i$. As shown in Appendix B of the main text, assuming the perturbation is small enough so that $|\epsilon^{(c)}|\ll 1$ and $|\delta^{(c)}_i|\ll 1$, we can express the free energy $\tilde{\f}$ of the perturbed state as 
\begin{equation}
\tilde{\f} = \f + \frac{1}{2}\sum_{c=1}^Cw^{(c)} \sum_{ij}\delta^{(c)}_i\delta^{(c)}_{j}h_{ij} , 
 \label{eq:tildeF}
\end{equation} 
where $\f$ is the free energy of the target homogeneous state and $w^{(C)}=1-\sum_{c<C}w^{(c)}$. 
This equation follows from the Eq. (B6) of the main text and using that in the homogeneous phase  $\vec{\phi}^{(c)}=\vec{\phi}$, $\partial_i f^{(c)} = \partial_i f$, and $h^{(c)}_{ij} = h_{ij}$ for all compartments $c$.  Notice that in  \eqref{eq:tildeF} we neglected contributions to the free energy that are of third order or higher in the perturbation variables.       

There are two interesting points to observe from Eq.~\eqref{eq:tildeF}.  First, there is no linear term in the perturbation.   This implies that all homogeneous states are stationary, as claimed before.   Second,  the second order term in the perturbed free energy is positive if and only if the Hessian $\mathbf{h}$ of the homogeneous state is positive definite.   Consequently, if $\mathbf{h}$  is positive definite, then  the free energy increases for all possible physical perturbations that respect the constraints Eqs. (8) and (9) of the main text. This proves our second claim that the homogeneous state is metastable if and only if the hessian of the free energy, $\mathbf{h}$, is positive definite. 

Note that we did not consider the nucleation of phases  with entirely different composition to the homogeneous state.   In the study of metastable states, such perturbations are discarded, as a new phase requires a  finite nucleation size, and thus these perturbations are not small.   In physical systems, the nucleation size is determined by a balance between surface tension and the bulk free energy of the phase.

\subsection{Algebraic approach to metastability}\label{app:border}

Besides the geometric approach developed in the Main Text, there is also   an alternative algebraic formalism to metastable phase separation.  As the distinct formalisms can be advantageous in different contexts,   we summarize here  for completeness the algebraic approach.

The starting point of this formalism is to define a  potential $\g$ that extends $\f$ by including Lagrange multipliers to account for the constraints in Eqs. (8) and (9) of the main text. We thus write
\begin{align}\label{eq:Ftot}
\g = \f -\sigma\left(\sum^C_{c=1}w^{(c)}-1\right)- \vec{\lambda}\cdot\left(\sum^C_{c=1} w^{(c)} \phic-\vec{\phi}\right),
\end{align}
where the Lagrange multipliers $\sigma$ and $\vec{\lambda}$ enforce the constraints in Eqs. (8) and (9) of the main text, respectively. In total, $\g$ has $N+1$ variables in addition to the original $C(N+1)$ of $\f$. Note that, when evaluated at its minimum value compatible with the constraints, $\g$ turns into the grand potential.

The stationarity conditions for a phase separated state are directly obtained from setting the gradient of $\g$  to zero.  Setting ${\partial \g}/{\partial \phi_i^{(c)}}=0$, we find $CN$ equations 
\begin{align}\label{eq:mui}
\lambda_i = \left(\frac{\partial  f}{\partial \phi_i}\right)_{T,\vec{\phi}=\vec{\phi}^{(c)}},
\end{align}
which are  statements of chemical balance between the phases. These conditions imply that at stationarity the exchange chemical potentials of each species $i$ is the same across all compartments, i.e. $\bar{\mu}_i^{(c)}=\bar{\mu}_i^{(c')}$ for all pairs of compartments $c$ and $c'$. Setting ${\partial{\g}}/{\partial{w^{(c)}}}=0$ yields the $C$ conditions
\begin{align}\label{eq:Pi}
\sigma = f^{(c)}-\sum^N_{i=1}\lambda_i\phi_i^{(c)},  
\end{align}
which are statements of mechanical balance between the phases. This equation implies that the osmotic pressures of all compartments are identical~\cite{rubinstein2003polymer, adame2020liquid}, and so $\Pi^{(c)} = \Pi^{(c')}$ for all pairs of compartments $c$ and $c'$. Setting to zero the derivatives with respect to $\lambda_i$ and $\sigma$ yields the $N+1$ constraints in Eqs. (8) and (9) of the main text.

Studying  stability requires to analyze the convexity of the free energy, $\f$, subject to the constraints  in Eqs. (8) and (9) of the main text. To this end, we use the  bordered Hessian (see Chapter 4 of Ref.~\cite{colley2011vector}) of $\f$, denoted by ${\bf bH}$, which is the Hessian of  $\g$ towards its variables $(\{w^{(c)}\},\{\phic\},\vec{\lambda},\sigma)$. To obtain the bordered Hessian we compute it by blocks. The non-null terms are 
\begin{align}
\frac{\partial^2 \g }{\partial \phi^{(c)}_i\partial \phi^{(c')}_{i'}}&= \delta_{c,c'}   w^{(c)}\frac{\partial^2 f }{\partial \phi^{(c)}_i\partial \phi^{(c')}_{i'}},\\
\frac{\partial^2 \g }{\partial w^{(c)}\partial \lambda_i} &=  -\phi^{(c)}_i,\\
\frac{\partial^2 \g }{\partial w^{(c)}\partial  \sigma} &=  -1,\\
\frac{\partial^2 \g }{\partial \phi^{(c)}_i\partial \lambda_j} &=  -w^{(c)} \delta_{i,j}.
\end{align}
We can write the bordered hessian matrix  ${\bf bH}$ as a block matrix with blocks of dimension $(N+1)\times (N+1)$, viz., 
\begin{align}\label{eq:bordered}
{\bf bH}=
\begin{pmatrix}
  {\bf 0}&-{\bf B}^{(1)}&-{\bf B}^{(2)}&\ldots&-{\bf B}^{(C)}\\
  -({\bf B}^{(1)})^{\top}  &{\bf C}^{(1)}&{\bf 0}&\ldots&{\bf 0}\\
  -({\bf B}^{(2)})^{\top}  &{\bf 0}&{\bf C}^{(2)}&\ldots&{\bf 0}\\
  \vdots  &\vdots&\vdots&\ddots&{\bf 0}\\
  -({\bf B}^{(C)})^{\top}  &{\bf 0}&{\bf 0}&\ldots&{\bf C}^{(C)}\\
\end{pmatrix}
\end{align}
where ${\bf 0}$ is the square null matrix of size $N+1$, and the remaining blocks are defined as
\begin{align}
{\bf B}^{(c)}&=
\begin{pmatrix}
  1&\vec{0}\\
  (\phic)^{\top}& w^{(c)}{\bf 1}
\end{pmatrix},
\\
{\bf C}^{(c)}&=
\begin{pmatrix}
  0&\vec{0}\\
  \vec{0}^{\,\top}& w^{(c)}{\bf h}^{(c)}
\end{pmatrix},
\end{align}
where $\vec{0}$ denotes a  column vector of size $N$ with null entries, ${\bf 1}$ denotes the identity matrix of size $N$, and the matrix ${\bf h}^{(c)}$ is the hessian of $f$ evaluated at compartment~$c$.

The stability conditions for stationary points in a constrained function in terms of the bordered Hessian are different from the more familiar conditions for an unconstrained function in terms of the Hessian \cite{colley2011vector}. For a problem with $n$ variables and $m$ constraints (here $n=(N+1)(C+1)$ and $m=N+1$), the stability conditions are determined by the signs of  the determinants of the submatrices ${\bf bH}_{2m+k}$ with $k=1,2,\ldots,n-m$, where  ${\bf bH}_s$ is the $s-${th} leading principal minor of ${\bf bH}$, i.e., the upper left square submatrix of size $s\times s$.  Specifically, a stationary state is a local minimum if 
\begin{align}\label{eq:bHcond}
(-1)^{m}\det({\bf bH}_{2m+k})>0
\end{align}
for $k=1,\ldots, n-m$, a local maximum  if the  determinant of ${\bf bH}_{2m+1}$ is negative and all subsequent ones alternate sign, and a saddle point when neither of these conditions are satisfied.

While the bordered Hessian formalism is mathematically straightforward, analyzing the conditions in Eq.~\eqref{eq:bHcond} poses practical challenges. In particular, it is not {\it a priori} clear what is the relationship between these conditions, which determine stability of a phase separated state, and conditions on the stability of each of the constituting phases, which are determined by  $P$ hessians. The following table relates this algebraic formalism to the geometric formalism followed in the Main Text.

\begin{center}
\begin{tabular}{c | c | c} \label{table:Sum}
 Thermodyn. & Algebraic & Geometrical  \\ 
 \hline\hline
 state & point in $\mathbb{R}^{C(N+1)}$ & $C$ points in $\mathbb{R}^{N+1}$  \\ 
 \hline
 potential & $\g$ & $\f$  \\ 
 \hline
 stationarity & zero gradient & tangent simplex \\
 \hline
 local stability & bordered  hessian & constrained convex  \\
 \hline
 global stability & lowest free energy & convex hull\\
 \hline
\end{tabular}
\end{center}

\subsection{Binary mixture with quadratic interactions}\label{app:binmixbook}
For completeness, we  review the physics of binary mixtures with quadratic interactions~\cite{rubinstein2003polymer}. In this case the free energy of Eq. (15) of the main text has the energetic contribution 
\begin{align}
    u(\phi) = -\frac{b}{2}\left(\phi-\frac12\right)^2, \label{eq:ubinaryQ}
\end{align}
where we have assumed that the free-energy is symmetric around $1/2$. The spinodal manifold solves $f''(\phi)=0$, which for  $b>4$ and $\nu_0=1$ yields
\begin{align} 
\phi^\pm_{\rm sp} = \frac{1}{2} \pm  \sqrt{1/4-1/b}, \label{eq:rhosp}
\end{align}
which determines the two branches of the spinodal line.

For a phase separated state, the stationarity conditions are those in Eqs. (10) and (11) of the main text. Due to the symmetry of the free-energy around $\phi=1/2$ , we look for solutions of the form  $\phi_{\rm bin} = \phi^{(1)} = 1-\phi^{(2)}$, where the subindex indicates that these solutions determine the binodal line. Inserting on the stationarity conditions, and after some algebra, this yields 
\begin{align}
\phi_{\rm bin} = (1-\phi_{\rm bin}) \exp\left(b(\phi_{\rm bin}-1/2)\right).
\end{align}
Defining now $\phi_{\rm bin} =  (1+\gamma)/2$ we are left to solve
\begin{align}\label{eq:tanh}
\gamma = \tanh(b\gamma/4). 
\end{align}
Hence, the binodals are given by 
\begin{align}
\phi^{\pm}_{\rm bin} = (1 + \gamma_{\pm})/2 \label{eq:rhoBin}
\end{align}
where $\gamma_{\pm}$ are the positive  and negative solutions of Eq.~\eqref{eq:tanh}.

\begin{figure}[h!]\centering
     \includegraphics[width=0.5\textwidth]{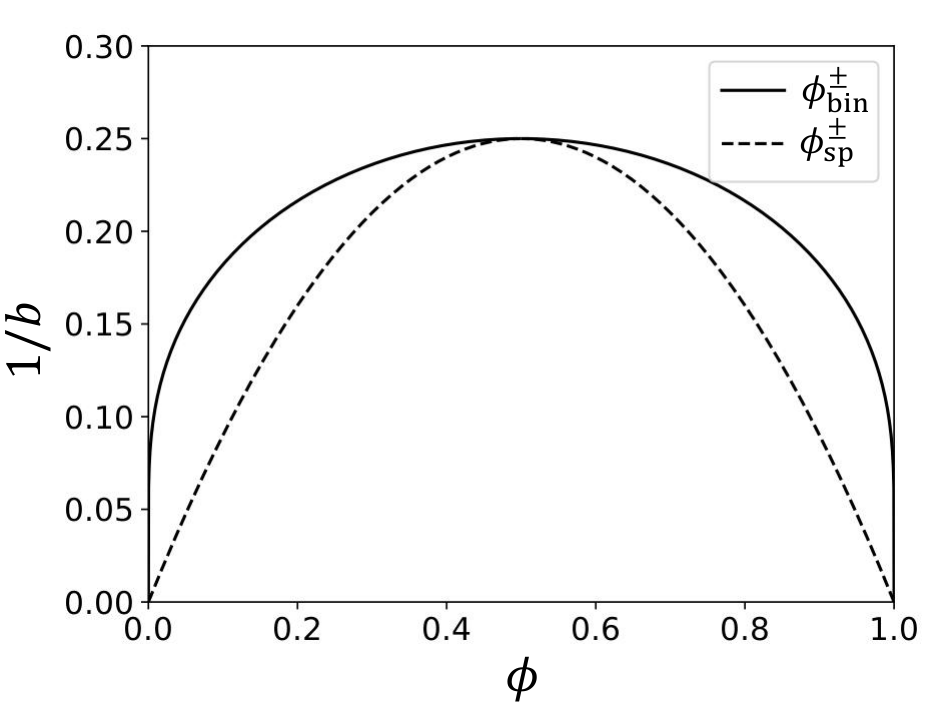}
\caption{Plot of the spinodal and binodal lines in a binary mixture with the quadratic energy function given by Eq.~(\ref{eq:ubinaryQ}).   The functions plotted are given by the Eqs.~\eqref{eq:rhosp}  and    \eqref{eq:rhoBin} for the spinodal and the binodal, respectively.}\label{fig:a}  
\end{figure}

In Figure~\ref{fig:a} we plot the binodal and spinodal lines. Note that, while the above only concerns stationarity, the phase separated state for quadratic interactions is always stable. This is so because $f''(\phi^{(1)})>0$ and $f''(\phi^{(2)})>0$.

\subsection{Binary mixture with quartic interactions}\label{SI:quartic}
As for the case of quadratic interactions (see Sec.~\ref{app:binmixbook}), the spinodal manifold can be computed analytically  for the quartic potential, Eq.~(\ref{eq:ubin}) in the main text.    Solving $f''(\phi)=0$ yields the solution
\begin{align}\label{eq:spinquart}
\phi_{\rm sp} &= \frac{1}{4}\left(2\mp \sqrt{2\pm 2\sqrt{\frac{-64+c}{c}}}\right)\quad.
\end{align}

Let us now focus on the binodal.
According to Gibbs phase rule, the phase separated states of a binary mixture consist of two compartments. The stationarity conditions Eqs. \eqref{eq:chem_bal}, and \eqref{eq:press_bal} for the phase separated states are then given by 

\begin{align}\label{eq:chem_bal_quartic}
\bar{\mu}^{(1)}=\bar{\mu}^{(2)},
\end{align}
and 
\begin{align}\label{eq:press_bal_quartic}
\Pi^{(1)} = \Pi^{(2)}.
\end{align}
The osmotic pressure and exchange chemical potential of each compartment is 
\begin{align}
\Pi^{(1)}=f(\phi^{(1)})-\bar{\mu}^{(1)}\phi^{(1)} \quad {\rm and} \quad
\Pi^{(2)}=f(\phi^{(2)})-\bar{\mu}^{(2)}\phi^{(2)}\quad.
\end{align}

\begin{align}
\bar{\mu}^{(1)}=f'(\phi^{(1)}) \quad  {\rm and} \quad 
\bar{\mu}^{(2)}=f'(\phi^{(2)})\quad.
\end{align}

Using Eq. \eqref{eq:chem_bal_quartic} we can simplify condition  \eqref{eq:press_bal_quartic} in the following way

\begin{align}\label{eq:press_bal_quartic_simp}
-\bar{\mu}^{(1)}+\frac{f(\phi^{(2)})-f(\phi^{(1)})}{\phi^{(2)}-\phi^{(1)}} = 0 .
\end{align}

Using the free-energy in Eq. (15) of the main text with the quartic energy function in Eq. (16) of the main text, the osmotic pressure and exchange chemical potential for a compartment $j$ becomes

\begin{align}
\bar{\mu}^{(j)} &= -\frac{c}{3\nu_0} \left( \phi^{(j)} - \frac{1}{2} \right)^3 + \frac{1}{\nu_0} \log \left( \frac{\phi^{(j)}}{1 - \phi^{(j)}} \right)
\\
\Pi^{(j)} &= -\frac{c}{12\nu_0} \left( \phi^{(j)} - \frac{1}{2} \right)^4 
+ \frac{1}{\nu_0} \phi^{(j)} \log \phi^{(j)} 
+ \frac{1}{\nu_0} (1 - \phi^{(j)}) \log (1 - \phi^{(j)}) 
+ \frac{c}{3\nu_0} \left( \phi^{(j)} - \frac{1}{2} \right)^3 \phi^{(j)} 
- \frac{1}{\nu_0} \phi^{(j)} \log \left( \frac{\phi^{(j)}}{1 - \phi^{(j)}} \right)
\label{eq:stat_quart1}
\end{align}

We solve Eqs.~\eqref{eq:chem_bal_quartic}, and \eqref{eq:press_bal_quartic_simp}  for $(\phi^{(1)},\phi^{(2)})$ to obtain the spinodal lines in Fig. 3 of the main text. We can write these conditions explicitly in the following way

\begin{align*}
0 &= - \frac{c}{3} \left[ \left( \phi^{(1)} - \frac{1}{2} \right)^3 - \left( \phi^{(2)} - \frac{1}{2} \right)^3 \right] + \log \left( \frac{\phi^{(1)} (1 - \phi^{(2)})}{\phi^{(2)} (1 - \phi^{(1)})} \right)
 \\
\\
0 &= \frac{c}{3\nu_0} \left( \phi^{(1)} - \frac{1}{2} \right)^3 
- \frac{1}{\nu_0} \log \left( \frac{\phi^{(1)}}{1 - \phi^{(1)}} \right) 
- \frac{c}{12\nu_0} \cdot \frac{ \left(\phi^{(2)} - \frac{1}{2}\right)^4 - \left(\phi^{(1)} - \frac{1}{2}\right)^4 }{\phi^{(2)} - \phi^{(1)}} 
\\
\\ + &\frac{1}{\nu_0} \cdot \frac{ \phi^{(2)} \log \phi^{(2)} - \phi^{(1)} \log \phi^{(1)} + (1 - \phi^{(2)}) \log (1 - \phi^{(2)}) - (1 - \phi^{(1)}) \log (1 - \phi^{(1)}) }{\phi^{(2)} - \phi^{(1)}}
\end{align*}

Fig. 3C of the main text shows the regions of stability for the different phase separated states in  a binary mixture with a quartic potential.  A corresponding phase diagram showing the binodals is plotted in Fig.~\ref{fig:si_fig3} by selecting the lowest free-energy amongst the metastable phase separated states (this is the free energy Eq. (7) of the main text for two compartments).

\subsection{Non-dimensional formulation of Cahn-Hilliard equations}
\label{si:nondim}
We now discuss how to transform the Cahn-Hilliard equation into  their dimensionless form. We will here consider the general case of a multicomponent mixture. Consider the $i$-th component: its density field evolves according to (see Eq. (E9) of the main text)
\begin{equation}
    \frac{\partial \phi_i(\mathbf{r},t)}{\partial t} = \ell \nu_0 \nabla^2\bigg[\frac{\partial f}{\partial\phi_i}(\vec{\phi}(\mathbf{r},t)) - k\nabla^2 \phi_i(\mathbf{r},t)\bigg]
\end{equation}
and this equation is to be solved for spatial coordinates $\mathbf{r}\in [0,L] \times [0,L]$ with periodic boundary conditions. Here, $\nu_0$ has the dimensions of an area, $f$ has dimensions of energy per unit area, $k$ has dimensions of energy, and $\ell$ has dimensions of area per unit energy and per unit time. We have set $\nu_i = \nu_0$ for all species $i$.  As we aim to derive an adimensional form of the Cahn-Hilliard equations, we are \textit{not}  yet assuming $k_B T = 1$. In fact, as a first step, we are going to multiply and divide the r.h.s. by $k_B T$, thus obtaining
\begin{equation}
\frac{\partial \phi_i(\mathbf{r},t)}{\partial t} = \ell k_B T \nu_0 \nabla^2\bigg[\frac{\partial \beta f}{\partial\phi_i}(\vec{\phi}(\mathbf{r},t)) - k \beta \nabla^2 \phi_i(\mathbf{r},t)\bigg],
\end{equation}
where $\beta = 1/k_B T$. Then, since we will solve the equation on a square domain of size length $L$, we introduce the rescaled position vectors $\tilde{\mathbf{r}} = \mathbf{r}/L$: in these new dimensionless variables, the equation is solved in the region $\tilde{\mathbf{r}}\in [0,1]\times[0,1]$ and it reads

\begin{equation}
\frac{\partial \phi_i(\tilde{\mathbf{r}},t)}{\partial t} = \frac{\ell k_B T}{L^2} \tilde{\nabla}^2\bigg[\nu_0\frac{\partial \beta f}{\partial\phi_i}(\vec{\phi}(\tilde{\mathbf{r}},t)) - \frac{k \beta \nu_0}{L^2} \tilde{\nabla}^2 \phi_i(\tilde{\mathbf{r}},t)\bigg],
\end{equation}
where $\tilde{\nabla}$ indicates the gradient towards $\tilde{\mathbf{r}}$. We call $\kappa = k\beta\nu_0/L^2$, which is an adimensional parameter. Also, we introduce $\tau = L^2/(\ell k_B T)$, which has the dimensions of time. Notice that $\tau$ can be interpreted as the time required for a free particle to diffuse over a distance $L$.  As a final step, we rescale time by introducing $\tilde{t} = t/\tau$. The equation is reduced into its adimensional form 
\begin{equation}\label{eq:nondim_CH_final}
\frac{\partial \phi_i(\tilde{\mathbf{r}},\tilde{t})}{\partial \tilde{t}} =  \tilde{\nabla}^2\bigg[\nu_0\frac{\partial \beta f}{\partial\phi_i}(\vec{\phi}(\tilde{\mathbf{r}},\tilde{t})) - \kappa\tilde{\nabla}^2 \phi_i(\tilde{\mathbf{r}},\tilde{t})\bigg],
\end{equation}
where $\tilde{\mathbf{r}}\in[0,1]\times [0,1]$.
In this  form, the effect of the surface tension $k$ is observed by modifying the adimensional parameter $\kappa$ (instead, the width (or height) of the box  is fixed to $1$).  The effect of  the mobility $\ell$  on the phase separation kinetics can  be analysed by simulating the system at different time scales, as changing $\ell$ changes  the  time unit $\tau$.

\subsection{Numerical implementation}\label{sec:NI}

\subsubsection{Solving the stationarity conditions}
The stationarity conditions in Eqs. (10) and (11) of the main text (explicit expressions for the non-linear binary mixture in \ref{SI:quartic} and liquid Hopfield model in \ref{si:lhm_stat}) are solved using \texttt{scipy.optimize.fsolve}.

\subsubsection{Simulating the multicomponent mixture}\label{SI:numerics}
The dynamics of phase separation is modeled by the Cahn-Hilliard system of equations (E9) in the main text. This system of equations (in its adimensional form, see Eq.~\eqref{eq:nondim_CH_final}) is solved numerically in the $[0, 1] \times [0, 1]$ region, discretized into a $64 \times 64$ lattice, with periodic boundary conditions. We employ a semi-implicit integration scheme~\cite{Chen1998IMEX, mao2019phase, brenner}, in which the density field of component $i$ at a given lattice site evolves according to
\begin{equation}
\frac{\phi_{i}^{n+1} - \phi_{i}^{n}}{\Delta \tilde{t}} = \tilde{\nabla}^2 G[\{\phi_i\}^{n}] -\kappa \tilde{\nabla}^4 \phi_i^{n+1},
\end{equation}
where $\Delta\tilde{t}$ is the timestep, the superscript $n$ denotes the $n$-th timestep, and we have defined 
\begin{equation}
G[\{\phi_i\}^{n}] = \nu_{0}\beta \frac{\partial f}{\partial \phi_i}(\{\phi_i\}^{n}).
\end{equation}

By indicating with $\,\, \widehat{   }\,\,$ the Fourier transform, the equation in Fourier space becomes
\begin{equation}  \frac{\widehat{\phi}_i^{n+1} - {\widehat{\phi}_{i}}^{n}}{\Delta \tilde{t}} = -q^2\widehat{G}[\{\phi_i\}^{n}] - \kappa q^4{\widehat{\phi}_{i}}^{n+1}
\end{equation}
where $q$ indicates the wave vector. Rearranging and solving for ${\widehat{\phi}_{i}}^{n+1}$ we get
\begin{equation}
\widehat{\phi}_{i}^{n+1} = \frac{\widehat{\phi}_{i}^{n} - \Delta \tilde{t}\, q^2\,\widehat{G}[\{\phi_i\}^{n}]}{1 + \Delta \tilde{t}\,\kappa\,q^{4}}
\end{equation}
The Fast-Fourier-Transform method is used to switch between real and Fourier space.

\subsubsection{Algorithm to identify phases}\label{SI:pca_method}

The procedure is that described in \cite{mao2019phase, brenner}. It begins with a gradient-based filtering step. Specifically,  the gradients $\nabla\phi_i$ of the local volume fractions of each of the $N$ components,   $i = 1, \dots, N$,  are computed.   Any spatial point for which $\max_i |\nabla \phi_i|^2$ exceeds a prescribed threshold is discarded, as it is considered to lie at an interface between two  phases.

A principal component analysis in composition space is then performed on the remaining lattice sites. Owing to the  symmetry between retrieval phases and anti-retrieval phases, the total number of phases is estimated as twice the number of significant eigenvalues obtained from the principal component analysis. Here “significant” refers to eigenvalues exceeding the Marchenko–Pastur threshold~\cite{Marvcenko1967, brenner}, which accounts for the non-zero eigenvalues that would arise  from a purely random configuration (details on the values of the gradient and Marchenko-Pastur thresholds can be found in~\ref{SI:figs_details}).
Once the number of phases has been estimated, a K-means clustering algorithm is performed, assigning each spatial point to a corresponding phase. Finally, the overlaps of each identified phase with each of the targets are computed.

\subsubsection{Extended figure captions}\label{SI:figs_details}
In what follows, we list the simulation details for Figs.4, 6, 7 and 8 of the main text. In all cases, results were obtained by solving  a non-dimensional version of Eq. (17) of the main text (see SI~\ref{si:nondim}).  We recall that simulations are performed on the $[0, 1] \times [0, 1]$ region, which has been discretized into a $64 \times 64$ grid. 

\begin{itemize}
    \item \textbf{Fig.4} - The initial volume fractions $\phi_i(\tilde{\mathbf{r}})$ of the solute components are  sampled for each lattice point $\tilde{\mathbf{r}}$ independently  from a Gaussian distribution with a standard deviation $\sigma_{\rm init} = 0.02$ and a mean   $\mu_{\rm init}$ that depends on the panel.   For Panel \textbf{A.} $\sigma_{\rm init} = 0.02$ and $\mu_{\rm init} = 0.75$; for Panel \textbf{B.} $\sigma_{\rm init} = 0.02$ and $\mu_{\rm init} = 0.25$; for Panel \textbf{C.}   the initial condition contains a square region of linear size $28$ lattice sites that is enriched in the solute, namely,  $\mu_{\rm init} = 0.85$.  Outside the square,   the mean $\mu_{\rm init}$ is determined by the condition $\phi = 0.25$.  
    All simulations were run for a total time $\tilde{t} = 1.0$ with integration step $\Delta \tilde{t} = 10^{-8}$ and  $\kappa = 10^{-3}$.
    
    \item \textbf{Fig.6} - The initial configuration contains a circular-shaped seed that is enriched (or depleted)  
    in the first four components of the mixture based on one of the three prescribed target vectors $\vec{\gamma}^{(\alpha)}$ (for Panel A this is $\alpha=1$ and for Panel B this is $\alpha=2$).   In particular,  for the first four components $i=1, 2, 3, 4$,  the $\phi_i(\tilde{\mathbf{r}})$ are enriched if  $\gamma^{(\alpha)}_i = 1$ ($\phi_i(\tilde{\mathbf{r}}) = 0.17$) and  the $\phi_i$ are depleted  ($\phi_i(\tilde{\mathbf{r}}) = 0.03$) if  $\gamma^{(\alpha)}_i = -1$. The four remaining components $i=5, 6, 7, 8$ are set equal to the uniform value $\phi_i(\tilde{\mathbf{r}})= \phi/N =0.1$.
    The radius of the seed is $1/3$ of the total length of the system.   The density of the  components outside the circular region are set by the condition that the total $\phi_i = \int d\tilde{\mathbf{r}}\: \phi_i(\tilde{\mathbf{r}}) = 0.1$.     A gaussian noise with zero mean and standard deviation $10^{-3}$ is added to each lattice site independently. Simulations were run for a total time $\tilde{t} = 0.5$, with $\Delta\tilde{t} = 10^{-7}$ and $\kappa = 5\times10^{-3}$.
    
    \item \textbf{Fig.7} - The initial $N$-component composition vectors $\vec{\phi}(\tilde{\mathbf{r}}, 0)$ are sampled as    $\vec{\phi}(\tilde{\mathbf{r}},0) = \phi  \: \vec{x}(\tilde{\mathbf{r}})$, where the $\vec{x}(\tilde{\mathbf{r}})$ are  independent and identically distributed random vectors drawn from the symmetric Dirichlet distribution, 
    \begin{equation}
    p(\vec{x}) =
    \Gamma(N\alpha)\prod^{N}_{i=1}\frac{x^{\alpha-1}_i}{\Gamma(\alpha)},   
    \end{equation}
    defined on the standard simplex $x_i\in [0,1]$ for all $i$ and with $\sum^N_{i=1}x_i = 1$.  The  $\Gamma(\alpha) = \int^{\infty}_0 t^{\alpha-1}e^{-t}{\rm d}t$ is the Gamma function.   This guarantees that the sum of all composition vectors is exactly $\phi$ for each value of $\tilde{\mathbf{r}}$, and the average density of each component is $\phi/N$. The standard deviation of the density of each component is given by $\phi/N \sqrt{(N - 1)/(N\alpha + 1)}$ (in the specific case of Figure 7, $\phi=0.8$ and  $\alpha=5$). The targets used in Fig. 7 are $\vec{\gamma}^{(1)} = (+ + + - - - + -)$, $\vec{\gamma}^{(2)} = (+ - - - + - + +)$ and $\vec{\gamma}^{(3)} =(+ + - - - + - +)$. Simulations were run for a total time $\tilde{t} = 0.1$, with $\Delta \tilde{t} = 10^{-8}$ and $\kappa = 10^{-3}$.

    \item \textbf{Fig.8} - The initial conditions are the same as in Fig. 7. The $p$ target vectors for a given $N$ are chosen by randomly picking $p$ columns from a $N \times N$ Hadamard matrix generated using Paley's construction~\cite{paley1933orthogonal}: this ensures both $q = 1/2$ and orthogonality. In order to make such construction possible, we used $N = 8, 12, 20$ and $24$. Simulations are run for a total time of $\tilde{t} = 0.1$ for $N = 8, 12, 20$ and $\tilde{t} = 0.05$ for $N = 24$, with $\Delta \tilde{t} = 10^{-8}$ and $\kappa = 0.001$. The average overlap is calculated as follows: for each of the $P$ identified phases, the matrix $a^{(\nu)}(\vec{\phi}^{(\alpha)})$ is computed, with $\nu = 1, \dots, p$ and $\alpha = 1, \dots, P$. Then, the value $\max_{\nu} |a^{(\nu)}(\vec{\phi}^{(\alpha)})|$ is selected for each $\alpha$. Finally, a weighted average of these overlaps is performed, using the  volume fractions of the $P$ identified phases as weights.
\end{itemize}

Concerning Figs. 7 and 8, we provide in Table~\ref{tab:threshold} the values adopted for the gradient threshold (to distinguish between phases and interfaces) and for the Marchenko-Pastur threshold (to discard principal component analysis eigenvalues due to noise).

\begin{table}
\centering
\begin{tabular}{|c|c|c|c|c|}
\hline
 & $N = 8$ & $N = 12$ & $N = 20$ & $N = 24$ \\
\hline
gradient threshold & $1.0 \times 10^{-3}$ & $7.5 \times 10^{-4}$ & $3.5 \times 10^{-4}$ & $3.0 \times 10^{-4}$ \\
\hline
Marchenko - Pastur threshold & $5.0 \times 10^{-3}$ & $2.0 \times 10^{-3}$ & $8.0 \times 10^{-4}$ & $5.0 \times 10^{-4}$ \\
\hline
\end{tabular}
\caption{Values of the gradient threshold and Marchenko - Pastur threshold used for the data analysis of Fig. 7 and 8. The listed values are the same for every $p$.}
\label{tab:threshold}
\end{table}

The gradient threshold was set manually based on  the heat maps for  $\max_i|\nabla \phi_i|^2$  obtained from  the final configurations of the Cahn-Hilliard simulations. The Marchenko-Pastur threshold was set initially  to  the value $(\phi/N)^2 a_\ast^2$, which is the variance  in the components of a $\vec{\phi}$ vector that has half of its components equal to  $\phi_+$ and  the other half equal to $\phi_-$.  The value of the threshold was then slightly adjusted (if needed) upon direct visual inspection of some snapshots in individual trajectories.

\subsection{Derivation of stationarity conditions in the liquid Hopfield model}
\label{si:lhm_stat}
{\it Preliminaries.} Here we provide some details to the calculations of stationarity among target phases in the liquid Hopfield model for the case of arbitrary $q$. First, we provide the explicit expressions of the exchange chemical potentials of the individual components
\begin{equation}
\bar{\mu}_i = \nu_0\frac{\partial f}{\partial \phi_i}
= - v_2 \sum_{j=1}^{N} J_{ij} \phi_j
+ \frac{v_3}{2} (N \phi_i)^2
+ \log \phi_i - \log(1 - \phi)
\end{equation}
and the osmotic pressure
\begin{align}\label{eq:press}
\Pi &= \sum_{i=1}^{N} \phi_i \mu_i - f 
  = -\frac{ v_2}{2} \sum_{i=1}^{N} \sum_{j=1}^{N} J_{ij} \phi_i \phi_j
  + \frac{ v_3}{3} \sum_{i=1}^{N} \phi_i^3 N^2
  - \log(1 - \phi).
\end{align}
These are general expressions for the liquid Hopfield model, where we have not considered any particular values for the densities.

We consider a system with $C=2$ compartments, let us denote the compositions of these compartments by $\vec{\phi}^{(\alpha)}$ and $\vec{\phi}^{(-\alpha)}$. The two compartments correspond to enrichment and depletion according to a particular target phase. The conditions for stationarity are then given by chemical balance
\begin{equation}
\bar{\mu}_i^{(\alpha)} = \bar{\mu}_i^{(-\alpha)}
\end{equation}
and pressure balance
\begin{equation}
\Pi^{(\alpha)} = \Pi^{(-\alpha)} .
\end{equation}
In addition, we require that
\begin{equation}\label{eq:SI_const_dens}
 w^{(\alpha)} \phi_j^{(\alpha)}  +  w^{(-\alpha)} \phi_j^{(-\alpha)}= \phi_j,
\end{equation}
which is a statement of mass conservation. 

We begin by establishing the following ansatz for the compositions:
\begin{align}\label{eq:ansatz_q}
\phi_i^{(\alpha)} &= \frac{\phi_+}{N} \left( 1 + a_+ \gamma_i^{(\alpha)} \right), \\\label{eq:ansatz_q2}
\phi_i^{(-\alpha)} &= \frac{\phi_-}{N} \left( 1 - a_- \gamma_i^{(\alpha)} \right),
\end{align}
where $\phi_+ = \sum_{i=1}^N \phi_i^{(\alpha)}$ and $\phi_- = \sum_{i=1}^N \phi_i^{(-\alpha)}$ are the total volume fractions of the retrieval phase and the anti-retrieval phase, and $a_+ = \sum^N_{i=1} \phi_i^{(\alpha)} \gamma_i^{(\alpha)}/\phi_+$ and $a_- = -\sum^N_{i=1} \phi_i^{(-\alpha)} \gamma_i^{(\alpha)}/\phi_-$ are the overlap parameters that quantify the enrichment and depletion of components in these compartments in accordance to the given target. For the volume of the compartments we have that $w^{(\alpha)} = 1 - w^{(-\alpha)}$, and simplify notation by writing $w^{(\alpha)}=w$.
\newline

{\it Density constraints.} Using this ansatz in the abundance constraints of Eq.~\eqref{eq:SI_const_dens} we obtain that 
\begin{align}
\phi&=w \phi_+ \left(1 + a_+ \frac{1 - q}{n}\right) + (1 - w) \phi_- \left(1 - a_- \frac{1 - q}{n}\right)\label{eq:asymm_phase}\\
\phi&=w \phi_+ \left(1 - a_+ \frac qn\right) + (1 - w) \phi_- \left(1 + a_- \frac qn\right).\label{eq:asymm_antiphase}
\end{align}
for components in which $\xi_i^{(\alpha)}=1$ and $\xi_i^{(\alpha)}=0$, respectively.
Notice that if we multiply the first equation by $q$ and the second equation by $1 - q$, and if we then add up the two equations, then we get the simpler equation
\begin{equation}
w \phi_+ + (1 - w) \phi_- = \phi,\label{eq:balbal}
\end{equation}
and thus
\begin{equation}
\phi_- = \frac{\phi - w \phi_+}{1 - w}.\label{eq:rho_min_exp}
\end{equation}
Substitution of $\phi_-$ in Eq.~\eqref{eq:asymm_phase} gives
\begin{equation}
w \phi_+ (1 + a_+ [1 - q]/n) + (\phi - w \phi_+) (1 - a_- [1 - q]/n) = \phi,
\end{equation}
which can be solved towards $w \phi_+$ to yield the simple equation
\begin{equation}\label{eq:rho_p}
w \phi_+ = \phi \, \frac{a_-}{a_+ + a_-},
\end{equation}
and thus analogously,
\begin{equation}\label{eq:rho_m}
(1 - w) \phi_- = \phi \, \frac{a_+}{a_+ + a_-}.
\end{equation}
Equations~\eqref{eq:rho_p} and \eqref{eq:rho_m} relate the parameters $\phi_+$ and $\phi_-$ to the other three parameters, $a_+$, $a_-$ and $w$, which must be determined from the chemical and pressure balance conditions
\newline

{\it Pressure balance.} We establish the conditions for pressure balance. Following Eq.~\eqref{eq:press} we see that it is helpful to first compute the following statistical properties of the density states: 
\begin{align}
\sum^N_{i=1}\sum^N_{j=1}J_{ij}\phi^{(\alpha)}_i\phi^{(\alpha)}_j &= (\phi_+ a_+)^2\\
\sum^N_{i=1}\sum^N_{j=1}J_{ij}\phi^{(-\alpha)}_i\phi^{(-\alpha)}_j &= (\phi_- a_-)^2\\
N^2\sum^N_{i=1}\left(\phi^{(\alpha)}_i\right)^3  &= \phi^3_+ \left(1 + 3a_+^2 + a_+^3 \frac{1-2q}{n} \right)\\
N^2\sum^N_{i=1}\left(\phi^{(-\alpha)}_i\right)^3  &= \phi^3_- \left(1 + 3a_-^2 - a_-^3 \frac{1-2q}{n} \right) .
\end{align}
In deriving this, we simply have used the properties $\sum_i\gamma_i^{(\alpha)}=0$ as well as the two values of $\gamma_i^{(\alpha)}$ for arbitrary $q$.

Having established the above, it is now straightforward to obtain
\begin{align}
-\frac{ v_2}{2} \left((\phi_+ a_+)^2-(\phi_- a_-)^2\right)
 + \frac{ v_3}{3} \left( \phi^3_+ \left(1 + 3a_+^2 + a_+^3 \frac{1-2q}{n} \right) -  \phi^3_- \left(1 + 3a_-^2 - a_-^3 \frac{1-2q}{n} \right) \right) = \log \frac{1-\phi_+}{1-\phi_-}
\end{align}
Using the expressions for $\phi_-$ and $\phi_+$ we can turn this into an equation to be solved towards $w$ for given values of $a_-$ and $a_+$, which yields
\begin{align}
&-\frac{ v_2 \phi^2(a_-a_+)^2}{2(a_++a_-)^2} \left(\frac{1-2w}{(1-w)^2w^2}\right)
+ \frac{ v_3 \phi^3}{3(a_-+a_+)^3} \left( \frac{a^3_-}{w^3}\left(1 + 3a_+^2 + a_+^3 \frac{1-2q}{n} \right) -  \frac{a^3_+}{(1-w)^3} \left(1 + 3a_-^2 - a_-^3 \frac{1-2q}{n} \right) \right) 
\nonumber\\ 
&= \log \left( \frac{1-w}{w}\frac{w(a_++a_-)-\phi a_-}{(1-w)(a_++a_-)-\phi a_+}\right). \label{eq:mechBal}
\end{align}
This last equation is the same as Eq. (F9) of the main text.
\newline

{\it Chemical balance.} Next, we consider the conditions for chemical balance. For each of the two coexisting phases there are two different values of the chemical potential depending on whether the corresponding component is enriched or depleted in the target. For 
 $\xi^{\alpha}_i=1$ we get 
\begin{equation}
\bar{\mu}^{(\alpha)}_i =  \mu^{\uparrow} \quad {\rm and } \quad \bar{\mu}^{(-\alpha)}_i = \mu^{\uparrow}\label{eq:mu1}
\end{equation}
and for $\xi^{\alpha}_i=0$
\begin{equation}
\bar{\mu}^{(\alpha)}_i = \mu^{\downarrow}  \quad {\rm and} \quad  \bar{\mu}^{(-\alpha)}_i = \mu^{\downarrow}. \label{eq:mu2}
\end{equation}
We can  express the first two equations as  
\begin{align}
- v_2 \phi_+ a_+  [1-q]/n  +\frac{ v_3\phi^2_+}{2}(1 + a_+ [1-q]/n)^{2} +\log(\phi_+ (1 +  a_+ [1-q]/n)) -\log(1-\phi_+)&=   \mu^{\uparrow}  + \log (N)  \label{eq:36}\\
 v_2 \phi_- a_-  [1-q]/n  +\frac{ v_3\phi^2_-}{2}(1 - a_- [1-q]/n)^{2}  +\log(\phi_- (1- a_- [1-q]/n)) -\log(1-\phi_-)&=   \mu^{\uparrow}  + \log (N) .\label{eq:37}
\end{align}
and the last two equations as 
\begin{align}
 v_2 \phi_+ a_+  q/n  +\frac{ v_3\phi^2_+}{2}(1 - a_+ q/n)^{2}
 +\log(\phi_+ (1 -  a_+ q/n)) -\log(1-\phi_+) &=   \mu^{\downarrow}  + \log (N) \label{eq:38}\\
- v_2 \phi_- a_-  q/n  +\frac{ v_3\phi^2_-}{2}(1 + a_- q/n)^{2} +\log(\phi_- (1 + a_- q/n)) -\log(1-\phi_-)&=   \mu^{\downarrow}  + \log (N) .\label{eq:39}
\end{align}
This constitutes a system of four equations, but we have introduced two new unknowns: $\mu^{\downarrow}$ and $\mu^{\uparrow}$. Next, we eliminate these.

Subtracting \eqref{eq:37}  from \eqref{eq:36}  gives the equation
\begin{eqnarray}
v_2 (\phi_+ a_+ + \phi_-a_-)  (1-q)/n  - \frac{ v_3}{2} \left(\phi^2_+(1+a_+[1-q]/n)^2   - \phi^2_-(1-a_-[1-q]/n)^2 \right)  
\nonumber\\ 
= \log \frac{\phi_+(1-\phi_-) (1+a_+[1-q]/n)}{(1-\phi_+)\phi_- (1-a_-[1-q]/n)}.
\end{eqnarray}
and  analogously, subtracting \eqref{eq:39}  from \eqref{eq:38} yields the equation, 
\begin{align}
- v_2 (\phi_+ a_+ + \phi_-a_-)  q/n  - \frac{ v_3}{2} \left(\phi^2_+(1-a_+q/n)^2   - \phi^2_-(1+a_-q/n)^2 \right) = \log \frac{\phi_+(1-\phi_-) (1-a_+q/n)}{(1-\phi_+)\phi_- (1+a_-q/n)}.
\end{align}
Above we showed that $\phi_+$ and $\phi_-$ are dependent variables of $a_+$, $a_-$ and $w$. Therefore, we now use the expressions in Eq.~\eqref{eq:rho_p} and \eqref{eq:rho_m} to remove the dependencies in $\phi_+$ and $\phi_-$. This results in
\begin{align}
& v_2 \phi \frac{a_+a_-}{a_-+a_+}\left(\frac{1}{w(1-w)}\right)  \left(\frac{1-q}{n}\right)   
- \frac{ v_3\phi^2}{2(a_-+a_+)^2} \left(\frac{a^2_-}{w^2}(1+a_+[1-q]/n)^2   - \frac{a^2_+}{(1-w)^2}(1-a_-[1-q]/n)^2 \right)  
\nonumber\\ 
&= \log \left(\frac{a_-((1-w)(a_-+a_+)-\phi a_+) }{a_+ (w(a_-+a_+)-\phi a_- )}\right) + \log \left(\frac{1+a_+[1-q]/n}{1-a_-[1-q]/n}\right).
\end{align} 
and 
\begin{align}
&- v_2\phi\frac{a_-a_+}{a_-+a_+} \left(\frac{1}{w(1-w)}\right)  \frac{q}{n}  - \frac{ v_3\phi^2}{2(a_-+a_+)^2}  \left(\frac{a^2_-}{w^2}(1-a_+q/n)^2   - \frac{a^2_+}{(1-w)^2}(1+a_-q/n)^2 \right)  
\nonumber\\ 
&= \log \left(\frac{a_-((1-w)(a_-+a_+)-\phi a_+) }{a_+ (w(a_-+a_+)-\phi a_- )}\right) + \log \left(\frac{1-a_+q/n}{1+a_-q/n}\right).
\end{align}
These two equations should be solved towards $a_+$ and $a_-$ for given fixed value of~$w$. The simplified form that appears in Appendix F of the main text, Eqs. (F5) and (F6), can be derived by using on the previous two equations the following form of the pressure balance:
\begin{align}
\log \left( \frac{(1-w)(a_++a_-)-\phi a_+}{ w(a_++a_-)-\phi a_-}\right) &= \log \left(\frac{1-w}{w}\right)  +\frac{ v_2 \phi^2(a_-a_+)^2}{2(a_++a_-)^2} \left(\frac{1-2w}{(1-w)^2w^2}\right)\nonumber\\ 
&-\frac{ v_3 \phi^3}{3(a_-+a_+)^3} \left( \frac{a^3_-}{w^3}\left(1 + 3a_+^2 + a_+^3 \frac{1-2q}{n} \right) -  \frac{a^3_+}{(1-w)^3} \left(1 + 3a_-^2 - a_-^3 \frac{1-2q}{n} \right) \right)\nonumber\\
\end{align}
\newline

{\it Recovery of $q=1/2$ case.} From the above expressions we can recover the particular case of $q=1/2$ that is presented in the main text. To see this note that the functions $f$ and $g$ in Appendix F of the main text for $q=1/2$ simplify into 
\begin{align}
f(a_+,a_-,w) &= \kappa + \delta \\
g(a_+,a_-,w)  &=  - \kappa + \delta
\end{align}
with 
\begin{eqnarray}
 \kappa =  v_2 \phi \frac{a_-a_+}{a_-+a_+}\left(\frac{1}{w(1-w)}\right)  
-\frac{ v_3\phi^2}{2(a_-+a_+)^2} \left(\frac{2a^2_-a_+}{w^2}  + \frac{2a^2_+a_-}{(1-w)^2} \right) 
\end{eqnarray}
and 
\begin{align}
\delta &=  - v_2 \phi^2 \frac{(a_-a_+)^2}{2(a_-+a_+)^2}\left(\frac{1-2w}{w^2(1-w)^2}\right) 
- \frac{ v_3\phi^2}{2(a_-+a_+)^2} \left(\frac{a^2_-(1+a_+^2) }{w^2}  - \frac{a^2_+(1   +   a_-^2)}{(1-w)^2} \right) 
\nonumber\\ 
& +\frac{ v_3 \phi^3}{3(a_-+a_+)^3} \left( \frac{a^3_-\left(1 + 3a_+^2  \right)}{w^3} -  \frac{a^3_+\left(1 + 3a_-^2  \right) }{(1-w)^3} \right).
\end{align}
Substitution of these values of $f$ and $g$ into Eqs. (F5) and (F6) of the main text we find the simpler set of equations 
\begin{align}
a_+ &= n\frac{(1-w)(e^{\kappa}-e^{-\kappa})}{(1-w)[(1-q)e^{-\kappa}+e^{\kappa}q] +w e^{\delta}} \label{eq:aP3}\\
a_- &=    n\frac{w(e^{\kappa}-e^{-\kappa})}{(1-w)e^{-\delta} + w[(1-q)e^{\kappa}+qe^{-\kappa}]}, \label{eq:aM3}
\end{align}
In addition, we can readily verify that $w=1/2$, $a_-=a_+=a$, solves the equation for mechanical balance, Eq.~\eqref{eq:mechBal}. As a consequence, we have that both equations \eqref{eq:aP3} and \eqref{eq:aM3} become 
\begin{equation}
a = \frac{e^{2 v_2\phi a - 2 v_3  \phi^2a}-e^{-2 v_2\phi a + 2 v_3  \phi^2a}}{2+e^{2 v_2\phi a - 2 v_3  \phi^2a}+e^{-2 v_2\phi a + 2 v_3  \phi^2a}} = \tanh\left( (v_2-v_3\phi) \phi a\right) ,
\end{equation}
which is the same as the equation shown in the main text for $q=1/2$, i.e. Eq. (29).

\subsection{Derivation of stability conditions in the liquid Hopfield model}
\label{si:lhm_stab}
In order to discuss stability of the stationary phase separated state, we first compute the general expression of the hessian:
\begin{align}
H_{ij}
= - \frac{v_2}{\nu_0} J_{ij}
+  \frac{v_3}{\nu_0} N^2 \phi_i \delta_{i,j}
+ \frac{1}{\nu_0\phi_i} \delta_{i,j}
+ \frac{1}{\nu_0(1 - \phi)}.
\end{align}
Using then the ansatz in Eq.~\eqref{eq:ansatz_q} and \eqref{eq:ansatz_q2} we  find the Hessian of each of the two phases (we drop, for simplicity, the factor of $\nu_0$): 
\begin{align}
H^{(\alpha)}_{ij}  &= - v_2 J_{ij}  + \frac{1}{1-\phi_+}  +  N\delta_{i,j} \left[  v_3  \phi_+ (1+a_+\gamma^{(\alpha)}_i) + \frac{1}{\phi_+}\frac{1}{1+a_+\gamma^{(\alpha)}_i}  \right]\\
H^{(-\alpha)}_{ij}  &= - v_2 J_{ij}  + \frac{1}{1-\phi_-}  +  N\delta_{i,j} \left[  v_3  \phi_- (1-a_-\gamma^{(\alpha)}_i) + \frac{1}{\phi_-}\frac{1}{1-a_-\gamma^{(\alpha)}_i}  \right]. 
\end{align}

In \cite{braz2024liquid} we derived sufficient conditions so the target phases are stable. These are defined in terms of the following coefficients for phase $\alpha$,
\begin{align} 
c^{(\alpha)}_1 &=  v_3 \phi_+ (1-a_+ q/n) + \frac{1}{\phi_+}  \frac{1}{1-a_+q/n}\\
c^{(\alpha)}_2 &= -\frac{ v_3\phi_+ a_+ }{n} + \frac{a_+}{n\phi_+}\frac{1}{(1+a_+(1-q)/n)(1-a_+q/n)},
\end{align}
and the corresponding coefficients for phase $-\alpha$,
\begin{align} 
c^{(-\alpha)}_1 &=  v_3 \phi_- (1+a_- q/n) + \frac{1}{\phi_-}  \frac{1}{1+a_-q/n}\\
c^{(-\alpha)}_2 &= \frac{ v_3\phi_- a_- }{n} - \frac{a_-}{n\phi_-}\frac{1}{(1-a_-(1-q)/n)(1+a_-q/n)}. 
\end{align}
The sufficient stability condition for the $\alpha$-phase are then: 
\begin{align}
&{\rm if}\quad c^{(\alpha)}_2 \geq 0\quad:\quad\quad- v_2 + c^{(\alpha)}_1 - c^{(\alpha)}_2 >0\\
&{\rm if}\quad c^{(\alpha)}_2 < 0\quad:\quad\quad- v_2 + c^{(\alpha)}_1 >0 .
\end{align}
Analogously, for the $-\alpha$-phase: 
\begin{align}
&{\rm if}\quad c^{(-\alpha)}_2 \geq 0\quad:\quad\quad 
- v_2 + c^{(-\alpha)}_1 - c^{(-\alpha)}_2 >0 \\
&{\rm if}\quad  c^{(-\alpha)}_2 < 0\quad:\quad\quad 
- v_2 + c^{(-\alpha)}_1 >0 .
\end{align}

\subsection{Unstable modes in the Liquid Hopfield model  that preserve the total number of phases $P$}\label{sec:unstableP}
In the main text we have shown that a phase separated state is unstable if  one of the phases it is composed of has a negative Hessian.   In this case, the  liquid mixture is unstable towards a mode that generates a new phase, i.e., the number of phases $P$ increases by one.    
In this Supplementary section, we analyse unstable modes that do not change the number of phases.   These, phase preserving instabilities arise due to changes in the compositions and the compartment volumes of the phases, while keeping $P$ fixed.      
  We derive in what follows the region in parameter space for which stationary retrieval states  are stable towards such phase preserving instabilities.   Interestingly, even for $v_3=0$  there is a parameter region for which stationary retrieval states cannot be destabilized by modes that preserve  the number of phases $P$.

\begin{figure}[h!]\centering
\includegraphics[width=0.5\textwidth]{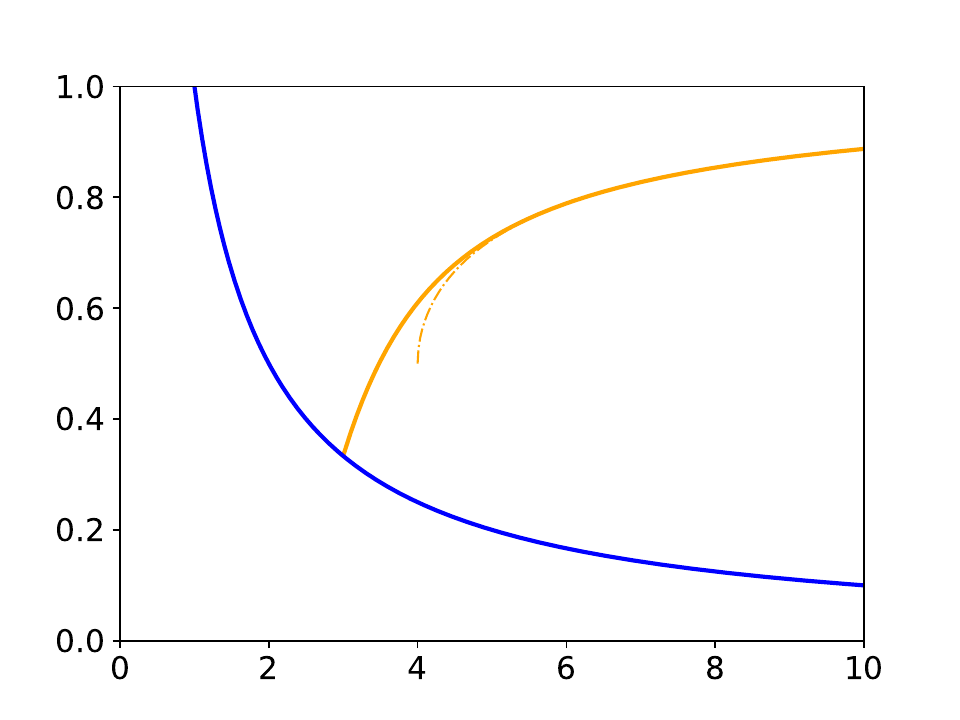}
 \put(-200,40){homogeneous}
  \put(-170,140){Retrieval}
    \put(-100,100){Disordered}
 \put(-255,90){\Large$\phi$}
 \put(-130,-4){\Large$v_2$}
\caption{{\it Stability of retrieval states towards modes that preserve the number of phases $P$ for $v_3=0$ and arbitrary $N$. }  The blue line is the spinodal of the homogeneous state, $\phi v_2=1$ [Eq. (30) of the main text for $v_3=0$].   Above the solid orange line, stationary   retrieval states are stable towards phase preserving modes.  This line is given by  $\phi v_2 =1/(1-a^2_\ast\phi)$ [Eq.~(\ref{eq:orangeline}) for $v_3=0$]. 
 The dashed-dotted line gives the asymptotics of the latter for large $v_2$, and takes the form   $\phi=1/2(1+\sqrt{1-4/v_2})$.  The blue and orange line meet at the ``triple" point $(\phi,v_2)=(1/3,3)$.  } \label{fig:stab}   
\end{figure} 

\subsubsection{Stability of the  phase separated phase  ($C=2$)}\label{app:stabHetero}
We determine the stability of the phase separated state  $\left\{w^{(\alpha)},w^{(-\alpha)},\vec{\phi}^{(\alpha)},\vec{\phi}^{(-\alpha)}\right\}$ that consists of two phases, namely,   
\begin{equation}
\phi_i^{(\alpha)} = \frac{\phi}{N}\left(1+a_\ast\gamma_i^{(\alpha)}\right) 
\end{equation}
with volume fraction $w^{(\alpha)}$, and 
\begin{equation}
\phi_i^{(-\alpha)} = \frac{\phi}{N}\left(1-a_\ast\gamma_i^{(\alpha)}\right) 
\end{equation}
and volume fractions $w^{(-\alpha)}=1-w^{(\alpha)}=1/2$.  These two phases are in   a stationary state when  $a_\ast$ solves 
\begin{align}
a_\ast = \tanh( a_\ast \phi (v_2 -v_3\phi) ) .\label{eq:easy}
\end{align}  

To determine the stability of this stationary state, we  determine  whether the matrix $\mathbf{A}$, as defined in Appendix B of the main text, is positive definite.  In the present case, the matrix $\mathbf{A}$ takes the form 
\begin{equation}
\mathbf{A} = \left(\begin{array}{cc} \frac{1}{2}(\mathbf{h}^{(\alpha)}+\mathbf{h}^{(-\alpha)}) & \delta_1 \vec{\gamma}^{(\alpha)} + \delta_2 \vec{1}\\ \left[\delta_1 \vec{\gamma}^{(\alpha)} + \delta_2 \vec{1}\right]^{\rm T}  &   4\phi a_\ast \delta_1\end{array}\right).
\end{equation}
Here $\mathbf{h}^{(\alpha)}$ and $\mathbf{h}^{(-\alpha)}$ are the hessians of the two phases, i.e.,  
\begin{eqnarray}
[\mathbf{h}^{(\pm\alpha)}]_{ij} &=& -v_2 J_{ij} + \frac{1}{1-\phi} \nonumber\\ 
&& + N \delta_{i,j} \left(v_3 \phi (1\pm a^\ast \gamma^{(\alpha)}_i) + \frac{1}{\phi } \frac{1}{1\pm a^\ast \gamma^{(\alpha)}_i}\right) \nonumber\\
\end{eqnarray}
and the two constants 
\begin{eqnarray}
\delta_1 &=&  2\phi a_\ast\left( -v_2+v_3\phi   +  \frac{1}{\phi} \frac{1}{1-a^2_\ast}\right)  
\end{eqnarray}
and
\begin{eqnarray}
\delta_2 &=&  2\phi a_\ast  \left(- v_3 
\phi  a_\ast  + \frac{1}{\phi} \frac{a_\ast}{1-a^2_\ast}\right). 
\end{eqnarray}
Note that $\mathbf{A}$ is a square matrix of order $N+1$.  

The matrix $\mathbf{A}$ has five distinct eigenvalues as we summarise below: 

\begin{itemize}

\item There is the eigenvalue 
\begin{equation}
\lambda_+ = N\left(-v_2 + v_3\phi + \frac{1}{\phi}\frac{1}{1-a^2_\ast}\right) \label{eq:lambdaP}
\end{equation}
with degeneracy $p-1$.  The eigenspace of $\lambda_+$ is  spanned by the target states that are not the retrieved one, i.e.,  $(\vec{\gamma}^{(\eta)} 0)^{\rm T}$ with $\eta \in \left\{1,2,\ldots,p\right\}\setminus \left\{\alpha\right\}$.    
\item  There is the eigenvalue 
\begin{equation}
\lambda_-  = N\left( v_3\phi + \frac{1}{\phi}\frac{1}{1-a^2_\ast}\right).
\end{equation} 
with degeneracy $N-p-1$.  The eigenspace of $\lambda_-$ is spanned by vectors of the form $(\vec{e}^{(i)} 0)^{\rm T}$ with the $\vec{e}_i$ vectors that are orthogonal to all the target states $\vec{\gamma}^{(\eta)}$ (including $\vec{\gamma}^{(\alpha)}$) and the all ones vector $\vec{1}$.    

\item  The remaining three eigenvalues are the eigenvalues of the matrix 
\begin{equation}
\mathcal{M} =\left( \begin{array}{ccc}\lambda_+& 0 &\sqrt{N}\delta_1 \\ 0&  \lambda_0&  \sqrt{N}\delta_2\\ \sqrt{N}\delta_1 &  \sqrt{N} \delta_2& 4 \phi a_\ast \delta_1\end{array}\right), \label{eq:M}
\end{equation}
where 
\begin{equation}
\lambda_0 = N\left(\frac{1}{1-\phi} + v_3\phi + \frac{1}{\phi}\frac{1}{1-a^2_\ast} \right).
\end{equation} 
The corresponding eigenspace is spanned by the vectors $\vec{\zeta}_1 = (\vec{\gamma}^{(\alpha)} 0)^{^{\rm T}}$,  $\vec{\zeta}_2 =  (\vec{1} 0)^{^{\rm T}}$ and $\vec{\zeta}_3 =  (\vec{0} 1)^{\rm T}$.   In fact, these three vectors correspond with rows and columns of $\mathcal{M}$ with corresponding numbering.

The eigenvalues of this matrix  $\mathcal{M}$ solve a cubic equation of the form
\begin{equation}
\alpha_1 + \alpha_2  \lambda +  \alpha_3 \lambda^2 + \lambda^3 = 0
\end{equation}
that can be solved explicitly to obtain three additional eigenvalues $\lambda_1(\mathcal{M})$, $\lambda_2(\mathcal{M})$, and  $\lambda_3(\mathcal{M})$.  

  For what follows, the main relevant constant in the cubic equation is  
\begin{equation}
\alpha_1 = N \delta^2_1 \lambda_0+  N\delta^2_2 \lambda_+ - \lambda_+\lambda_0  4 \phi a_\ast \delta_1.
\end{equation}

\end{itemize}

\subsubsection{Spinodal lines: general case}\label{app:spinodallines}
The spinodal lines are found by setting the smallest eigenvalue of   $\mathbf{A}$ equal to zero.  
  As $\lambda_->0$, there are two possibilities,  either $\lambda_+=0$  or  one of the eigenvalues of $\mathcal{M}$ equals zero, which is equivalent to the condition $\alpha_1=0$.  Note that also the condition $\alpha_1=0$ is independent of $N$, and thus the stability  lines are independent of $N$ (we can set $N=1$ in what follows).

The inequality $\lambda_+ \geq 0$ yields the   condition 
 \begin{equation}
v_2 \leq  v_3\phi + \frac{1}{\phi}\frac{1}{1-a^2_\ast} \label{eq:phaseSepCond}
\end{equation} 
for the stability of the phase separated phase; note that for $a_\ast=0$ we recover the stability condition for the homogeneous state $C=1$. If $a_\ast\neq 0$, then the right-hand side of (\ref{eq:phaseSepCond}) implies the inequality Eq. (30) of the main text, and thus a stationary retrieval state  is metastable in a larger region of parameter space than the homogeneous state.  
  The corresponding spinodal line is  given by the equality in (\ref{eq:phaseSepCond}).  

Next we discuss $\alpha_1=0$.  This yields after some calculations two solutions for $v_2$, namely,
\begin{eqnarray}
v_2&=& \phi v_3 +\frac{1}{\phi} \frac{1}{1-a^2_\ast} 
\end{eqnarray}
which is the same spinodal line as found for $\lambda_+$ (namely,  the equality in Eq.~(\ref{eq:phaseSepCond})), and there is a second solution 
\begin{eqnarray}
v_2= 
\frac{
    1 + \phi^2 v_3 [2 - \phi + a^2_\ast \left(2 - 3 \phi\right)  + \left(-1 + a^2_\ast \right)^2 \left(1 - \phi\right) \phi^2 v_3]
}{
    \phi [1 - a^2_\ast  \phi + \left(-1 + a^2_\ast \right) \left(-1 + \phi\right) \phi^2 v_3]
} .\nonumber\\ \label{eq:cnd2}
\end{eqnarray}
We have verified numerically and by analysing relevant limiting cases that the corresponding stability condition is  
\begin{eqnarray}
v_2 \leq  
\frac{
    1 + \phi^2 v_3 [2 - \phi + a^2_\ast \left(2 - 3 \phi\right)  + \left(-1 + a^2_\ast \right)^2 \left(1 - \phi\right) \phi^2 v_3]
}{
    \phi [1 - a^2_\ast  \phi + \left(-1 + a^2_\ast \right) \left(-1 + \phi\right) \phi^2 v_3]
} .\nonumber\\ \label{eq:cnd3}
\end{eqnarray}

For $v_3=0$ the spinodal lines simplify into 
\begin{equation}
\phi v_2 = \frac{1}{1-a^2_\ast}. \label{eq:orangeline}
\end{equation}
and 
\begin{equation}
\phi v_2-1 = \frac{a^2_\ast\phi}{1-a^2_\ast\phi}.
\end{equation} 

\subsubsection{Spinodal of the retrieval phase when $v_2$ is large}\label{app:limitingCase}
We consider the limiting case of  $a^\ast\approx 1$ corresponding with large values  $v_2\gg v_3$.  For $a^\ast\approx 1$, the inequality  (\ref{eq:phaseSepCond})  is satisfied, and 
the  spinodal line (\ref{eq:cnd2}) simplifies into 
\begin{equation}
v_2 = \frac{1+4v_3 \phi^2 (1-\phi)}{\phi(1-\phi)}.\label{eq:asymptotic}
\end{equation}
This equation is plotted as  the dashed lines in Fig.~\ref{fig:stab}.   Note that    (\ref{eq:asymptotic}) admits two branches of $\phi$ as a function of $v_2$, one that is increasing and the other decreasing.    Only the increasing one is relevant, and it is the one plotted in Fig.~\ref{fig:stab}.  For example, when $v_3=0$ we find the two branches 
\begin{equation}
\phi = \frac{1}{2}\left(1+ \sqrt{1-\frac{4}{v_2}}\right),
\end{equation}
and 
\begin{equation}
\phi = \frac{1}{2}\left(1- \sqrt{1-\frac{4}{v_2}}\right),
\end{equation}
only  the former one being relevant and plotted in Fig.~\ref{fig:stab}.

\subsubsection{Minimal value $\phi_{\rm min}$ for which retrieval through liquid-liquid demixing is possible}\label{app:triple}

The minimal value $\phi_{\rm min}$ is found as the intersection point of the spinodal of the homogeneous and retrieval phases.  

Setting $a_\ast = \pm \sqrt{3(\phi v_2-1)}$,  yields 
\begin{equation}
\phi v_2 -1 = \frac{3(\phi v_2-1)\phi}{1-3(\phi v_2-1)\phi}.  
\end{equation}
Next, we set $v_2=1/\phi + \epsilon$, and expand in $\epsilon$, 
\begin{equation}
\epsilon = 3\epsilon \phi + O(\epsilon^2)
\end{equation}
and thus $\phi=1/3$.  

For $v_3>0$, we can do a similar trick.  In this case, $a_\ast = \sqrt{3[\phi_\ast (v_2-v_3\phi_\ast)-1]}$.   Setting $v_2 = \frac{1}{\phi_\ast} + v_3\phi_\ast +  \epsilon$, we get 
\begin{equation}
\epsilon =  \frac{3\phi (1 + 3 \phi v_3 - 3 \phi^2 v_3 - \phi^3 v_3^2 + \phi^4 v_3^2)}{1 + \phi^2 v_3 - \phi^3 v_3}\epsilon.
\end{equation}
We thus get the equation 
\begin{equation}
\frac{3\phi (1 + 3 v_3  \phi (1-\phi)- v_3^2  \phi^3  (1-\phi))}{1 + \phi^2 v_3(1-\phi) } = 1
\end{equation}
This equation does not have a closed form solution, but numerically solving it gives one solution in $[0,1]$ (see Fig.~\ref{fig:stab}).

\subsubsection{Stable retrieval phase at $v_3=0$ }\label{app:analytv30}
To demonstrate that there exists a parameter regime at $v_3=0$ for which the target-antidroplet state is metastable, we will consider the eigenvalues of $\mathbf{A}$ in the limit of  large $\phi(v_2-v_3\phi)$, so that $a^\ast\approx 1$.   Specifically, we show that  for large enough $v_2$, left of the dashed-dotted line in Fig.~\ref{fig:stab}, all eigenvalues of $\mathbf{A}$ are positive. 

Using the asymptotic series
\begin{equation}
\tanh(x) = 1 - 2e^{-2x} + O(e^{-4x})
\end{equation}
in the right-hand side of  (\ref{eq:easy})  we obtain the explicit expression
\begin{equation}
a^\ast = 1 - 2 e^{-2 \phi (v_2-\phi v_3)} + O\left(e^{-4 \phi (v_2-\phi v_3)}\right) .  \label{eq:explicit}
\end{equation}
Using this formula in the expression for  $\lambda_+$ in (\ref{eq:lambdaP}) we get the stability condition  
\begin{equation}
v_2 \leq v_3\phi +  \frac{e^{2 \phi (v_2-\phi v_3)} }{4\phi}  + O\left(e^{-4 \phi (v_2-\phi v_3)}\right),
\end{equation}
which can clearly be satisfied when $v_2$ is large enough.  

Using (\ref{eq:explicit}) in the expression (\ref{eq:M}) for the matrix $\mathcal{M}$, we find that 
\begin{eqnarray}
\mathcal{M} &=&e^{2\phi(v_2-\phi v_3)} \left( \mathcal{M}_0 +e^{-2\phi(v_2-\phi v_3)}  \mathcal{M}_1 \right)+ O\left(e^{-4 \phi (v_2-\phi v_3)}\right) \nonumber\\
\end{eqnarray}
where 
\begin{equation}
\mathcal{M}_0 = \left( \begin{array}{ccc}N/(4\phi)& 0 &\sqrt{N}/2 \\ 0&  N/(4\phi)&  \sqrt{N}/2\\ \sqrt{N}/2 &  \sqrt{N}/2 & 2 \phi \end{array}\right)
\end{equation}
and 
\begin{eqnarray}
 \lefteqn{\mathcal{M}_1 }&& 
 \nonumber\\ 
&&  =  \left( \begin{array}{ccc}N(-v_2+v_3\phi)& 0 &\sqrt{N}(2\phi(-v_2+v_3\phi)-1)\\ 0&  N(\frac{1}{1-\phi}  + v_3\phi)& \sqrt{N} ( -2v_3\phi^2 -2  )\\ \sqrt{N}(2\phi(-v_2+v_3\phi)-1)& \sqrt{N}( -2v_3\phi^2 -2) & 8 \phi  (-v_2\phi+v_3\phi^2-1)\end{array}\right).
\end{eqnarray}
The eigenvalues of $\mathcal{M}_0$ are $\left\{0, N/(4\phi), (8 \phi^2 + N)/(4\phi)\right\}$, and thus non-negative.  The zero eigenvalue of  $\mathcal{M}_0$ may yield a negative eigenvalue of $\mathcal{M}$, and thus an instability, but for this we need to study the perturbation from $\mathcal{M}_1$.   The eigenvector associated with the zero eigenvalue is 
\begin{equation}
\vec{e}_0 = \left(\begin{array}{c}-2\phi/\sqrt{N} \\ -2\phi/\sqrt{N} \\ 1\end{array}\right).
\end{equation}

The perturbative contribution to the zero eigenvalue takes the form 
\begin{equation}
\frac{\vec{e}_0 \mathcal{M}_1 \vec{e}_0 }{\vec{e}_0\cdot\vec{e}_0} 
\end{equation}
and thus we find the condition
\begin{equation}
\vec{e}_0 \mathcal{M}_1 \vec{e}_0  \geq 0
\end{equation}
that reads
\begin{equation}
v_2 \leq  \frac{1+4v_3 \phi^2 (1-\phi)}{\phi(1-\phi)},\label{eq:asymptotic2}
\end{equation}
consistent with the equation (\ref{eq:asymptotic}) we found before.   

Hence, for large enough $v_2$ so that $\lambda_+>0$, it holds that left of the dashed-dotted orange lines in Fig.~\ref{fig:stab} we have a stable retrieval phase, as all eigenvalues of $\mathbf{A}$ are  then positive.

\subsection{Additional Figures}
\begin{figure}[htbp!]\centering
\includegraphics[width=0.6\textwidth]{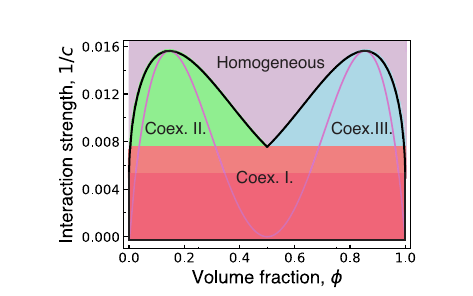}
\caption{{\bf Phase diagram for a binary mixture with quartic potential}. Solutions of Eqs.~\eqref{eq:chem_bal_quartic}, and \eqref{eq:press_bal_quartic_simp} towards the interaction parameter $c^{-1}$ as a function of the solute volume fraction, $\phi^{(1)} =\phi$. Only the solutions with the minimal free energy $\mathfrak{f}$ are plotted, representing the thermodynamic equilibrium. Each color represents a different stable state as labeled. As $c^{-1}$ increases, the stable phase separated state switches from coexistence between low and high volume fraction (family I) to a coexistence between intermediate and high volume fraction (family III and II). For high values of $c^{-1}$  the homogeneous state is the stable state.} 
\label{fig:si_fig3}
\end{figure}

\begin{figure}[htbp!]\centering
\includegraphics[]{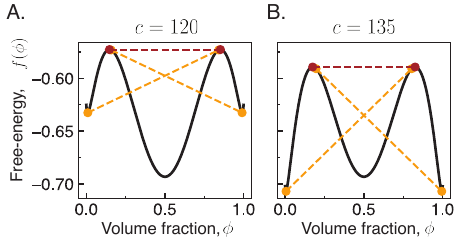}
\caption{{\bf Saddle and unstable stationary states for a binary mixture with quartic potential}. {\bf A}. Free energy density $f(\phi)$, in units of $\nu_0^{-1},$ from Eq. (15) with the energy function in Eq. (16) of the main text for $c=120$. Tangent dashed lines correspond to stationary states satisfying chemical and mechanical balance, but not satisfying the stability conditions. The orange lines correspond to saddle points, since one of the two coexisting phases is stable (convex $f(\phi)$) and the other unstable (concave $f(\phi)$). The red line corresponds to a maximum, since both coexisting phases are unstable. {\bf B}. Same as in {\bf A}, but for $c=135$.} \label{fig:si_saddle} 
\end{figure}

\begin{figure}[htbp!]\centering
\includegraphics[width=0.6\textwidth]{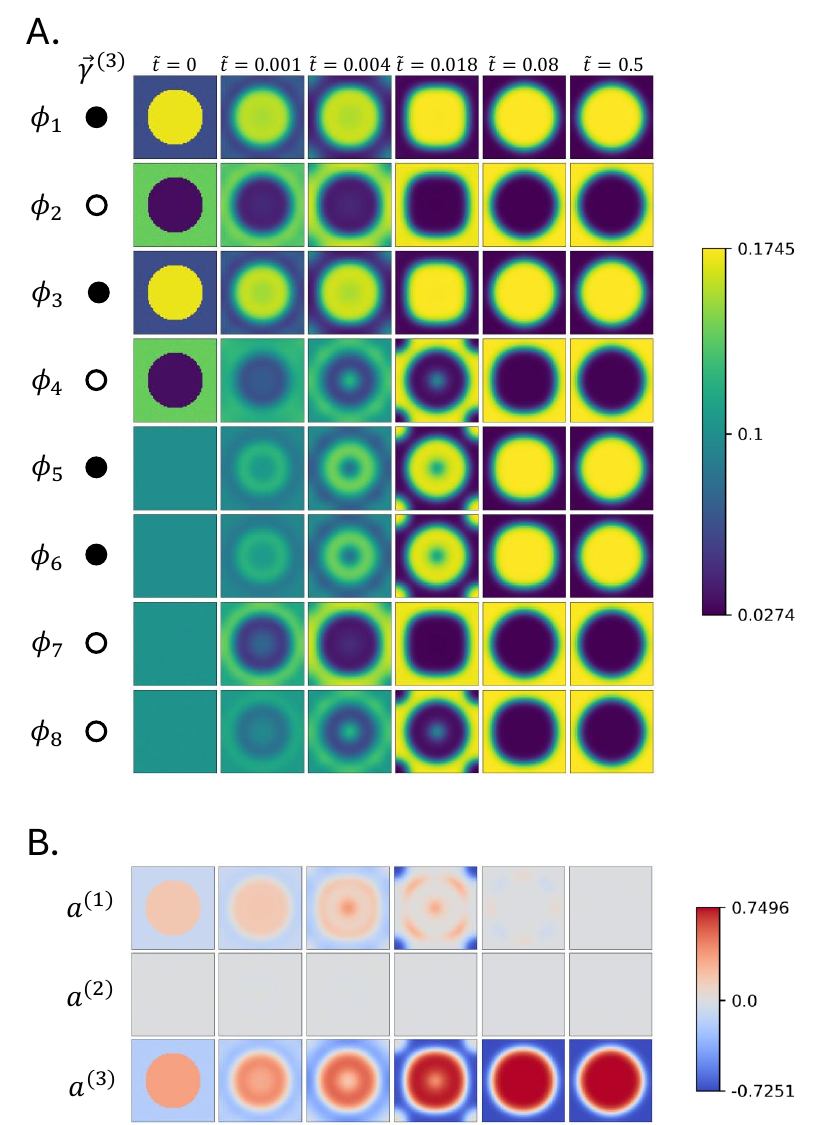}
\caption{{\it Retrieval of pattern $\alpha=3$ in the liquid Hopfield model} {\bf A} Concentration-field dynamics showing demixing into the retrieval and anti-retrieval phase $\alpha=3$. The initial state contains a seed for the target in the first four components; in these components, the background is set to the homogeneous concentration minus the contribution of the anti-retrieval state, and the remaining components are set to the homogeneous value. Spatial white noise is added in each site outside the seed. The system evolves towards the expected enrichment–depletion pattern for $\alpha=3$ (see black and white scheme on the left). {\bf B} Snapshots from the dynamics displayed using the spatial overlap color scheme explained in Fig. 6 of the main text (color bars to the right); the spatial overlap of target $\alpha=3$ approaches the theoretical prediction, confirming retrieval. Parameters as in Fig. 6.}
 \label{fig:si_ret_p3} 
\end{figure}

\begin{figure}[h!]\centering
\includegraphics[width=\textwidth]{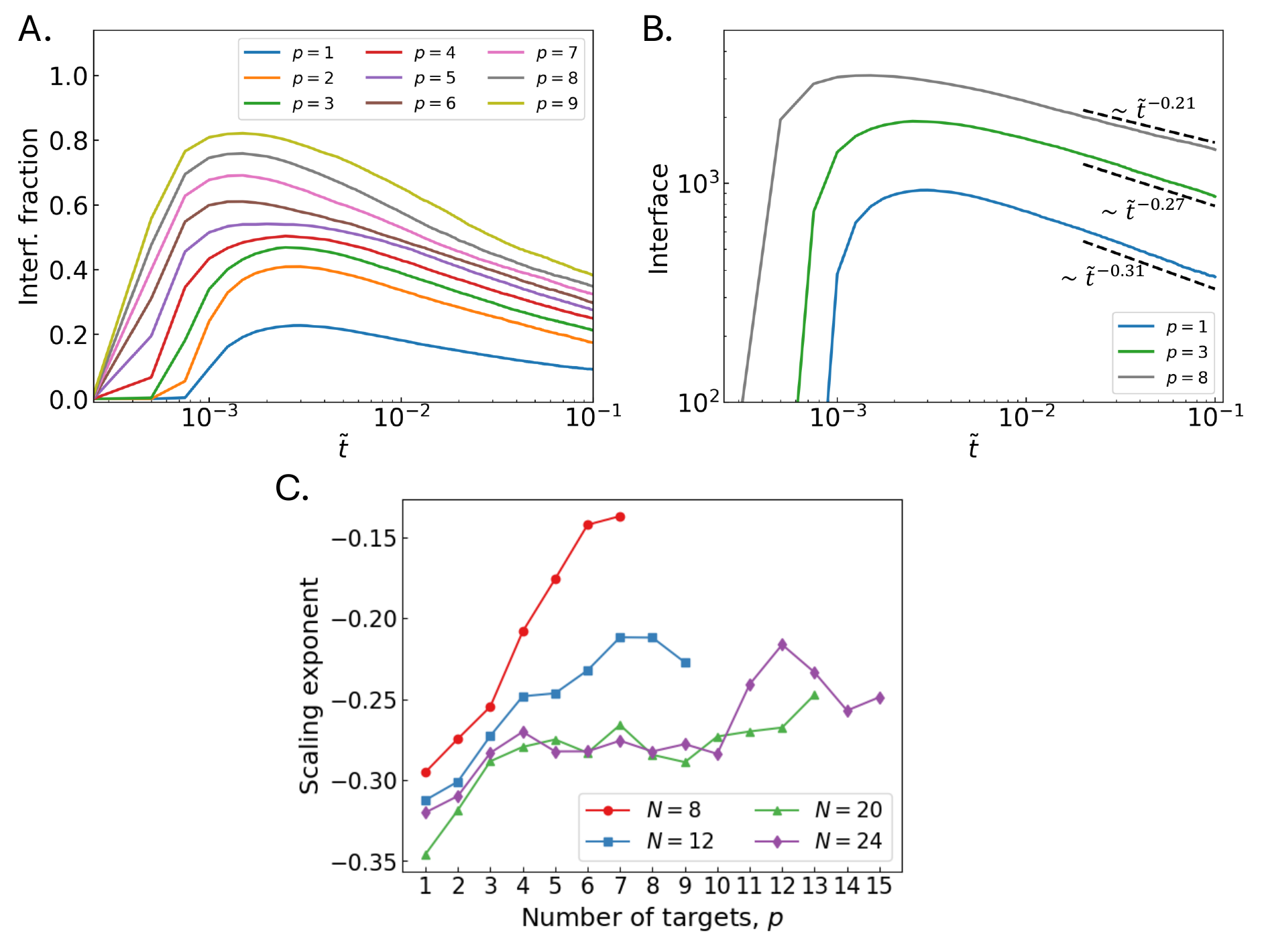}
\caption{Interface dynamics for $N = 12$ (same simulations of Fig. 8 of the main text, with interface points identified through the gradient thresholding procedure described in~\ref{SI:pca_method} using the gradient threshold reported in Table~\ref{tab:threshold}). \textbf{A.} Fraction of interface points as a function of the dimensionless time $\tilde{t}$ for different values of $p$. Each curve represents an average over 30 simulations. The results are shown starting from $\tilde{t} = 2.5\times 10^{-4}$: in the short time interval before this instant, the high gradients in the density fields due to the (artificial) initial noise rapidly disappear. At this point, spinodal decomposition begins: as more and more phases nucleate, the fraction of interface points increases. After this initial increase, the curves reach a maximum and start to slowly decrease: this regime is associated to Ostwald ripening. \textbf{B.} Total number of interface points as a function of $\tilde{t}$ for different values of $p$. For the sake of clarity, only the curves corresponding to $p = 1, 3, 8$ are reported. During Ostwald ripening, the number of interface points decreases with time following a power law with an exponent that depends on $p$: the larger $p$, the slower the decrease. Black dashed lines indicate the slopes obtained from the fitting procedure. \textbf{C.} Scaling exponent of the number of interface points as determined in Panel B for different values of $p$ and $N$. Overall, the qualitative trend observed for $N = 12$ remains valid also for other values of $N$.}
\label{fig:SIinter}
\end{figure}

\begin{figure}[h!]\centering
\includegraphics[width=0.5\textwidth]{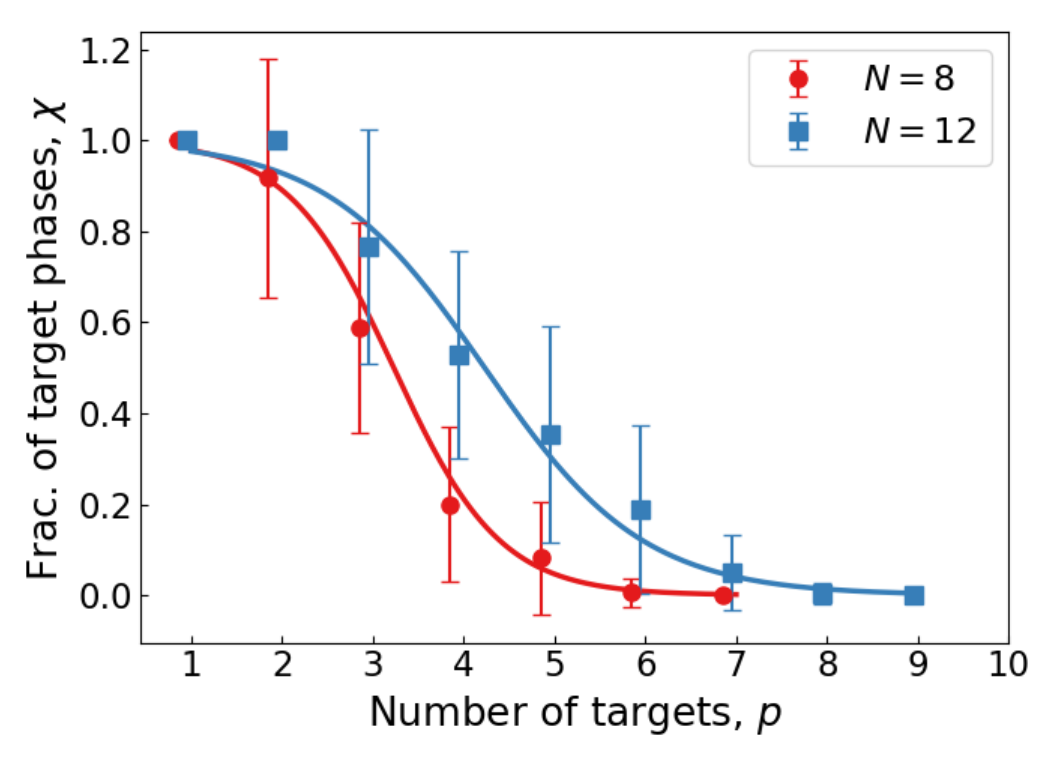}
\caption{Fraction of target phases $\chi$ (see definition in Sec. VIF and in the caption of Fig. 8 of the main text) for non-orthogonal, linearly independent targets.  The simulations were run for $N = 8$ and $N = 12$ components,  confirming  qualitatively that the capacity of the Hopfield liquid increases with increasing number of components $N$. Compared to the case of orthogonal targets presented in Panel B of Fig. 8, the mid-points of the sigmoids appear shifted to the left, highlighting a slightly less efficient retrieval. The parameters are the same as for the simulations of Fig. 8.}
\label{fig:SIcapacity_nonortho}
\end{figure}

\begin{figure}[h!]\centering
\includegraphics[width=1\textwidth]{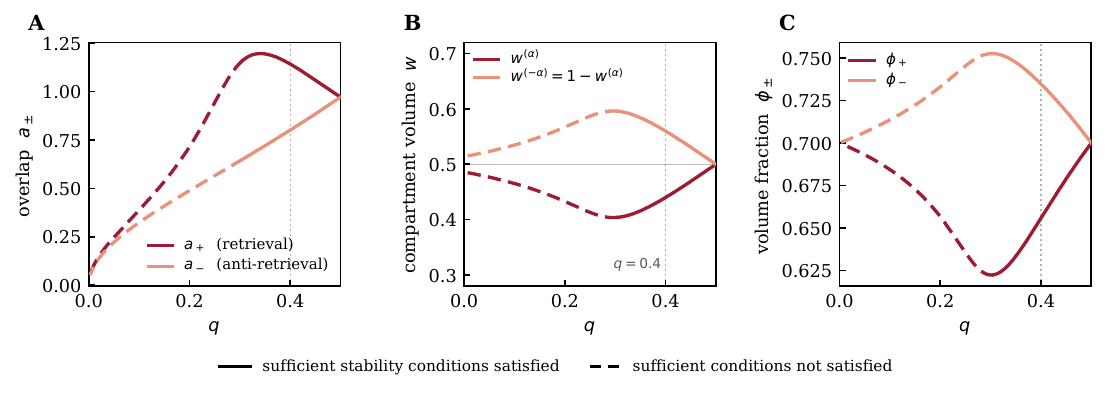}
\caption{Plots of the stationary solutions obtained by solving the equations (F5), (F6), and (F8).   The densities $\phi_+$ and $\phi_-$ are obtained from Eqs.~(F3) and~(F4). Parameters used are $\phi=0.7$, $v_2=8$, and $v_3=7$. Solid line indicates a solution that satisfies the sufficient conditions for stability (Eqs.~(G3) and~(G4)).  Dashed line is a stationary solution that does not satisfy the sufficient conditions.   The dotted line indicates the value of $q=0.4$ as used in Fig.~\ref{fig:q04}. } \label{fig:stabrhomin}   
\end{figure} 

\begin{figure}[h!]\centering
\includegraphics[width=\textwidth]{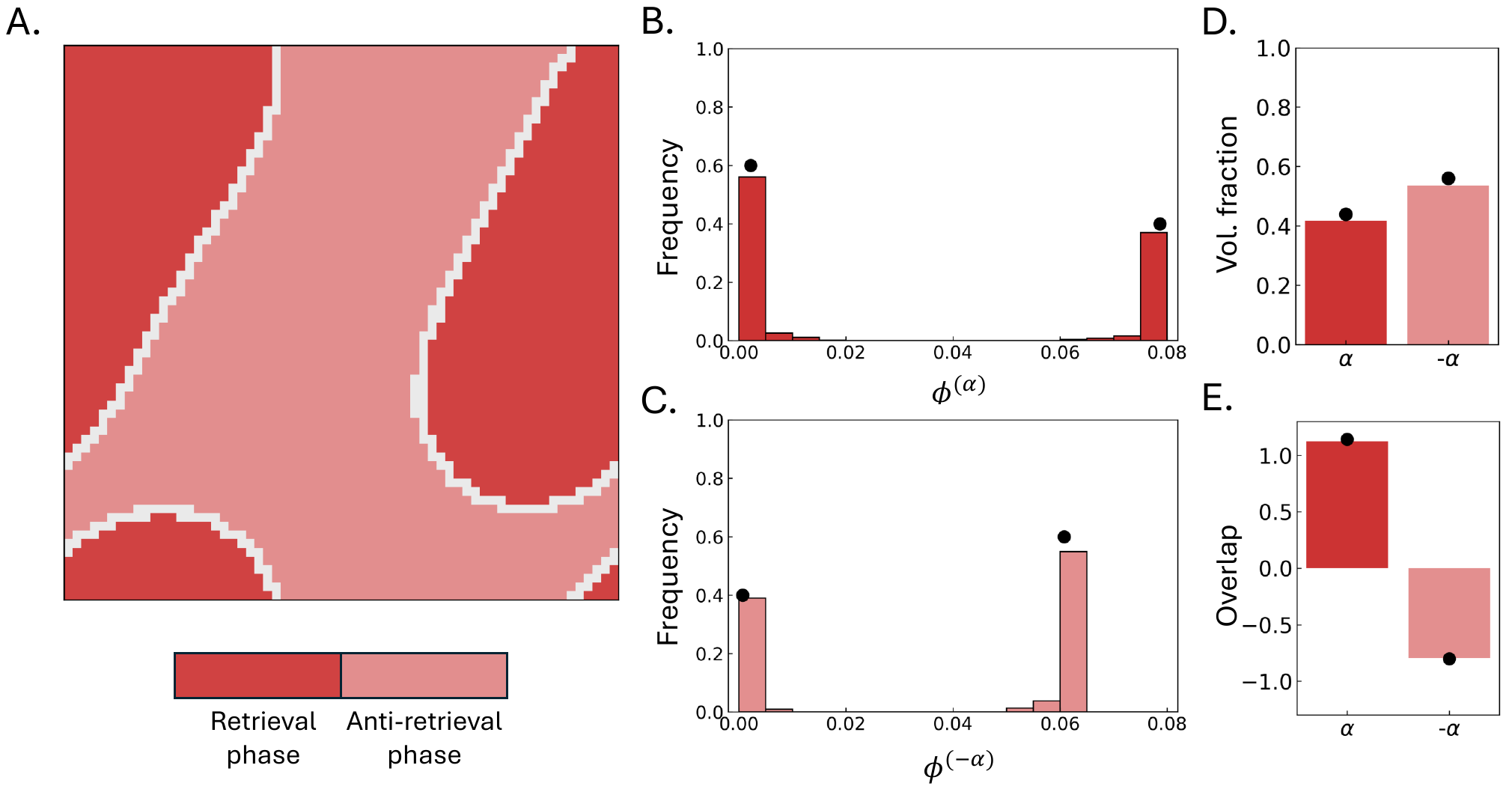}
\caption{Demixing in the liquid Hopfield model for  phases that are enriched in a fraction $q$ of components that is different from $1/2$.   Results are obtained by solving the adimensional Cahn-Hilliard equations   for $N = 20$, $p = 1$ and $q = 0.4$ (same procedure as for the other figures, and explained in~\ref{sec:NI}). The remaining parameters $\phi$, $v_2$ and $v_3$ are as in Fig.~\ref{fig:stabrhomin}. For these parameter values, the theoretical predictions from Eqs.~(F5),(F6) and~(F9) are $a_+ = 1.1420$, $a_- = 0.8007$, $w = 0.4400$. Using these values in Eqs.~(F3) and~(F4) leads to the predictions  $\phi_+ = 0.6557$ and $\phi_- = 0.7348$ for the total volume fraction of the retrieval and anti-retrieval phase, respectively.   The parameters used for the Cahn-Hilliard equations are  $\kappa = 0.005$, $\Delta \tilde{t} = 1.0 \times 10^{-8}$ and $\tilde{t} = 0.05$. \textbf{A}. Final snapshot of the Cahn-Hilliard simulation. A retrieval - anti-retrieval phase pair is identified, and colored in dark and light hues, respectively. Interface points are left white. \textbf{B.} Histogram for the composition of retrieval phase $(\alpha)$ (dark red). Black circles denote the theoretical prediction, \textit{i.e.} $(0.002215, 0.6)$ for depleted components and $(0.07864, 0.4)$ for enriched ones. \textbf{C.} Same as Panel B but for the anti-retrieval phase $(-\alpha)$: the theoretical predictions are in this case $(0.000711, 0.4)$ and $(0.06076, 0.6)$. \textbf{D.} Volume fraction of the retrieval phase $(\alpha)$ and anti-retrieval phase $(-\alpha)$. Black circles denote the theoretical prediction $w$ and $1-w$. \textbf{E.} Overlaps of the retrieval and anti-retrieval phase. Black circles denote analytical predictions, \textit{i.e.} $a_{+}$ for the retrieval phase $(\alpha)$ and $-a_-$ for the anti-retrieval phase $(-\alpha)$. Note that contrarily to the $q=1/2$ case, there is now an  asymmetry between the retrieval and the anti-retrieval phase.   The (non normalized)  target vector used in the simulation is $\vec{\xi} = (01010110001000110001)$.}
\label{fig:q04}
\end{figure}

\begin{figure}[h!]\centering
\includegraphics[width=0.9\textwidth]{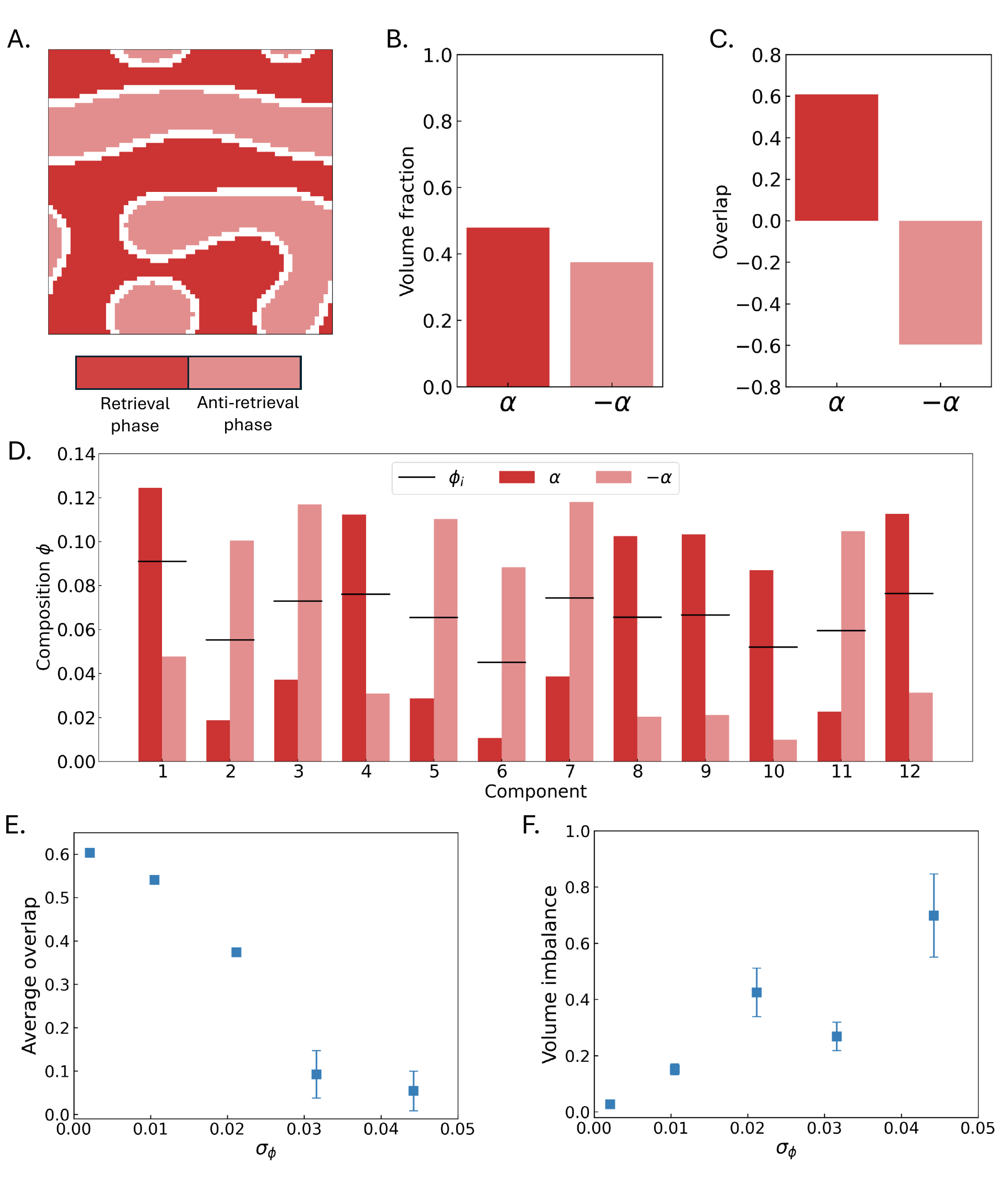}
\caption{\textit{Information retrieval in a non-equimolar mixture.} Cahn-Hilliard simulations for  mixtures composed of solute components that have distinct volume fractions, $\phi_i$, \textit{i.e.} the $\phi_i$ depend on $i$ (in the main text, $\phi_i=\phi/N$ for all $i$). The $\phi_i$ have been generated starting from a symmetric Dirichlet distribution: their average is $\phi/N$ and their standard deviation is $\sigma_{\phi}$. Simulations were performed using the same parameters as in Fig. 8 of the main text, \textit{i.e.} $v_2 = 4$, $v_3 = 3$ and $\phi = 0.8$; the number of components $N = 12$ and  the number of target vectors $p = 1$.   \textbf{A.} Final snapshot of a  simulation with $\sigma_{\phi} = 0.013$. The mixture is separated into a retrieval phase ($\alpha$) and an anti-retrieval phase, ($-\alpha$) colored respectively in dark and light red. Interface points are left white. \textbf{B.} Volume fractions of the retrieval and anti-retrieval phase. \textbf{C.} Overlap of the retrieval phase and the anti-retrieval phase with the target vector as defined in Eq.~(24) of the main text. \textbf{D.} Composition of  the retrieval phase and the anti-retrieval phase in the snapshot of Panel A. Solid black lines denote the value of $\phi_i$ for each of the 12 components. The target vector used in this simulation is $\vec{\gamma}^{(1)} = (+ - - + - - - + + + - +)$: this exact enrichment/depletion pattern is observed in the retrieval phase composition (dark red bars). \textbf{E.} Average overlap as a function of $\sigma_{\phi}$. The results are obtained by averaging over $30$ simulations for each value of $\sigma_{\phi}$, each characterized by a random composition. \textbf{F.} Volume imbalance, defined as $|w^{(\alpha)} - w^{(-\alpha)}|/w^{(\alpha)}$, as a function of $\sigma_{\phi}$.  Notice that, especially for larger values of $\sigma_{\phi}$, the mixture does not phase separate for every composition. The values used in Panels E and F are the ones pertaining to the cases in which the system actually undergoes spinodal decomposition. The error bars refer to the error of the mean. The simulation parameters are $\kappa = 0.001$, $\Delta \tilde{t} = 1.0 \times 10^{-8}$ and $\tilde{t} = 0.05$.}
\label{fig:SInonequim}
\end{figure}

\end{document}